\newif\ifsubmode
\submodefalse

\newif\ifprintfig
\printfigtrue

\newif\ifemulate
\emulatetrue

\ifsubmode
  \documentclass[12pt,preprint]{aastex}
  \received{}
  \accepted{}
  \journalid{}{}
  \articleid{}{}
\else
   \documentclass{emulateapj}
   \submitted{{\it Accepted for publication in the Astrophysical Journal}}
\fi

\usepackage{amsmath}

\newcommand{\kms}{\,km~s$^{-1}$}
\newcommand{\Msun}{\mbox{\,$M_{\odot}$}}

\def\lesssim{\mathrel{\hbox{\rlap{\hbox{\lower4pt\hbox{$\sim$}}}\hbox{$<$}}}}
\def\gtrsim{\mathrel{\hbox{\rlap{\hbox{\lower4pt\hbox{$\sim$}}}\hbox{$>$}}}}
\def\spose#1{\hbox to 0pt{#1\hss}}
\def\simlt{\mathrel{\spose{\lower 3pt\hbox{$\mathchar"218$}}
     \raise 2.0pt\hbox{$\mathchar"13C$}}}
\def\simgt{\mathrel{\spose{\lower 3pt\hbox{$\mathchar"218$}}
     \raise 2.0pt\hbox{$\mathchar"13E$}}}

\shorttitle{Diffuse Ionized Gas in NGC 891}
\shortauthors{Boettcher, Zweibel, Gallagher, and Benjamin}

\begin{document}

\title{Testing a Dynamical Equilibrium Model of the Extraplanar Diffuse
  Ionized Gas in NGC 891}

\author{Erin Boettcher$^{1,4}$, Ellen G. Zweibel$^{1,2}$,
  J. S. Gallagher III$^{1}$, and Robert A. Benjamin$^{1,3}$}

\affiliation{$^{1}$Department of Astronomy, University of Wisconsin - Madison,
  475 North Charter Street, Madison, WI 53706, USA; \texttt{boettche@astro.wisc.edu}} 
\affiliation{$^{2}$Department of Physics,
  University of Wisconsin - Madison, 475 North Charter Street, Madison, WI
  53706, USA} 
\affiliation{$^{3}$Department of Physics, University of
  Wisconsin - Whitewater, 800 West Main Street, Whitewater, WI 53190, USA}

\footnotetext[4]{Visiting astronomer, Kitt Peak
    National Observatory, National Optical Astronomy Observatory, which is
    operated by the Association of Universities for Research in Astronomy
    (AURA) under a cooperative agreement with the National Science
    Foundation.}


\ifsubmode\else
  \ifemulate\else
     \clearpage
  \fi
\fi


\ifsubmode\else
  \ifemulate\else
     \baselineskip=14pt
  \fi
\fi

\begin{abstract} 

  The observed scale heights of extraplanar diffuse ionized gas (eDIG) layers
  exceed their thermal scale heights by a factor of a few in the Milky Way and
  other nearby edge-on disk galaxies. Here, we test a dynamical equilibrium
  model of the extraplanar diffuse ionized gas layer in NGC 891, where we ask
  whether the thermal, turbulent, magnetic field, and cosmic ray pressure
  gradients are sufficient to support the layer. In optical emission line
  spectroscopy from the SparsePak integral field unit on the WIYN 3.5-meter
  telescope, the H$\alpha$ emission in position-velocity space suggests that
  the eDIG is found in a ring between galactocentric radii of $R_{min} \le R
  \le 8$ kpc, where $R_{min} \ge 2$ kpc. We find that the thermal
  ($\sigma_{th} = 11$ \kms) and turbulent ($\sigma_{turb} = 25$ \kms) velocity
  dispersions are insufficient to satisfy the hydrostatic equilibrium equation
  given an exponential electron scale height of $h_{z} = 1.0$ kpc. Using a
  literature analysis of radio continuum observations from the CHANG-ES
  survey, we demonstrate that the magnetic field and cosmic ray pressure
  gradients are sufficient to stably support the gas at $R \geq 8$ kpc if the
  cosmic rays are sufficiently coupled to the system ($\gamma_{cr} =
  1.45$). Thus, a stable dynamical equilibrium model is viable only if the
  extraplanar diffuse ionized gas is found in a thin ring around $R = 8$ kpc,
  and non-equilibrium models such as a galactic fountain flow are of interest
  for further study.

\end{abstract}


\section{Introduction}\label{sec_intro}

Multi-wavelength observations of nearby edge-on disk galaxies have revealed
multi-phase gaseous halos that include molecular, neutral, and warm and hot
ionized phases. The extraplanar diffuse ionized gas (eDIG) layers in the Milky
Way and other nearby edge-on disk galaxies are remarkable in that their
observed scale heights generally exceed their thermal scale heights by a
factor of a few \citep[e.g.,][]{Rand1997, Haffner1999, Collins2001,
  Gaensler2008, Voigtlander2013}. These warm ($T \sim 10^{4}$ K), diffuse
($\langle n_{e,0} \rangle \sim 0.1$ cm$^{-3}$) layers have a range of diffuse,
clumpy, and filamentary morphologies, a photoionization power requirement that
is satisfied by the O and B stars in the disk, and rotational velocity
profiles that suggest a probable disk origin \citep{Lehnert1995, Lehnert1996,
  Rossa2000, Tullmann2000, Miller2003b, Miller2003a, Rossa2003a, Rossa2003b,
  Heald2006b, Heald2006a, Heald2007}. Additionally, the detection of eDIG
layers is positively correlated with the star formation rate per unit area for
starburst, star-forming, and quiescent galaxies \citep{Rossa2003a}. It is also
spatially correlated with soft X-ray emission from hot halo gas
\citep{Strickland2004, Tullmann2006b, Tullmann2006a}, as well as with radio
continuum emission associated with extraplanar magnetic fields and cosmic rays
\citep{Dahlem1994, Collins2000, Tullmann2000, Li2016}.

The observation that eDIG layers are associated with a minimum star formation
rate per unit area is consistent with models of a star-formation driven
disk-halo flow. ``Superbubble'' \citep{MacLow1988}, ``galactic chimney''
\citep{Norman1989}, ``galactic fountain'' \citep{Shapiro1976}, and galactic
wind \citep[e.g.,][]{Veilleux2005} models all describe the local or global
circulation of gas between the disk and the halo due to star formation
activity in OB associations. There is observational evidence of bubbles, arcs,
and filaments in the halo that are spatially associated with HII regions as
well as ultraviolet continuum from young, hot stellar populations in the disk
\citep{Dettmar1990, Rand1990, Rand1996, Howk1997, Howk1999, Howk2000,
  Rossa2000, Rossa2003b, Tullmann2006b}. Thus, a general framework has emerged
in which eDIG layers are found in multi-phase gaseous, magnetic field, and
cosmic ray halos in galaxies with sufficient star formation rates per unit
area.

Within this framework, the vertical structure, support, and dynamical state of
these layers are not yet fully understood. A range of dynamical models exist
to explain the column densities, scale heights, and three-dimensional
kinematics of the extraplanar interstellar medium (ISM). One class of models
treats extraplanar gas layers as fluid disks that satisfy the hydrostatic
equilibrium equation \citep{Boulares1990, Barnabe2006, Henriksen2016}, while
another treats the layers as collections of clouds that travel ballistically
through the galactic gravitational potential \citep{Collins2002,
  Fraternali2006}. Some authors suggest that a combination of hydrodynamic and
ballistic effects may be closest to reality \citep[e.g.,][]{Benjamin2000},
while others seek to understand the effects of magnetohydrodynamics on
the disk-halo interface in a turbulent, star-forming ISM
\citep{Hill2012}. Discriminating between dynamical models for each phase of
the extraplanar ISM is important for undestanding how each phase is formed,
evolves, and participates in the transfer of mass and energy between the disk,
halo, and intergalactic environment. 

Here, we study the dynamical state of the eDIG layer in the nearby edge-on
disk galaxy NGC 891. This galaxy is an ideal candidate for this study due to
its proximity ($D = 9.9$ Mpc; $1" = 48$ pc; \citealt{Ciardullo1991}),
inclination angle ($i > 89^{\circ}$; \citealt{Oosterloo2007}), and
well-studied multi-phase gaseous halo. It is classified as an Sb galaxy in the
Third Reference Catalogue of Bright Galaxies \citep{deVaucouleurs1991}, but
there is evidence at multiple wavelengths for a bar \citep[e.g.,][]{Sofue1993,
  GarciaBurillo1995, SchechtmanRook2013}. Due to similarities in mass,
morphology, and bolometric luminosity, NGC 891 is often considered a Milky Way
analog \citep{vanderKruit1984}; however, the far-infrared star formation rate
is somewhat higher in the former at $3.8$ \Msun yr$^{-1}$
\citep{Popescu2004}. There is evidence in the HII region number density as
well as the far-infrared and radio emission morphology that the star formation
rate is highest in the northeast side of the disk
\citep[e.g.,][]{Wainscoat1987, Dettmar1990}. There is not evidence of a major
disturbance of the stellar disk in deep optical and near-infrared photometry
\citep{Morrison1997, SchechtmanRook2013}. However, \citet{Mapelli2008} show
that the slight lopsidedness of the disk suggests a mild flyby interaction
with the companion UGC 1807. \citet{Oosterloo2007} demonstrate that HI clouds
with counter-rotating velocities and an HI filament near systemic velocity
reaching over 20 kpc from the disk in projection towards the companion are
evidence of interaction and/or accretion. This system also includes
extraplanar dust \citep{Howk1997, Howk2000, Seon2014}, diffuse ionized gas
\citep{Dettmar1990, Rand1990, Rand1997}, and hot ionized gas
\citep{HodgesKluck2013}, as well as extraplanar magnetic fields and cosmic
rays \citep{Dahlem1994}.

The eDIG in NGC 891 is among the brightest, most spatially extended, and most
well-studied eDIG layers known. Discovered in H$\alpha$ narrowband imaging by
\citet{Dettmar1990} and \citet*{Rand1990}, the brightest and most vertically
extended eDIG is found on the northeast side of the galaxy, where it appears
to be spatially associated with the elevated star formation rate
\citep{Dettmar1990, Hoopes1999}. The morphology of the layer has both smooth
and filamentary components; \citet{Rossa2004} obtained high spectal resolution
($0.1'' = 4.8$ pc) H$\alpha$ narrowband imaging with the WFPC2 camera on the
\textit{Hubble Space Telescope} that revealed a diffuse background intersected
by filaments, arcs, plumes, bubbles, and supershells. Notably,
\citet{Howk2000} and \citet{Rossa2004} detect arcs and filaments that have
dimensions of tens of pc wide and several kpc long, are highly collimated to
large heights above the disk, and appear to have one or both ends in
star-forming regions. Qualitatively, these observations suggest that star
formation activity drives the warm ionized gas out of the disk by way of
galactic chimneys formed from the bursting of superbubbles associated with
spatially correlated supernovae \citep{Shapiro1976, Norman1989}.

The photoionization energy requirement of the eDIG layer is met by massive
stars in the disk if the ISM is sufficiently porous to UV photons
\citep[e.g.,][]{Dettmar1990}.  The emission line spectrum is broadly
consistent with a photoionized gas in the near-ultraviolet \citep{Otte2001,
  Otte2002}, optical \citep[e.g.,][]{Rand1997, Rand1998}, and infrared
\citep{Rand2008b, Rand2011}. However, in NGC 891, the Milky Way, and other
galaxies, the emission line ratios as a function of height above the disk
requires a supplemental source of heating and/or ionization; such sources may
include shocks \citep{Rand1998}, turbulent mixing layers \citep{Rand1998,
  Binette2009}, hot low-mass evolved stars \citep{Sokolowski1991,
  FloresFajardo2011}, and/or cosmic rays \citep{Wiener2013}.

A remarkable feature of the eDIG layer in NGC 891 is its considerable spatial
extent above and below the midplane. The vertical electron density
distribution is well-described by an exponential of the form $\langle n_{e}(z)
\rangle = \langle n_{e,0} \rangle e^{-|z|/h_{z}}$, where $\langle n_{e,0}
\rangle $ is the mean electron number density in the disk and $h_{z}$ is the
electron scale height. The eDIG layer in NGC 891 is well-fit by a scale height
of $h_{z} = 1.0$ kpc on the Northeast side of the disk \citep{Dettmar1990,
  Rand1990, Dettmar1991, Keppel1991}. An improved fit is found if the electron
density distribution is expressed as the sum of a thick disk component with
$h_{z,disk} = 1.0$ kpc and a halo component with $h_{z,halo} \sim$ a few kpc
\citep{Rand1997, Hoopes1999}. The thick disk and halo components may be
produced by different processes; for example, the former may be rising out of
the disk via galactic chimneys and supershells, while the latter may be
condensing onto the disk out of a hot halo phase \citep{Rand1997}. Changing
emission line ratios with distance from the midplane suggest that the large
scale height is due to true extraplanar emission and not to HII region
emission scattered by dust. Additionally, \citet{Ferrara1996} use Monte Carlo
radiative transfer simulations of H$\alpha$ photon propagation through the
dusty disk of NGC 891 to argue that scattered HII region emission is only 10\%
of eDIG emission at $z = 600$ pc.

Here, we test a dynamical equilibrium model of the eDIG layer in NGC
891. Although the observed scale height, lack of flaring, and general
inhomogeneity of the layer suggest a system out of dynamical equilibrium
\citep[e.g.,][]{Dettmar1990}, the various sources of vertical support have yet
to be fully quantified for any eDIG layer. Thus, we use optical emission line
spectroscopy from the SparsePak integral field unit (IFU;
\citealt{Bershady2004, Bershady2005}) on the WIYN 3.5-meter telescope at Kitt
Peak National Observatory, as well as radio continuum observations from the
CHANG-ES survey from \citet[][in preparation]{Schmidt2016}, to determine the
thermal, turbulent, magnetic field, and cosmic ray pressure gradients in the
eDIG layer. By comparing the observed and required pressure gradients to
support the eDIG layer at its observed scale height, we consider whether the
system is best characterized by equilibrium or non-equilibrium (i.e., galactic
fountain, galactic wind) models.

The paper is laid out as follows. In \S2, we create a mass model to determine
the galactic gravitational potential of NGC 891. We give a statement of the
problem and the model to be tested in \S3, and we discuss the collection and
reduction of optical emission line spectroscopy using the SparsePak IFU in
\S4. In \S5.1, we construct a model of the three-dimensional density
distribution of the eDIG layer, and we constrain the velocity dispersion along
the minor axis from the H$\alpha$, [NII] $\lambda$6583, and [SII]
$\lambda$6716, 6731 emission line widths in \S5.2. In \S5.3, we determine the
vertical magnetic field and cosmic ray pressure gradients in this system using
radio continuum observations from the CHANG-ES survey analyzed by \citet[][in
preparation]{Schmidt2016}, and in \S5.4 we assess whether a magnetized eDIG
layer in dynamical equilibrium is stable against the Parker instability. We
discuss our results in the context of multi-wavelength observations and our
knowledge of the Milky Way Galaxy in \S6. In \S7, we conclude that a dynamical
equilibrium model dominated by a magnetic pressure gradient is viable
for the eDIG layer in NGC 891 only over a limited range of
  galactocentric radii ($R \sim 8$ kpc). We include an Appendix to illustrate
the robustness of this result against variations in the assumed mass-to-light
ratio of the stellar disk.

\section{A Mass Model for NGC 891}\label{sec_mass_mod}

To create a mass model of NGC 891, we use the HI rotation curve of
\citet{Fraternali2011}, as well as the careful photometric deconstruction of
the disk by \citet{SchechtmanRook2014} (see also \citealt{Popescu2000,
  Popescu2004}). \citet{SchechtmanRook2014} use sub-arcsecond spatial
resolution imaging of NGC 891 in the near-infrared, as well as radiative
transfer modeling to perform a dust attenuation correction, to decompose the
stellar disk into five exponential disk components: a super-thin, thin, and
thick disk (truncated within $R = 2.8$ kpc), as well as a central disk and a
bar (truncated outside $R = 2.8$ kpc). A super-thin disk of stars and dust was
also found for this galaxy by \citet{Popescu2004}.

\begin{figure}[h]
\epsscale{1.2}\plotone{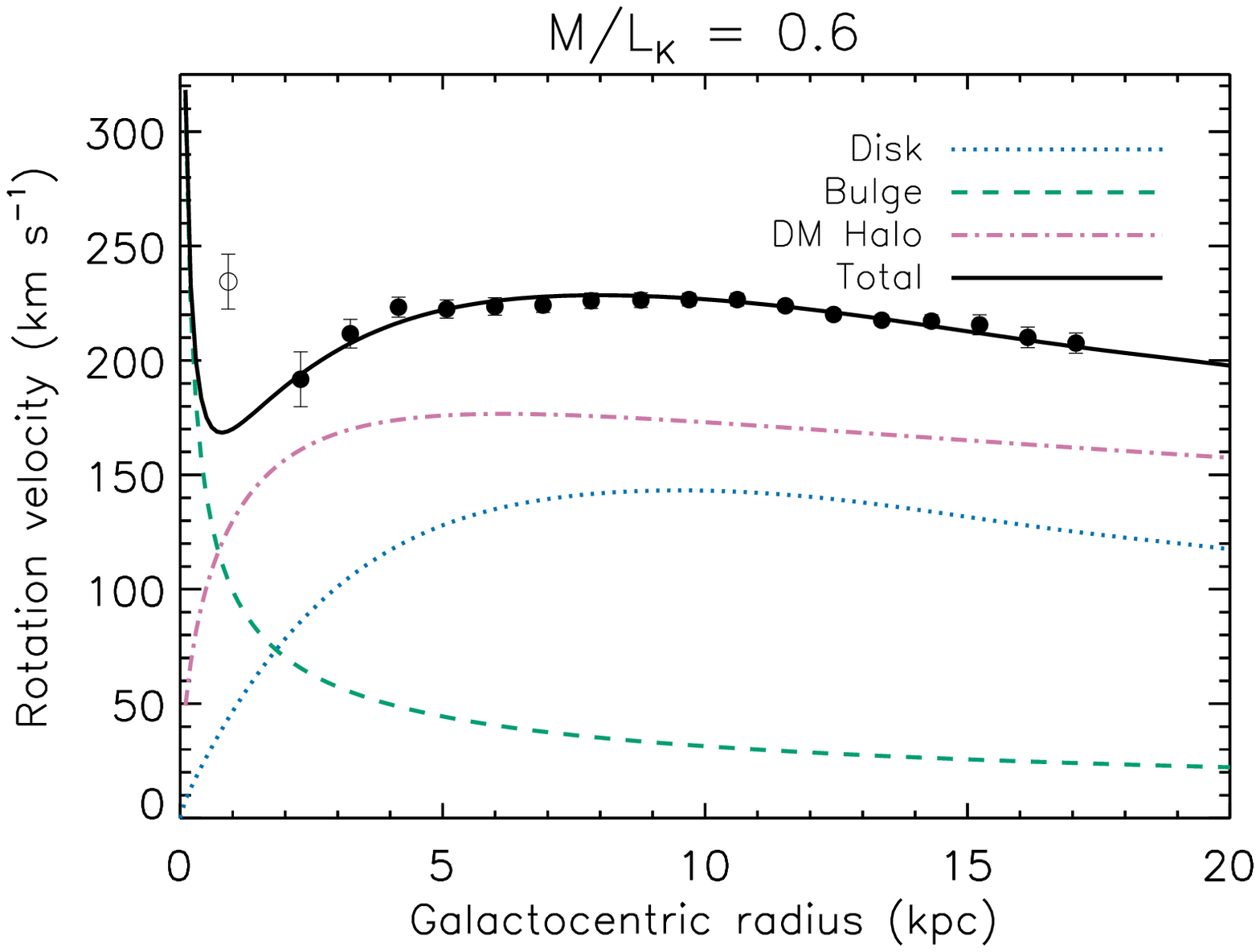}
\caption{The rotation curve resulting from our mass model of NGC 891 as
  compared to the HI rotation curve of this galaxy from
  \citet{Fraternali2011}. Our mass model consists of an exponential disk and
  bulge and an NFW dark matter halo whose properties are given in Table 1. The
  baryonic mass distribution is derived from the near-infrared photometry of
  \citet{SchechtmanRook2014}, and assumes a $K$-band mass-to-light ratio of
  $M/L_{K} = 0.6$. The data point at the smallest galactocentric radius is
  excluded from the fit because of uncertainties in the kinematics due to the
  bar at this location.}
\end{figure}

For the sake of computational simplicity, we reduce the five-component model
of \citet{SchechtmanRook2014} to a two-component, untruncated disk and
bulge/lens model of the form
\begin{equation}
  \rho(z, R) = \rho_{0,D}e^{-R/h_{R,D}}e^{-|z|/h_{z,D}} + \\
  \rho_{0,B}e^{-R/h_{R,B}}e^{-|z|/h_{z,B}},
\end{equation}
where $\rho$ is the baryonic mass density, $\rho_{0}$ is the central mass
density, $h_{R}$ is the radial scale length, and $h_{z}$ is the vertical scale
height of the disk (D) and bulge (B). This is done by fitting the
attenuation-corrected, $K$-band luminosity surface density of their model as
functions of $R$ and $z$ to determine the radial scale lengths, vertical scale
heights, and central luminosity densities of the disk and bulge. We then
assign a $K$-band mass-to-light ratio to determine the central baryonic mass
densities of each component. For the main mass model discussed in this paper
(the ``primary model''), we choose a $K$-band mass-to-light ratio of $M/L_{K}
= 0.6$, and we consider variations in $M/L_{K}$ in the Appendix
\citep{McGaugh2014}.

To determine the parameters of the dark matter halo, we fit the HI rotation
curve of \citet{Fraternali2011} using a reduced $\chi^{2}$ minimization. We
assume an NFW profile of the form
\begin{equation}
  \rho_{DM}(R) = \frac{\rho_{0,DM}}{\frac{R}{a_{DM}} \big(1 + \frac{R}{a_{DM}}
    \big)^{2}},
\end{equation}
where $\rho_{0,DM}$ is the central dark matter density and $a_{DM}$ is the
scale radius. Note that while gas is not explicitly included in our mass
model, this is a $\sim 10\%$ effect on the velocity dispersion required to
support the gas at a given scale height. See Table 1 for the parameters of our
mass model.

\begin{deluxetable}{lcccc}
\tabletypesize{\scriptsize}
\tablecaption{NGC 891 Primary Mass Model ($M/L_{K}$ = 0.6)}
\tablewidth{0pt}
\tablehead{ 
\colhead{Parameter} &
\colhead{Value} &
\colhead{Reference}
}
\startdata
$h_{R,D}$ & 4.1 kpc & 1\\
$h_{z,D}$ & 0.4 kpc & 1\\
$\rho_{0,D}$ & $0.09 \times 10^{10}$ \Msun kpc$^{-3}$ & 1\\
$h_{R,B}$ & 0.3 kpc & 1\\
$h_{z,B}$ & 0.1 kpc & 1\\
$\rho_{0,B}$ & $8.32 \times 10^{10}$ \Msun kpc$^{-3}$ & 1\\
$a_{DM}$ & 2.9 kpc & 2\\
$\rho_{0,DM}$ & $0.03 \times 10^{10}$ \Msun kpc$^{-3}$ & 2
\enddata
\tablecomments{The radial scale lengths, $h_{R}$, the vertical scale heights,
  $h_{z}$, and the central mass densities, $\rho_{0}$, determined for an
  exponential disk (D) and bulge (B) model of NGC 891. The results are derived
  from the near-infrared photometry of \citet{SchechtmanRook2014} assuming a
  $K$-band mass-to-light ratio of $M/L_{K} = 0.6$. We also list the scale
  radius, $a_{DM}$, and the central density, $\rho_{0,DM}$, of the dark matter
  halo required to reproduce the HI rotation curve of \citet{Fraternali2011}
  assuming an NFW profile.} 
\tablerefs{[1] \citet{SchechtmanRook2014}; [2] This work}
\end{deluxetable}

We adopt the approach of \citet{Cuddeford1993} to calculate the gravitational
potential $\Phi(R,z)_{D,B}$ of the exponential disk and bulge components as follows:
\begin{equation}
\begin{split}
  \Phi(R,z)_{D,B} = -\frac{4G\Sigma_{0}}{h_{R}} \int_{-\infty}^{\infty}
  \mathrm{d}z^{\prime} \, \Big(e^{-|z^{\prime}|/h_{z}} \times \\
  \int_{0}^{\infty} \mathrm{d}a \, \arcsin \Big(\frac{2a}{S_{+} +
    S_{-}} \Big)aK_{0} \Big(\frac{a}{h_{R}} \Big) \Big),
\end{split}
\end{equation}
where $\Sigma_{0}$ is the central mass surface density, $K_{0}$ is the zeroth
order modified Bessel function, and $S_{\pm} \equiv \sqrt{(z - z^{\prime})^{2}
  + (a \pm R)^{2}}$. The potential $\Phi(R,z)_{DM}$ of the dark matter halo is
given by:
\begin{equation}
  \Phi(R,z)_{DM} = -4 \pi G\rho_{0,DM}a_{DM}^{2}\frac{ln(1 + R/a_{DM})}{R/a_{DM}}.
\end{equation}
From the gravitational potential of the disk, bulge, and
dark matter halo, we determine the circular velocity due to each component,
and add these in quadrature to reproduce the HI rotation curve shown in
Figure 1.

Our mass model differs from that of \citet{Fraternali2006}, most noteably in
that the scale height of our disk ($h_{z,D} = 0.4$ kpc) is only $\sim 40\%$ of
theirs ($h_{z,D} = 1.05$ kpc). As the gravitational potential
within a few kpc of the disk is particularly important for the dynamical
equilibrium of the eDIG layer, we use the new near-infrared imaging and
radiative transfer modeling of \citet{SchechtmanRook2014} to find a more
conservative value of $h_{z,D}$.  

\section{The Scale Height Problem}\label{sec_thsh}

\begin{figure}[h]
\epsscale{1.2}\plotone{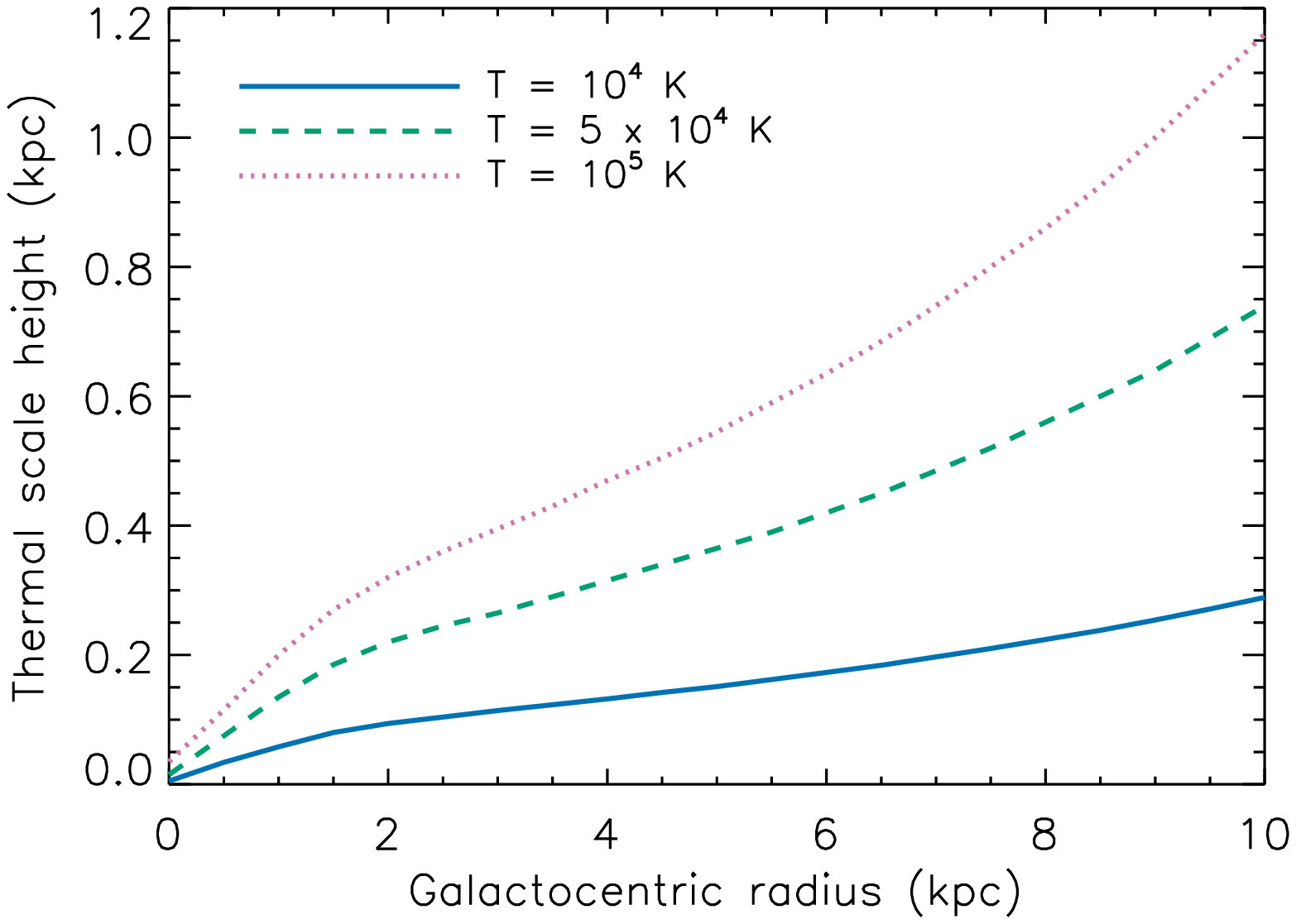}
\epsscale{1.2}\plotone{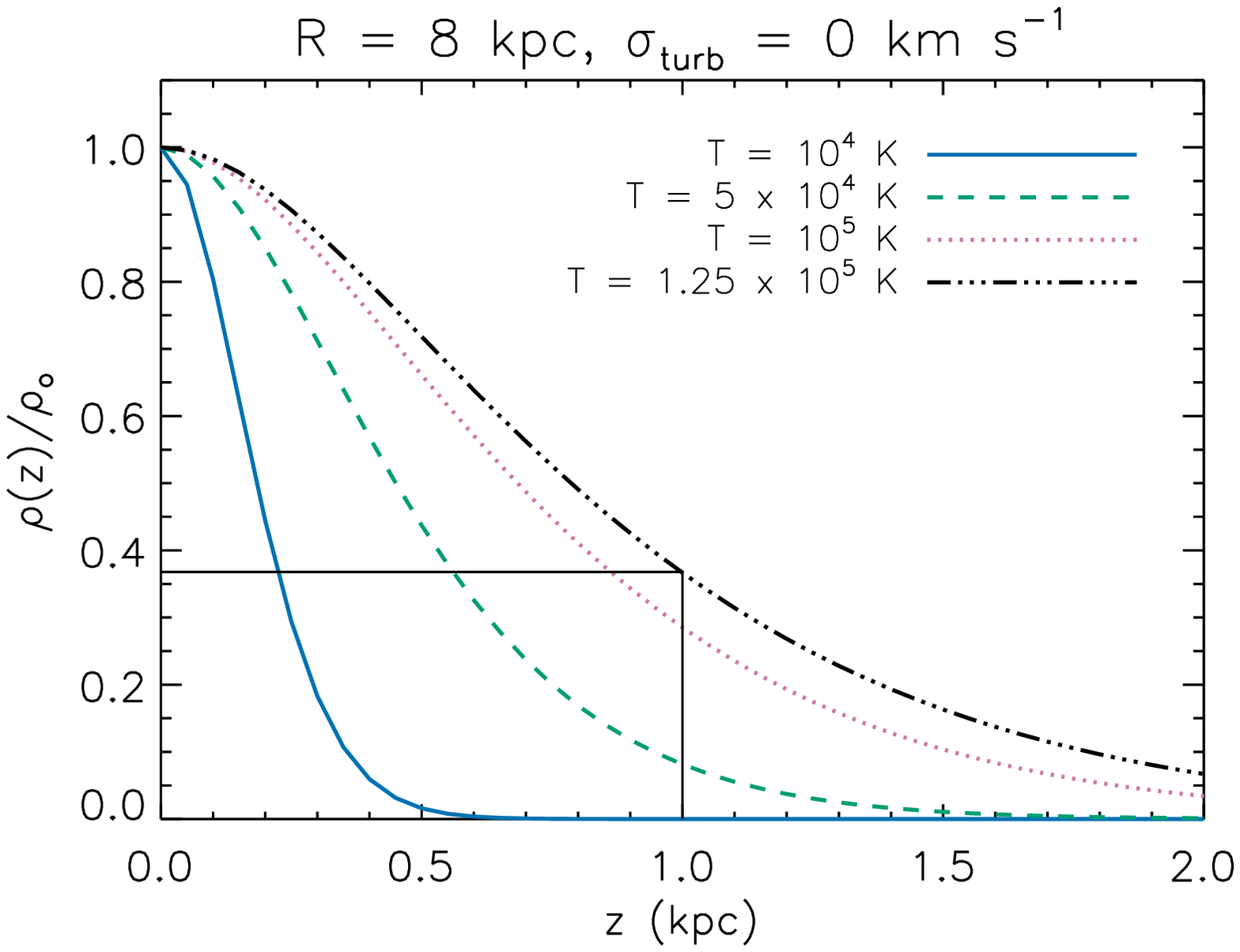}
\epsscale{1.2}\plotone{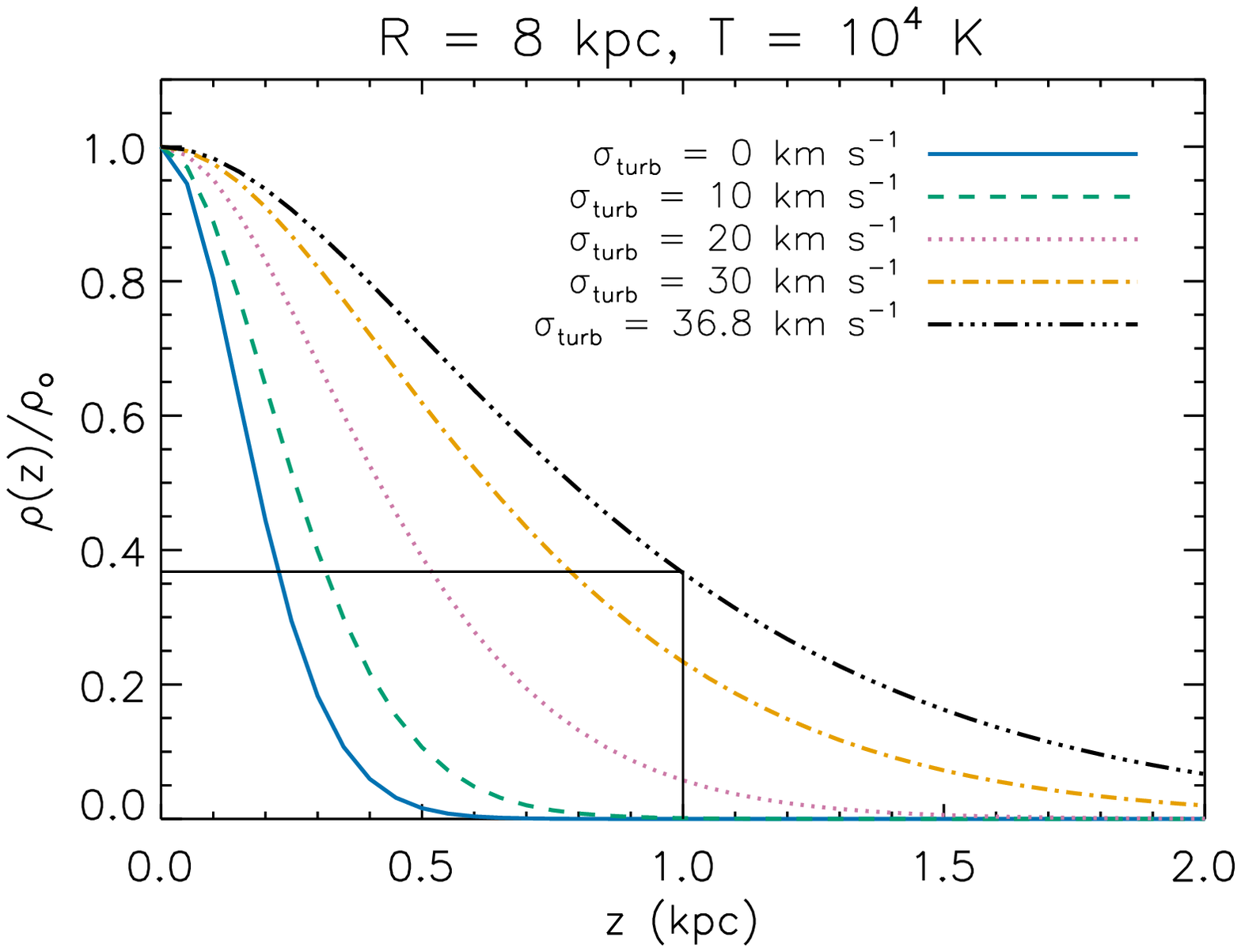}
\caption{In 2\textit{a}, the exponential electron scale height of an
  isothermal gas with a range of temperatures is shown as a function of
  galactocentric radius in NGC 891. The gas temperature must be $T \gtrsim
  10^{5}$ K to achieve a scale height $h_{z} = 1$ kpc comparable to the
  observed value for this system. In 2\textit{b} and 2\textit{c}, the
  isothermal gas density is considered for a thermally supported (\textit{b}),
  and thermally and turbulently supported (\textit{c}), gas layer as a
  function of height above the disk. A temperature of $T = 1.25 \times 10^{5}$
  K ($\sigma_{turb} = 0$ km s$^{-1}$), or a turbulent velocity dispersion
  $\sigma_{turb} = 36.8$ km s$^{-1}$ ($T = 10^{4}$ K) is required to produce a
  scale height $h_{z} = 1$ kpc at the observed cutoff radius of $R = 8$ kpc.}
\end{figure}

Throughout this study, we assume that the eDIG layer is in pressure balance,
and we solve the hydrostatic equilibrium equation given by
\begin{equation}
  \frac{dP(z,R)}{dz} = -\frac{d\Phi (z,R)}{dz}\rho (z,R),
\end{equation}
where $z$ and $R$ are the height and radial cyclindrical coordinates,
respectively. Here, $\rho (z,R)$ is the gas density, $\Phi (z,R)$ is the
galactic gravitational potential, and $\frac{d\Phi (z,R)}{dz} = g_z(z,R)$ is
the gravitational acceleration in the $z$ direction. The total pressure
$P(z,R)$ is the sum of the gas pressure, magnetic field pressure, and cosmic
ray pressure:
\begin{equation}
  P(z,R) = P_{g} + P_{B} + P_{cr}.
\end{equation}
The magnetic field pressure is assumed to be isotropic. We use an equation of
state of the form
\begin{equation}
  P_{g}(z,R) = \sigma^{2} \rho(z,R),
\end{equation} 
where $\sigma^{2} = \sigma_{th}^{2} + \sigma_{turb}^{2}$ is the quadrature sum
of the one-dimensional thermal and turbulent velocity disperisons. Note that
here and throughout the rest of the paper, the velocity dispersion refers to
the standard deviation, and not the full width at half maximum (FWHM), of a
Gaussian velocity distribution. We also use ``turbulence'' to refer simply to
random gas motions without assuming that they meet the strict definition of
turbulence.

Given a mass model of NGC 891, we solve Equation (5) to determine the vertical
scale height, $h_{z}$, of an isothermal gas with an equation of state given by
Equation (7). The general solution is:
\begin{equation}
  \frac{\rho(z)}{\rho(0)} = \frac{\sigma^{2}(0)}{\sigma^{2}(z)}exp \Big \{- \int_{0}^{z}
  \Big (\frac{\mathrm{d}z'}{\sigma^{2}(z')}\frac{d\Phi (z')}{dz'} \Big ) \Big \}.
\end{equation}
Assuming that $\sigma$ is independent of $z$ (see \S5.2), the simplified
solution is:
\begin{equation}
  \rho(z) = \rho(0) e^{-\sigma^{-2} \Phi (z')|_{z' = 0}^{z' = z}}.
\end{equation}
We define the scale height, $h_{z}$, as the distance from the midplane at which
the gas density has dropped by a factor of $e$:
\begin{equation}
  \frac{\rho(z = h_{z})}{\rho_{0}} = e^{-1}.
\end{equation}
Note that this calculation temporarily neglects any magnetic field and cosmic
ray pressure.

We can ask what the scale height of an isothermal gas layer is for a given
velocity dispersion as a function of galactocentric radius. First, we concern
ourselves only with the thermal velocity dispersion, $\sigma_{th} =
\sqrt{\frac{kT}{\alpha m_{p}}}$, where $k$ is the Boltzmann constant, $T$ is
the gas temperature, $m_{p}$ is the mass of a proton, and $\alpha$ is a
scaling factor based on the composition and the ionization state. For a
diffuse ionized gas with $T = 10^{4}$ K and $\alpha = 0.7$, we find
$\sigma_{th} = 11$ km s$^{-1}$ (this assumes a gas that is 9\% He by
number, with 90\% and 70\% of the H and He ionized, respectively;
\citealt{Rand1997, Rand1998}).

In Figure 2\textit{a}, we show the thermal scale height of an isothermal gas
layer for a range of temperatures as a function of galactocentric radius. Two
characteristics of a thermally-supported gas layer are immediately
apparent. First, the layer is highly flared. Second, the layer has a scale
height of only a few hundred parsecs within $R = 10$ kpc when $T = 10^{4}$ K,
a factor of a few smaller than the $h_{z} = 1$ kpc observed for the eDIG layer
in this galaxy \citep[e.g.,][]{Rand1997}. In fact, the scale height only
reaches $h_{z} = 1$ kpc if $T \sim 10^{5}$ K, an order of magnitude higher
than the temperature of the eDIG layer in this system.

This is further illustrated in Figure 2\textit{b}, where the density
distribution of the isothermal gas is shown as a function of height above the
disk for a range of temperatures at a radius of $R = 8$ kpc. A radius of $R =
8$ kpc is chosen because this is approximately the location of the observed
cutoff in H$\alpha$ intensity, and thus is the shallowest location in the
gravitational potential where significant H$\alpha$ emission is observed (see,
e.g., Figure 2 of \citealt{Dettmar1990}). A scale height of $h_{z} = 1$ kpc is
achieved only for a temperature of $T = 1.25 \times10^{5}$ K; thus, it is
clear that the eDIG layer in NGC 891 is not thermally supported, and requires
supplemental sources of pressure support if it is in dynamical equilibrium.

We consider one source of additional pressure support in Figure 2\textit{c},
where we now define the velocity dispersion of an isothermal gas as
$\sigma^{2} = \sigma_{th}^{2} + \sigma_{turb}^{2}$. Here, for a temperature of
$T = 10^{4}$ K, we consider the density distribution of an isothermal gas with
a range of turbulent velocity dispersions as a function of height above the
disk at $R = 8$ kpc. We see that a scale height of $h_{z} = 1$ kpc is achieved
only for a turbulent velocity dispersion of $\sigma_{turb} \sim 37$ km
s$^{-1}$, or turbulence with a sonic Mach number of 3 to 4 in a $T = 10^{4}$ K
gas. Thus, observationally constraining the velocity dispersion in the eDIG
layer of NGC 891 is a major goal of this work.

\begin{deluxetable*}{cccccccc}[t]
\tabletypesize{\scriptsize}
\tablecolumns{8}
\tablewidth{0pt}
\tablecaption{NGC 891 Observing Summary}
\tablehead{ 
\colhead{Pointing} &
\colhead{R.A. \tablenotemark{a}} &
\colhead{Decl. \tablenotemark{a}} &
\colhead{Exposure Time \tablenotemark{b}} &
\colhead{rms Continuum Noise \tablenotemark{c}} &
\colhead{Array P.A.} &
\colhead{$R'$ \tablenotemark{d}} &
\colhead{$z$ \tablenotemark {d}}\\
\colhead{Label} &
\colhead{(J2000)} &
\colhead{(J2000)} &
\colhead{(hr)} &
\colhead{($10^{-17}$ erg s$^{-1}$ cm$^{-2}$ \AA$^{-1}$)} &
\colhead{(deg)} &
\colhead{(kpc)} &
\colhead{(kpc)}
}
\startdata
p1 & 02 22 36.2 & +42 20 44 & 1.67 & 1.8 & +112 & -1.65 - 1.65 &  -3.2 - 0\\
p2 & 02 22 30.6 & +42 21 10 & 2.22 & 1.6 & -68 & -1.65 - 1.65 & 0 - 3.2\\
p3 & 02 22 34.8 & +42 23 00 & 2.22 & 1.5 & -68 & 4.05 - 7.35 & 0 - 3.2\\
p4 & 02 22 40.5 & +42 22 35 & 2.22 & 1.4 & +112 & 4.05 - 7.35 & -3.2 - 0
\enddata
\tablenotetext{a}{R.A. is measured in hours, minutes, and seconds; Decl. is
  measured in degrees, arcminutes, and arcseconds. The given R.A. and
  Decl. are for the central fiber in the near-integral core (fiber 52).}
\tablenotetext{b}{Total exposure time of a stack of four 1500 s or 2000 s
  exposures.}
\tablenotetext{c}{Median rms noise measured in the continuum between [NII]
  $\lambda$6583 and [SII] $\lambda$6716.}
\tablenotetext{d}{Range of projected radii and heights above the disk spanned
  by the fiber array (non-sky fibers only). A positive $R'$ is on the
  approaching (north) side of the disk; a positive $z$ is on the west
  side of the disk. Assumes a distance of $D = 9.9$ Mpc.}
\end{deluxetable*}

\section{Observations}\label{sec_data}

\subsection{Data Collection}\label{sec_datacoll}

Optical emission line spectroscopy of the eDIG layer in NGC 891 was obtained
on the WIYN\footnote[6]{The WIYN Observatory is a joint facility of the
  University of Wisconsin-Madison, Indiana University, the National Optical
  Astronomy Observatory, and the University of Missouri.} 3.5-meter telescope
at KPNO on 2014 November 15 - 17. The SparsePak IFU \citep{Bershady2004,
  Bershady2005} was used in conjunction with the STA1 CCD detector and the
Bench Spectrograph Camera. SparsePak is a fiber array consisting of 82 fibers;
the array sparsely samples the field around a nearly integral core. The 500
$\mu$m fibers span $4.7"$ on the sky, or $226$ pc at the distance of NGC
891. The 316@63.4 echelle grating was used at order 8 (grating angle =
62.840$^{\circ}$); this produces wavelength coverage from $\sim 6380$ \AA $\,
-$ 6805 \AA $\,$ with a dispersion of $0.21$ \AA/pixel. The spectral
resolution is $R = 7770$, or $\sigma = 0.36$ \AA $\,$ ($\sigma = 17$ km
s$^{-1}$) at H$\alpha$. The wavelength coverage includes the [NII]
$\lambda$6548, 6583, H$\alpha$, and [SII] $\lambda$6716, 6731 emission lines.

We obtained observations of NGC 891 at four SparsePak pointings. Two pointings
sample the extraplanar gas above and below the disk along the minor axis, and
the other two pointings are shifted along the major axis to a projected radius
of $R' = 5.7$ kpc on the Northeast side of the galaxy. These pointings are shown
projected onto an image of NGC 891 from the Digitized Sky Survey (Second
Generation) in Figure 3. In Table 2, the pointing centers, exposure times,
continuum noise, position angles, and ranges sampled in $R'$ and $z$ are given
(throughout this paper, $R'$ will be used to refer to projected radius, and
$R$ to galactocentric radius).

\begin{figure}[h]
\epsscale{1.2}\plotone{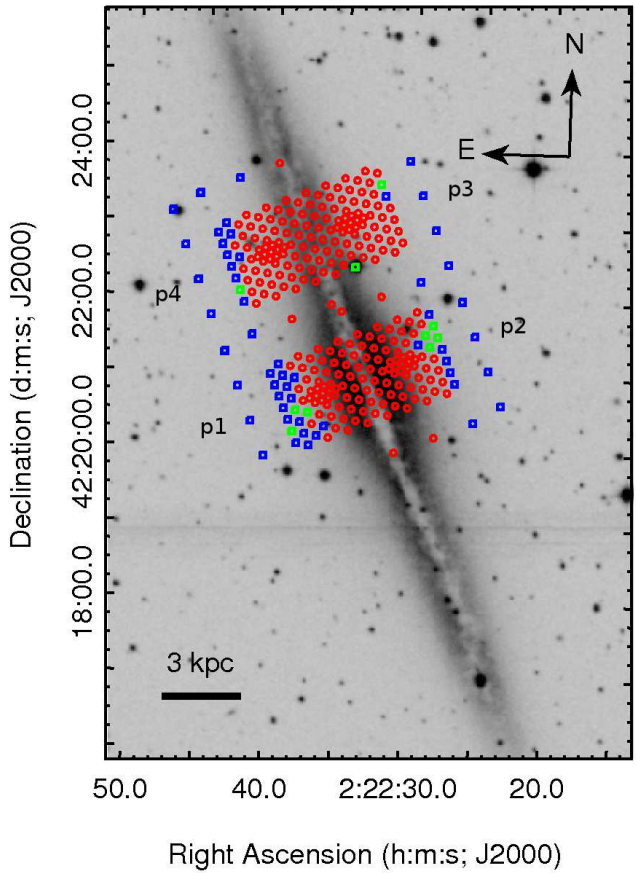}
\caption{The four SparsePak IFU pointings observed overplotted on a red image
  of NGC 891 from the Digitized Sky Survey (Second Generation). Two pointings
  sample the minor axis of the galaxy, while the other two are offset along
  the major axis by $R' = 5.7$ kpc. The red fibers indicate locations where
  spectra were obtained with positive eDIG detections, blue indicate locations
  with no eDIG detection, and green mark contamination by foreground stars or
  scattered light residuals (the fibers with non-detections or contamination
  are also surrounded by squares). An eDIG detection requires that at least
  one emission line is detected at the 5$\sigma$ level, or at least two
  emission lines are detected at the 4$\sigma$ level.}
\end{figure}

\subsection{Data Reduction}\label{sec_datared}

The data were reduced using standard procedures in IRAF\footnote[7]{IRAF is
  distributed by the National Optical Astronomy Observatories, which are
  operated by the Association of Universities for Research in Astronomy, Inc.,
  under cooperative agreement with the National Science Foundation.}. The
images were overscan, bias, and dark corrected using the \texttt{ccdproc}
task. Image stacking was performed using the \texttt{imcombine} task; images
of the same pointing were median combined on a given night, and then across
nights, when necessary. Cosmic ray removal was accomplished through image
stacking where possible, and through the package L.A.Cosmic on individual
images where necessary \citep{vanDokkum2001}. \texttt{dohydra} was used to
trace and extract the spectra, as well perform the flat field correction and
find the dispersion solution. Observations of a ThAr comparison lamp were used
to solve for the dispersion solution as well as the spectral resolution; these
observations were obtained before and after each set of object exposures. 

Sky subtraction was performed using a median stacking of the sky spectra in
each pointing in which no eDIG emission was detected. In deeper spectra,
\citet{Heald2006b} detect eDIG emission at the location of our sky spectra,
and we therefore compared our stacked sky spectra to that of
\citet{Osterbrock1996} to look for such contamination. The sky spectra for p1,
p2, and p4 (see Table 2) showed no evidence of eDIG emission, but the spectrum
for p3 showed a weak H$\alpha$ emission line ($I_{H\alpha} = 1.5 \times
10^{-16}$ erg cm$^{-2}$ s$^{-1}$) that was fit and removed using a single
Gaussian profile before sky subtraction was performed. Flux calibration was
then performed using observations of the standard stars PG0205+134 and Feige
110.  Since the sky and standard stars were observed only with certain fibers,
sky flats were used to perform a fiber-to-fiber throughput correction to
improve the quality of sky subtraction and flux calibration.

The data reduction presented an additional challenge due to a large amount of
instrumental scattered light in the data. After observations were concluded,
it was determined that improper closure of the CCD head electronics box after
servicing resulted in light leaking onto the CCD from exposed LEDs. This
resulted in a gradient of excess counts over the entire CCD that increases to
the blue, as well as a discrete feature primarily affecting fibers 8 - 18 near
the [NII] $\lambda$6583 emission line. The excess counts per (unextracted)
pixel range from a few percent to $100\%$ of the continuum counts per pixel in
the gradient, and are as high as $300\%$ of the continuum counts per pixel in
the discrete feature. The counts are present in all science and calibration
frames, although the magnitude and morphology of the counts varies somewhat
from night to night.

A scattered light correction was performed using the \texttt{apscatter} task
in IRAF. The scattered light between the spectra was identified, a model of
the excess counts over the CCD was constructed, and these counts were
subtracted to produce a cleaned version of the data. Specifically, a very high
order function (60th - 100th order spline) was fit to the scattered light
along and across the dispersion. A high order function was required to fit the
structure in the discrete features; however, a lower order function was
desired to fit the general gradient. Thus, a lower order function (5th order
spline) was fit across the dispersion of the higher order fit to characterize
the gradient. This latter fit was subtracted from the former to isolate the
discrete features, which were then subtracted from the data, leaving only the
smooth gradient to be removed. Finally, the gradient was fit and subtracted
using 45th and 5th - 8th order splines along and across the dispersion,
respectively.

Some residual evidence of the scattered light is apparent in several
fibers. This evidence is largely manifested as artificial structure in the
continuum around the [NII] $\lambda$6583 emission line. The fibers that were
rejected from the analysis due to the residual effects of scattered light are
marked as such in Figure 3.

\subsection{Error Analysis and eDIG Detection Criteria}\label{sec_error}

The Poisson (random) error was calculated for the raw data and calibration
frames, and propagated through the data reduction and analysis. The Poisson
error is generally comparable to the RMS error in the continuum near the
emission lines of interest. However, the scatter in a given
continuum-subtracted, emission-line masked spectrum does not have a strictly
Gaussian distribution; instead, a 3$\sigma$ Poisson error generally
corresponds to a $96 - 100\%$ confidence interval. Thus, we define our eDIG
detection criteria as follows: to have a detection, a fiber must have one or
more emission lines with at least a 5$\sigma$ detection, or two or more
emission lines with at least a 4$\sigma$ detection. The emission lines
considered in the detection criteria are the H$\alpha$, [NII] $\lambda$6583,
and [SII] $\lambda$6716, 6731 lines. The fibers with eDIG detections and
non-detections are marked in Figure 3, as well as those that were excluded
from the analysis due to contamination by foreground stars or scattered light
residuals. Detections were made in $81\%$ of the fibers, and to a
height above the disk of $|z| = 3.2$ kpc.

\begin{figure*}[h]
\epsscale{0.85}\plotone{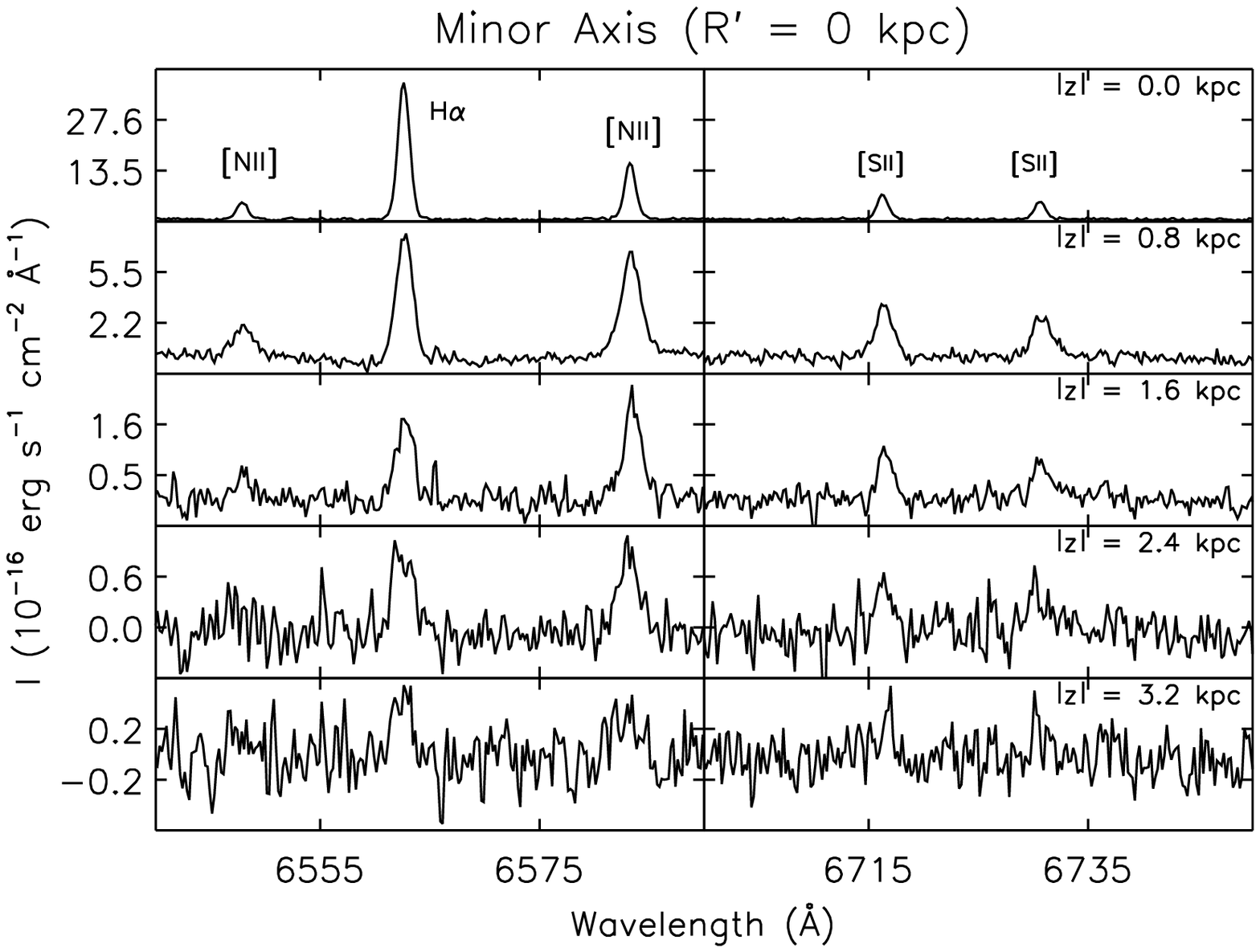}
\epsscale{0.85}\plotone{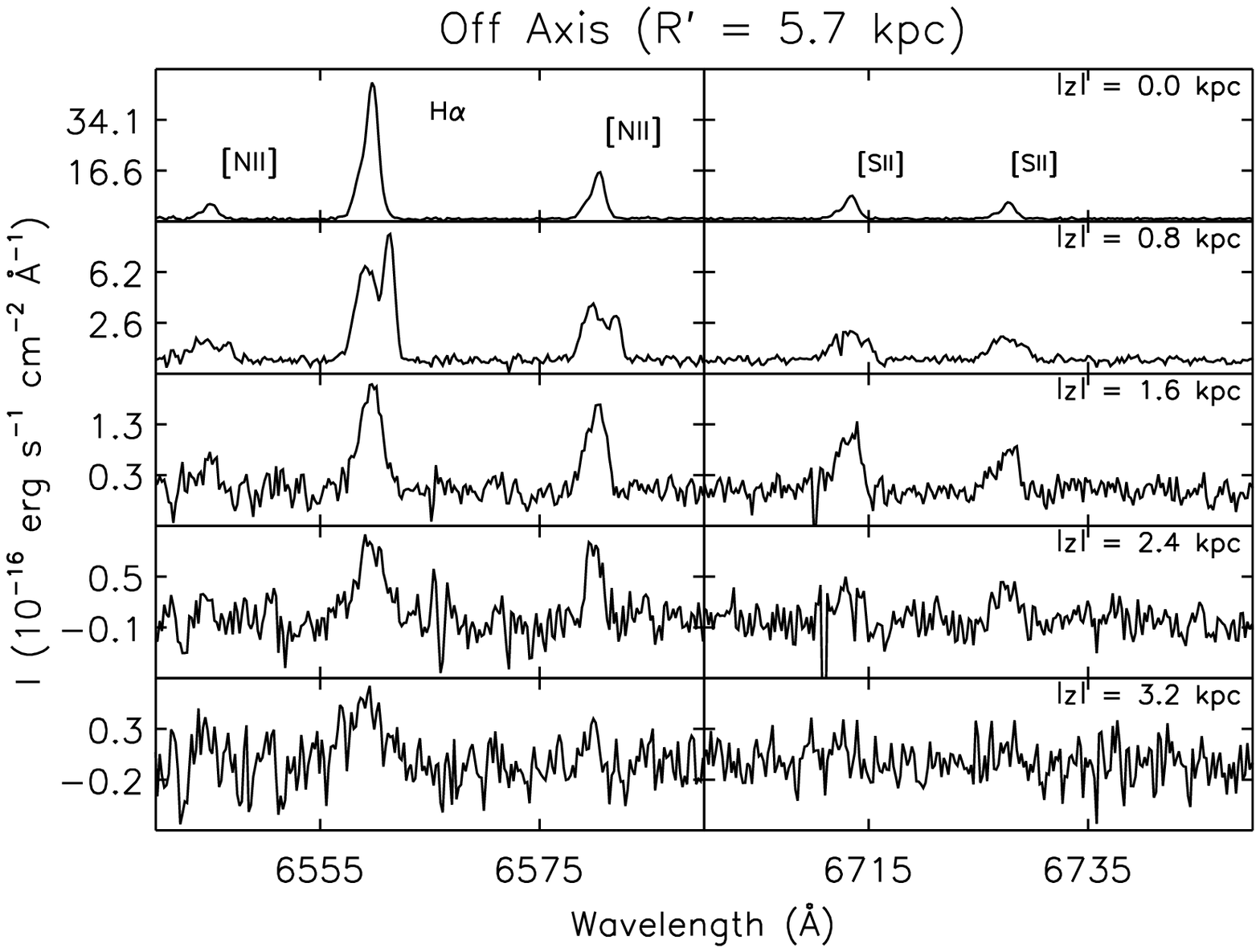}
\caption{Rest-frame, continuum-subtracted spectra along the minor axis of NGC
  891 ($R' = 0$ kpc; top panel, from p2 in Table 2), and offset along the
  major axis by $R' = 5.7$ kpc (bottom panel, from p3 in Table 2). The spectra
  sample a range of heights above the disk from $|z| = 0$ to $|z| = 3.2$ kpc
  as labeled. Near the midplane, the spectra are characteristic of HII region
  emission ([NII]$\lambda 6583$/H$\alpha$, [SII]$\lambda 6717$/H$\alpha < 1$),
  but they become indicative of diffuse emission by $|z| = 1$ kpc
  ([NII]$\lambda 6583$/H$\alpha > 1$). Along the minor axis, the emission
  lines are generally symmetric; however, off axis, the lines are asymmetric
  due to rotational broadening. Additionally, there is evidence of line
  splitting due to multiple velocity components along the line of sight, most
  clearly seen in H$\alpha$ at $|z| = 0.8$ kpc. This is qualitatively
  consistent with the eDIG being primarily found in discrete clouds and
  filaments over spiral arms.}
\end{figure*}

Some example spectra are shown at a range of $z$ values for both the minor
axis and off-axis pointings in Figure 4. In the disk ($z = 0$ kpc), the
spectra are consistent with HII region emission ([NII]$\lambda 6583$/H$\alpha
< 1$), while by $|z| = 1$ kpc the spectra are characteristic of diffuse
emission ([NII]$\lambda 6583$/H$\alpha \geq 1$). While the line profiles on
the minor axis are fairly symmetric, rotational broadening results in line
profiles off-axis that are clearly asymmetric. There is also evidence of line
splitting due to multiple velocity components along the line of sight (e.g.,
for $R' = 5.7$ kpc, $|z| = 0.8$ kpc).

To characterize the emission line properties further, we fit a fifth-order
Legendre polynomial to the continuum using the IRAF task
\texttt{continuum}. The stellar H$\alpha$ absorption is negligible for the
thick disk and halo, so we do not correct for any absorption. We then fit a
single Gaussian to each of the detected emission lines in the
continuum-subtracted spectra using the IDL function \texttt{mpfitfun}. (For
each spectrum that meets the detection criteria above, the detected emission
lines are those with at least 4$\sigma$ detections). The latter determines the
Gaussian parameters and their corresponding uncertainties that best fit the
data in a least-squares sense. These Gaussian fits yield the emission line
intensities and widths, as well as the uncertainties on these quantities; the
fiber locations and line intensities, widths, and velocities are given in
online Table 3. We use these emission line properties to determine the
three-dimensional density distribution and velocity dispersion of the eDIG
layer below.

\begin{deluxetable*}{cccccccc}[t]
\tabletypesize{\scriptsize}
\tablecolumns{8}
\tablewidth{0pt}
\tablecaption{Emission Line Spectroscopy}
\tablehead{ 
\colhead{Offset Along\tablenotemark{a}} &
\colhead{Offset Along\tablenotemark{a}} &
\colhead{$I_{H\alpha}$} &
\colhead{$\sigma_{I_{H\alpha}}$} &
\colhead{$\sigma_{H\alpha}$\tablenotemark{b}} &
\colhead{$\sigma_{\sigma_{H\alpha}}$} &
\colhead{...} &
\colhead{$v$\tablenotemark{c}}\\
\colhead{Major Axis ($"$)} &
\colhead{Minor Axis ($"$)} &
\colhead{($10^{-17}$ erg s$^{-1}$ cm$^{-2}$)} &
\colhead{($10^{-17}$ erg s$^{-1}$ cm$^{-2}$)} &
\colhead{(\kms)} &
\colhead{(\kms)} &
\colhead{...} &
\colhead{(\kms)}
}
\startdata
-10 & -34 & 15.2 & 2.4 & 0.0 & 0.0 & ... &  546\\
+30 & -96 & 0.0 & 0.0 & 0.0 & 0.0 & ... & 0\\
-34 & -59 & 0.0 & 0.0 & 0.0 & 0.0 & ... & 0\\
-25 & -42 & 0.0 & 0.0 & 0.0 & 0.0 & ... & 0\\
-30 & -51 & 0.0 & 0.0 & 0.0 & 0.0 & ... & 0\\
... & ... & ... & ... & ... & ... & ... & ...
\enddata
\tablenotetext{a}{Fiber offsets along the major and minor axes are
  measured with respect to the galactic center at RA, Decl. = 02 22 33.4, +42
  20 57.}
\tablenotetext{b}{All line widths are standard deviations; the line widths are
  corrected for the instrumental resolution.} 
\tablenotetext{c}{The velocity centroid of the H$\alpha$ emission line.}
 \tablecomments{Table 3 is published in its entirety in the machine-readable
   format. A portion is shown here for guidance regarding its form and
   content. Columns 7 - 12 of the online table contain the line widths and
   corresponding uncertainties of the [NII] $\lambda$6583 and [SII]
   $\lambda$6716, 6731 lines. Fibers without eDIG detections have null values,
   and emission lines with less than 4$\sigma$ detections have null line widths.}
\end{deluxetable*}

\section{Results}\label{sec_results}

\subsection {Density Distribution}

\subsubsection{Radial Density Distribution}\label{sec_emline}

Our first goal is to determine the three-dimensional density distribution of
the eDIG layer in NGC 891. We consider the radial density distribution by
examining constraints on the minimum and maximum galactocentric radius at
which the gas is found, $R_{min}$ and $R_{max}$. In H$\alpha$ imaging, there
is a sharp decrease in the eDIG emission around $R_{max} = 8$ kpc, and we
choose a radial cutoff to be consistent with this observation
\citep{Dettmar1990, Rand1990, Pildis1994}. \citet{Heald2006b} analyze the
position-velocity diagram of the eDIG layer in NGC 891 to argue that the gas
is found at $R \ge 15$ kpc. However, as discussed below, this may be due to
spiral structure or other deviations from the simplified symmetric disk
geometry assumed. Deep spectroscopic observations at large galactocentric
radii ($R \ge 10$ kpc) are needed to definitely determine the radial cutoff of
the eDIG in this galaxy. Here, we make the conservative assumption that the
gas is found within $R_{max} = 8$ kpc, and discuss the implications of a
larger $R_{max}$ in \S6.

We distinguish between a density distribution in which the gas fills the line
of sight ($R_{min} = 0$ kpc; a disk model) and one in which the gas is
preferentially found at moderate or large galactocentric radius ($R_{min} > 0$
kpc; a ring model). The motivation for testing such a model is
threefold. First, we expect to observe the eDIG at moderate galactocentric
radii over star-forming spiral arms due to galactic chimney mode
feedback. Second, the observed $I_{H\alpha}$ values as a function of $z$ are
remarkably similar for the on- and off-axis pointings, suggesting that the
pathlength through the gas is comparable in both locations (see
\S5.1.2). Third, as shown in Figure 2\textit{a}, the depth of the galactic
potential well at the center of the galaxy means that any eDIG in this region
is likely to remain close to the midplane.

In a rotating disk model ($R_{min} = 0$ kpc), a given line of sight samples
gas at a range of galactocentric radii, and thus a range of projected
velocities. In a rotating ring model ($R_{min} > 0$ kpc), however, a line of
sight generally samples gas at moderate or large galactocentric radii, where
the projected velocity of the gas is small. In the former case, we expect the
line profiles to be more asymmetric than in the latter, and the emission lines
may be characterized by low- or high-velocity wings. Note that this only
applies near the minor axis; at the edge of the gas distribution, the disk and
ring models are indistinguishable.

To explore this qualitatively, we compare the observed and modeled H$\alpha$
line profiles that arise from the disk and ring models given the following
assumptions. For the former model, the gas uniformly fills the disk between $0
\leq R \leq 8$ kpc; in the latter model, the gas is found between $R_{min}
\leq R \leq 8$ kpc, where $R_{min}$ is varied between $1 \le R_{min} \le 7$
kpc. We assume that the disk is oriented perfectly edge-on ($i = 90^{\circ}$),
and that the emission arises along a line of sight that passes all the way
through an optically thin disk or ring. We assume a flat rotation curve with
circular velocity $v_{c} = 226$ \kms in the disk \citep{Fraternali2011}, with
a vertical gradient in rotational velocity of $\Delta v_{c} = -15$ km s$^{-1}$
kpc$^{-1}$ starting at $|z| = 1$ kpc \citep{Heald2006b, Kamphuis2007}. Note
that \citet{Heald2006b} do not observe a rotational velocity gradient in the
southeast quadrant of the galaxy, and thus applying a single value of $\Delta
v_{c}$ to the northeast and northwest quadrants may be a simplification.

To construct the line profile at a given $R'$ and $z$, we assume that the line
profile from a cloud of gas moving with line-of-sight velocity $v_{los}$ with
an internal velocity dispersion $\sigma$ is a Gaussian of the form:
\begin{equation}
  I(v) = I_{peak}e^{\frac{-(v-v_{los})^{2}}{2\sigma^{2}}}.
\end{equation}
Here, $|v_{los}| = v_{c}cos(\theta)$, where $\theta$ is the angle between $R'$
and $R$, and the sign is determined by whether $R'$ is on the approaching or
receding side of the galaxy. We choose $I_{peak}$ necessary to reproduce the
observed intensity, and we define $\sigma^{2} = \sigma_{res,H\alpha}^{2} +
\sigma_{th,p}^{2} + \sigma_{turb}^{2}$, where $\sigma_{res} = 17$ \kms is the
instrumental resolution at $H\alpha$, $\sigma_{th,p} = 9$ \kms is the thermal
velocity dispersion of the protons, and $\sigma_{turb} = 25$ \kms is the
turbulent (random) velocity dispersion of the gas (see \S5.2). We
assume a gas temperature of $T_{4} = 1.0$ K, here and throughout the rest of
this paper \citep{Rand1997, Collins2001}. We construct the full line profile
by summing over the profiles of individual, identical, uniformly distributed
clouds of size $\Delta l = 1$ pc that fill the allowed range in $R$ along the
line of sight. We choose $\Delta l$ to be sufficiently small that
$\Delta v_{los}$ across a cloud is negligible.

In Figure 5, we compare the line profiles for the disk and ring models to the
observed line profiles for a range of $R'$ values. The off-axis pointings pass
close enough to the edge of the eDIG layer that the differences between the
disk and the ring models are small, and thus the minor-axis pointings are
where these differences are of interest. At $R' = 0$ kpc ($v_{los} = 0$ \kms),
the disk and ring models are indistinguishable. Moving to larger $R'$, the
profiles are fairly symmetric for $R_{min} \ge 2$ kpc, while the profiles
become increasingly asymmetric and develop low- or high-velocity wings for
$R_{min} < 2$ kpc. The observed line profiles appear much more consistent with
a ring model ($R_{min} \ge 2$) kpc than with a disk or ring with small
$R_{min} \sim 1$ kpc. However, the line profiles are qualitatively similar for
$R_{min} \ge 2$ kpc, and distinguishing definitively between larger $R_{min}$
is not possible. There are a small number of spectra that do have wings
consistent with a disk; an example of such a spectrum is shown at the bottom
left of Figure 5. Additionally, some spectra have significant asymmetries that
are inconsistent with both the disk and ring models, as shown at the bottom
right of the same Figure.

\begin{figure*}[h]
\epsscale{1.0}\plottwo{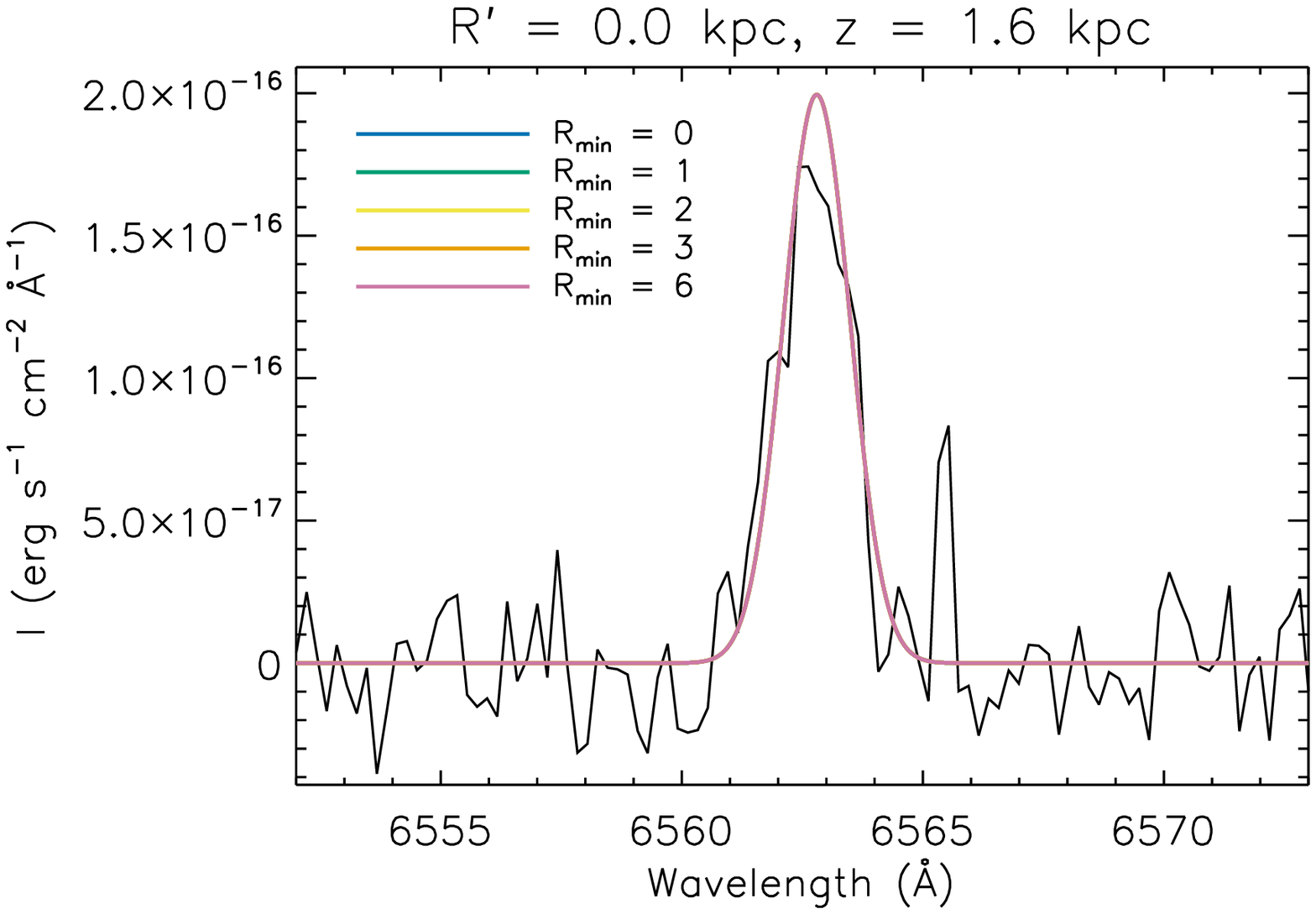}{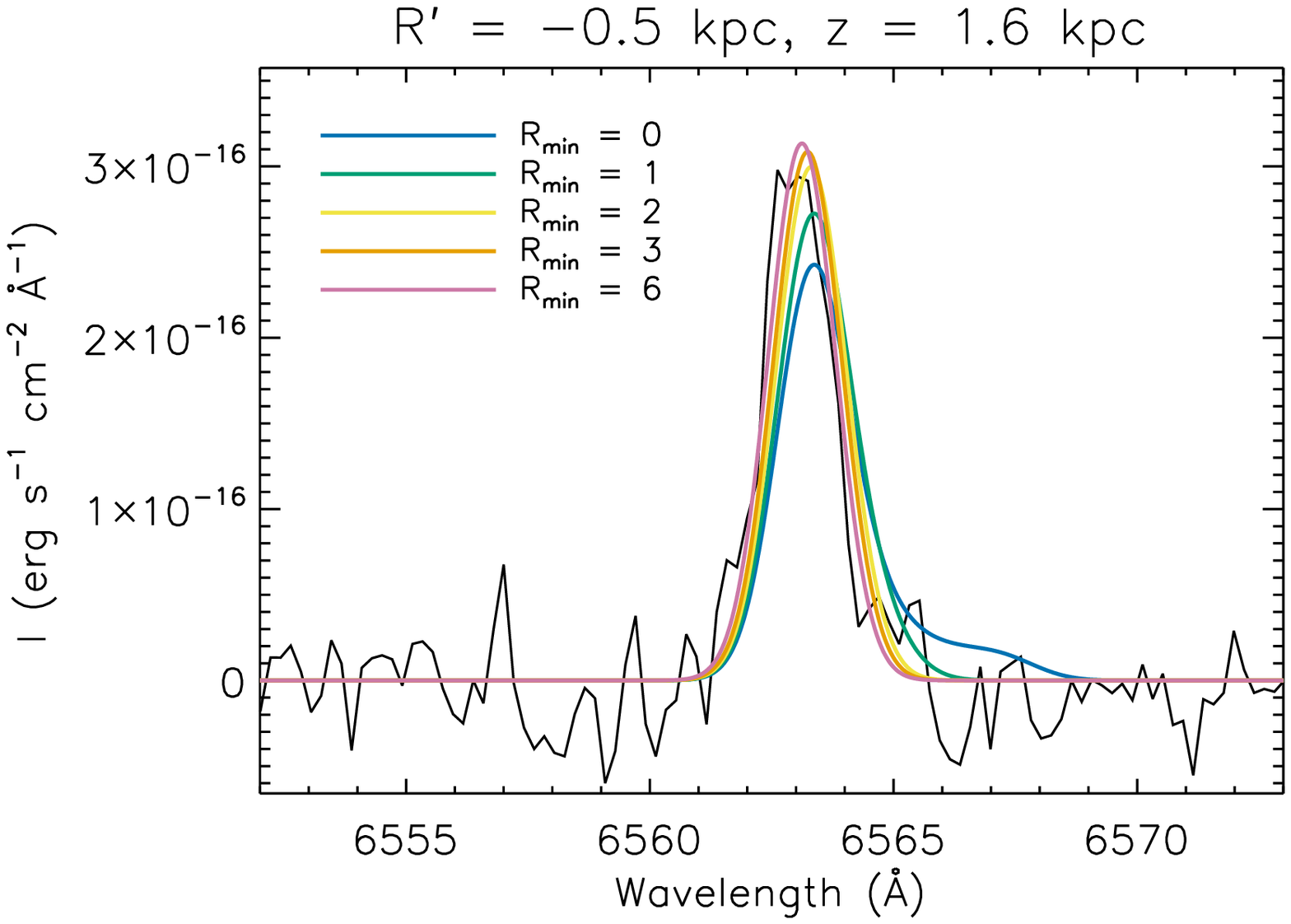}
\epsscale{1.0}\plottwo{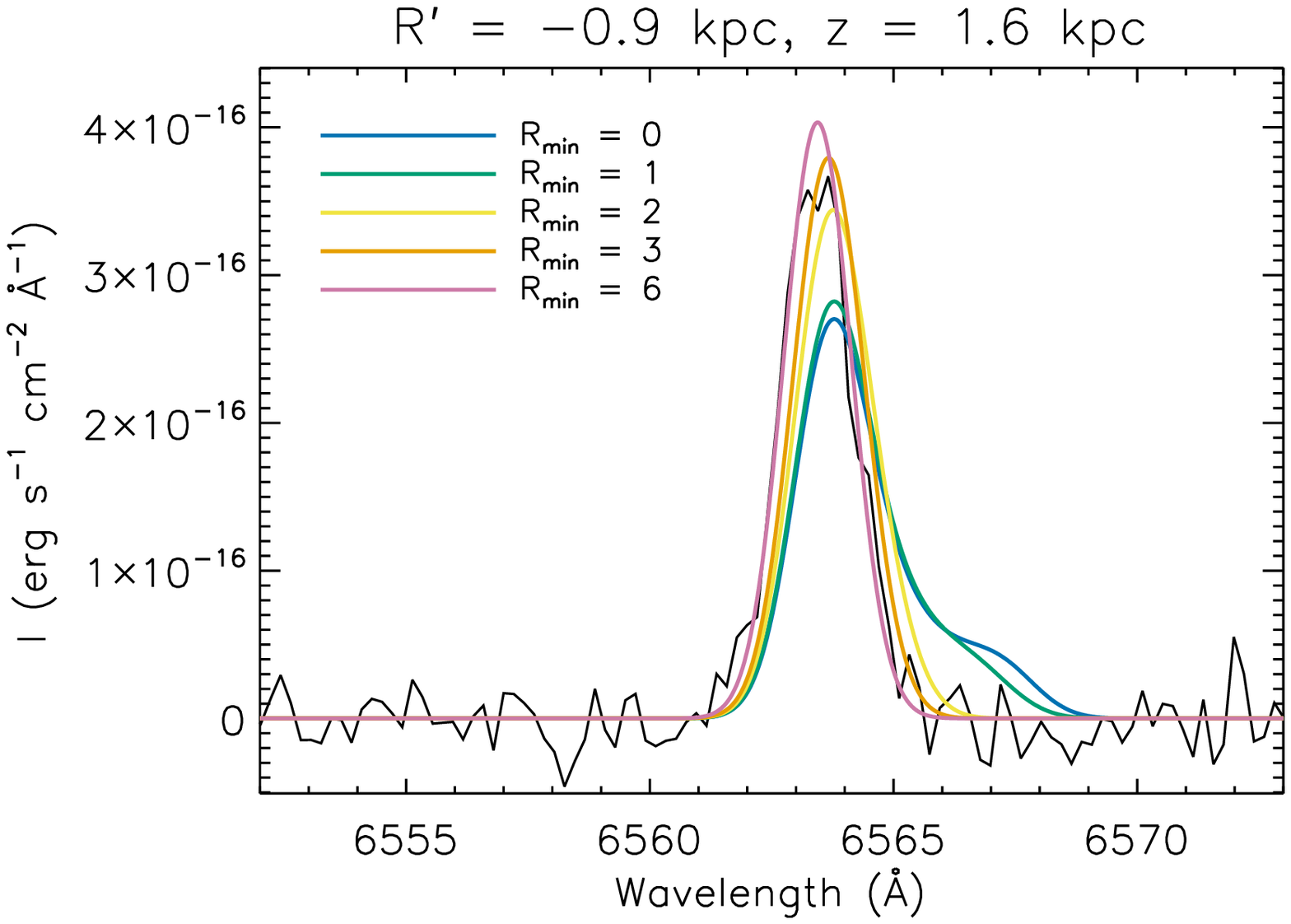}{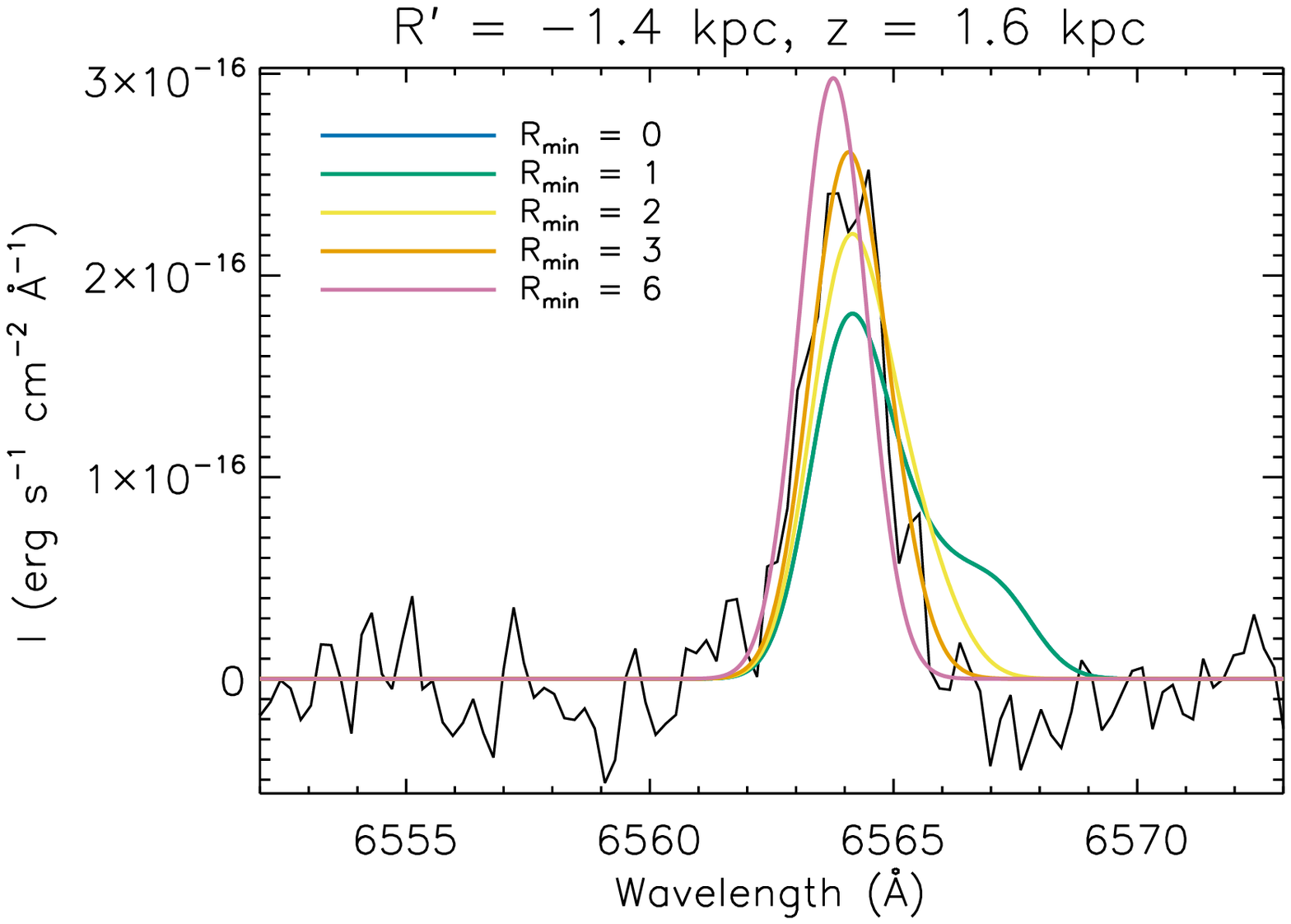}
\epsscale{1.0}\plottwo{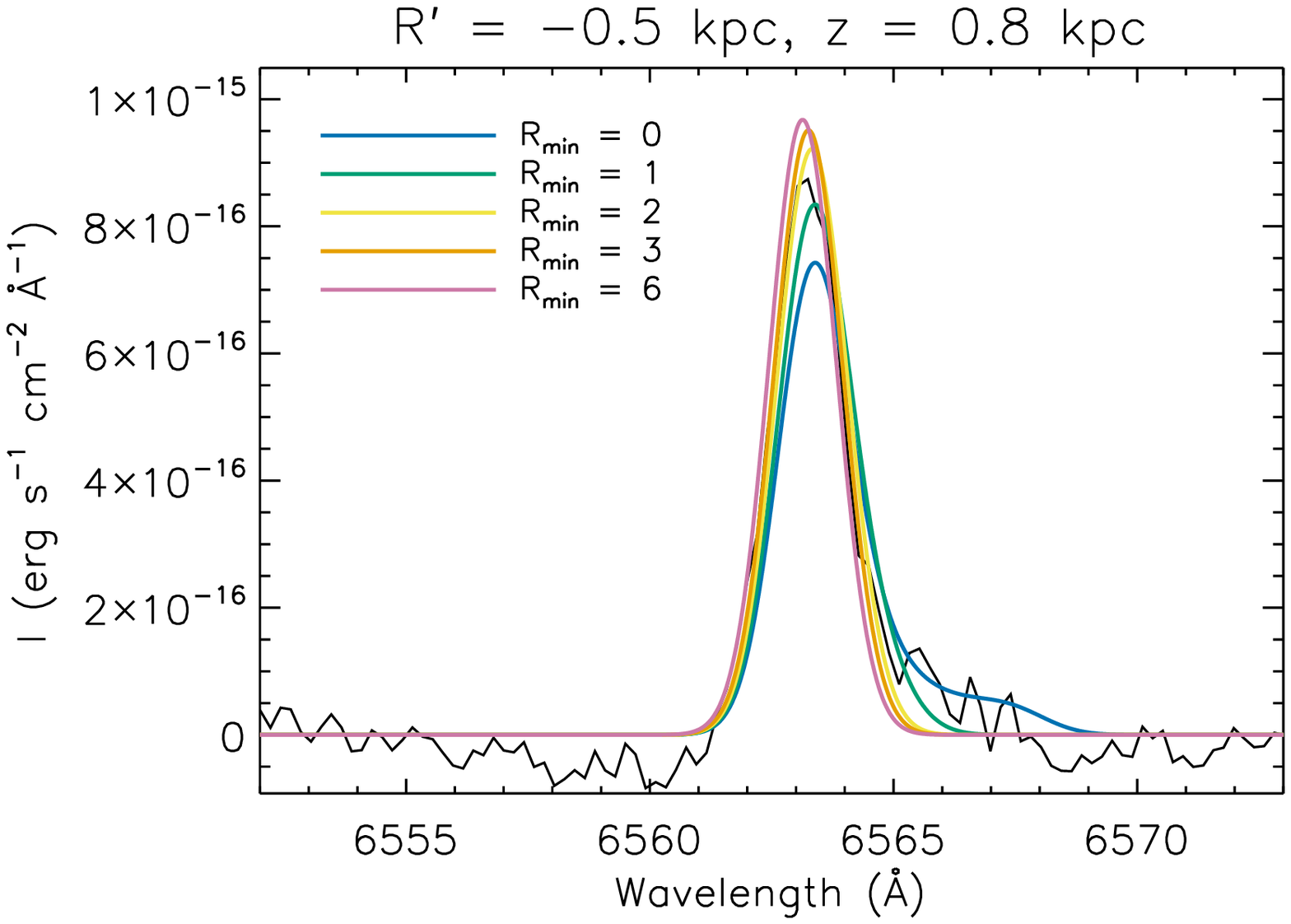}{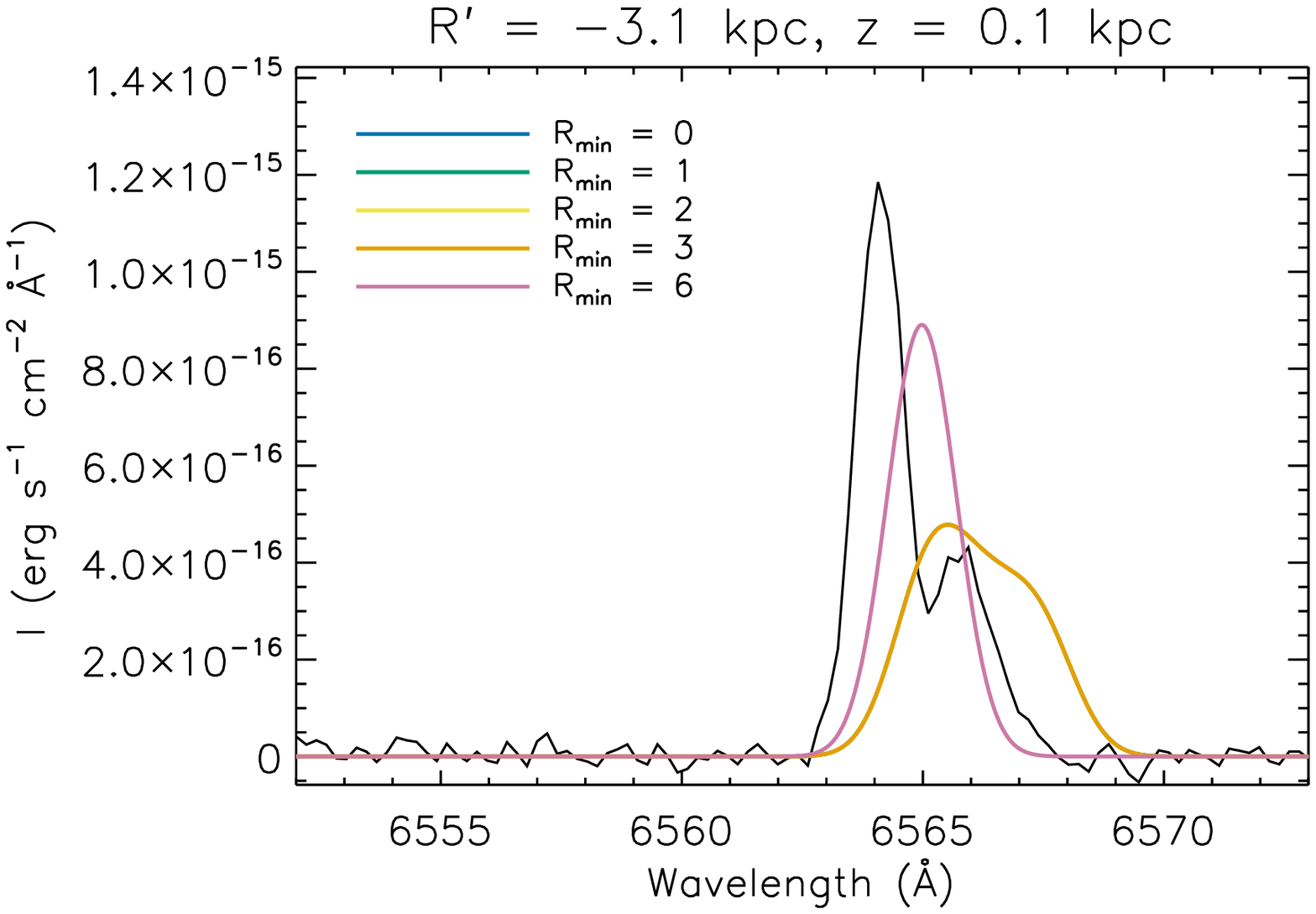}
\caption{Observed H$\alpha$ line profiles (black) compared to model line
  profiles with a range of $R_{min}$ values. In the disk model ($R_{min} = 0$
  kpc), the gas is uniformly distributed between $0 \le R \le 8$ kpc, whereas
  in the ring model ($R_{min} > 0$ kpc) it is restricted to $R_{min} \le R \le
  8$ kpc. The model line profiles are constructed to have the same integrated
  intensity $I_{H\alpha}$ as the observed profile, have a total velocity
  dispersion due to thermal motions ($\sigma_{th,p} = 9$ \kms), turbulent
  motions ($\sigma_{turb} = 25$ \kms), and instrumental resolution
  ($\sigma_{res} = 17$ \kms), and assume a flat rotation curve with $v_{c} =
  226$ \kms \citealt{Fraternali2011} and vertical rotational velocity gradient
  $\Delta v_{c} = -15$ km s$^{-1}$ kpc$^{-1}$ above $|z| = 1$ kpc
  \citep{Heald2006b, Kamphuis2007}. The observed line profiles do not show
  strong evidence of the low- and high-velocity wings that arise from emission
  near the center of the galaxy, and are qualitatively more consistent with a
  ring model than a disk model ($R_{min} \ge 2$ kpc). The spectra shown are
  from p2 in Table 2.}
\end{figure*}

\begin{figure}[h]
\epsscale{1.2}\plotone{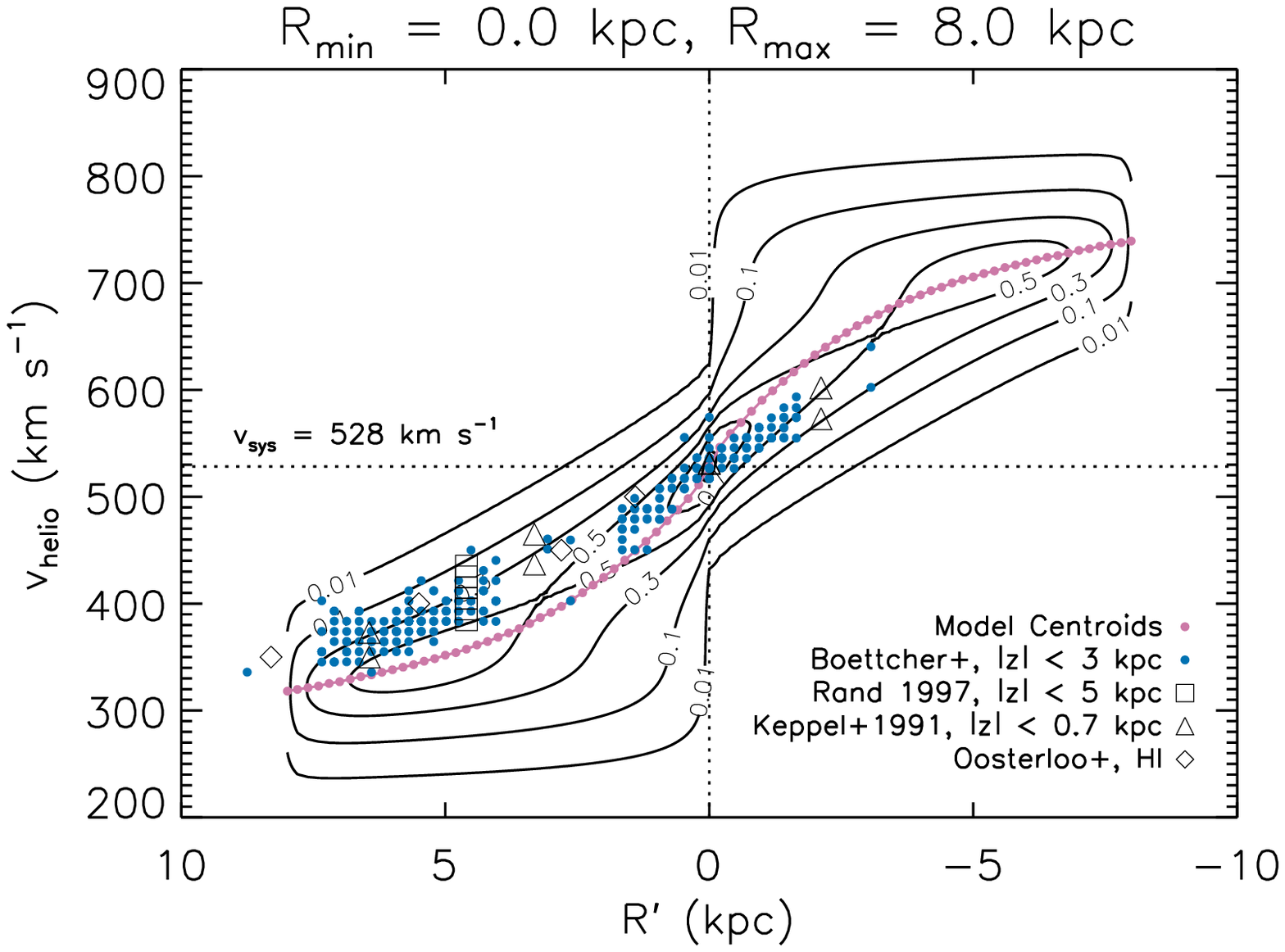}
\epsscale{1.2}\plotone{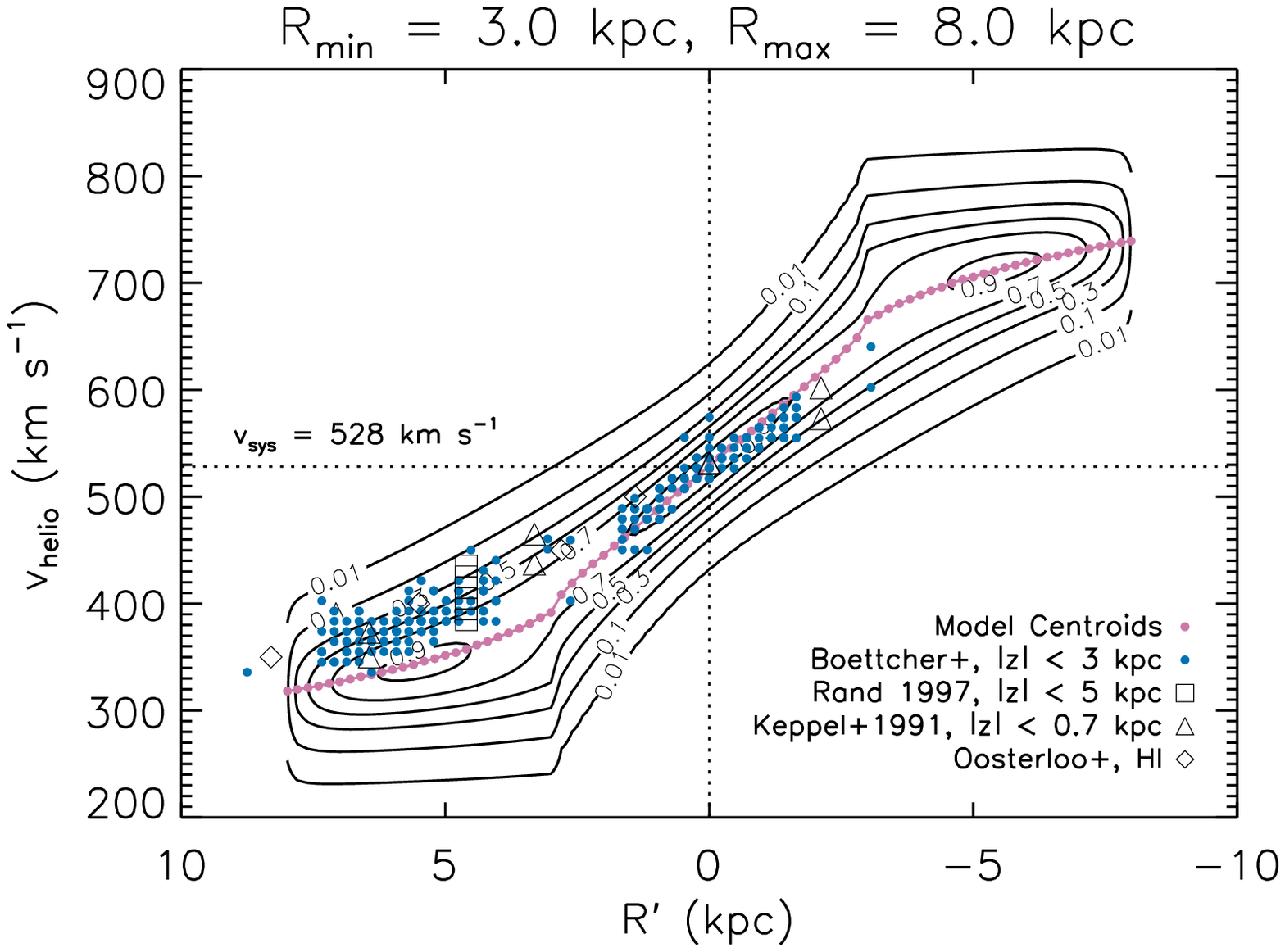}
\epsscale{1.2}\plotone{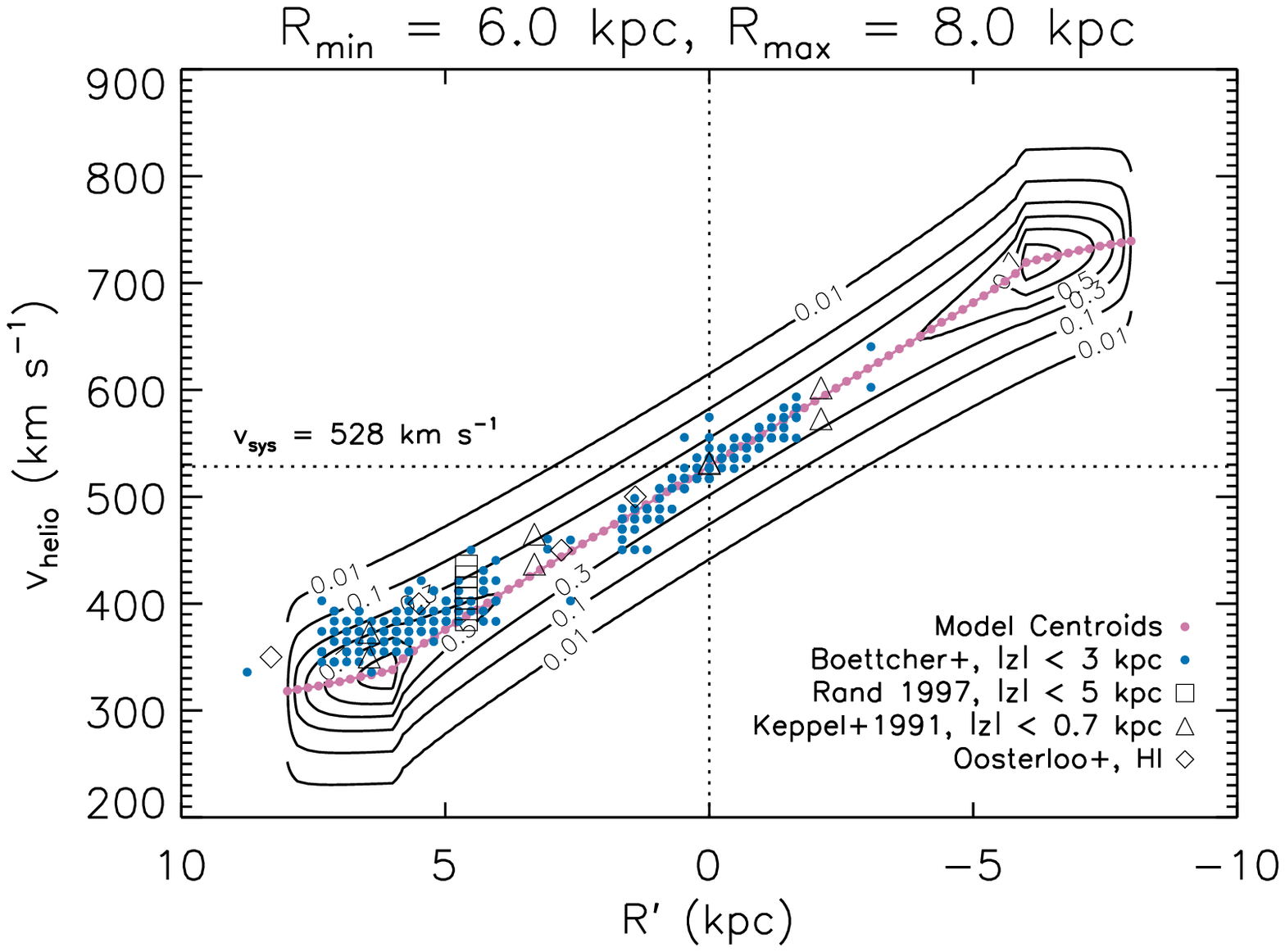}
\caption{We compare the position-velocity diagrams for a disk model ($R_{min}
  = 0$ kpc) and two ring models ($R_{min} = 3, 6$ kpc) of the eDIG layer in
  NGC 891. The PV diagram is modeled at a distance of $|z| = 2$ kpc from the
  midplane; the rotational velocity ($v_{c} = 226$ \kms;
  \citealt{Fraternali2011}) is adjusted by the observed velocity gradient
  $\Delta v_{c} = -15$ \kms kpc$^{-1}$ starting at $|z| = 1$ kpc
  \citep{Heald2006b, Kamphuis2007}. The pink points show the location of the
  model velocity centroids in the PV plane, while the blue points indicate the
  location of the observed H$\alpha$ velocity centroids. The open symbols
  denote eDIG and HI velocity centroids from the literature. Larger values of
  $R_{min}$ better reproduce the velocity centroids; however, the observed
  centroids are closer to the systemic velocity than is predicted by any model
  at large $R’$.}
\end{figure}

We also distinguish between a disk and a ring model by examining the
position-velocity (PV) diagram of each model. We cannot construct robust PV
diagrams from our data due to their low S/N. However, we can determine the
location of the H$\alpha$ velocity centroids in the PV plane, and compare
these to the velocity centroids of each model. The results are shown in Figure
6 for $R_{min} = $ 0, 3, and 6 kpc. Although neither the minor-axis or
off-axis centroids are well-reproduced for $R_{min} = 0$ kpc, the minor-axis
centroids are increasingly well-characterized with increasing $R_{min}$. The
off-axis centroids are closer to the systemic velocity than predicted by any
model; these can be made consistent by either increasing the radius of the
eDIG layer to $R \geq 10$ kpc, or decreasing the rotational velocity of the
gas by $\Delta v_{c} = -50$ \kms kpc$^{-1}$. However, the former is
inconsistent with $H\alpha$ imaging \citep{Dettmar1990, Rand1990, Pildis1994},
and the latter with PV diagram analysis of higher S/N data of the northeast
quadrant of the same system \citep{Heald2006b, Kamphuis2007}. It is
interesting to note that \citet{Heald2006b} observe a rotational velocity of
$v_{c} = 175$ \kms in the southeast quandrant without clear evidence of a
rotational velocity gradient. Alternatively, the centroids may be closer to
systemic velocity than expected due to spiral structure or other deviations
from our simple symmetric geometry. The observed velocity centroids are given
in online Table 3.

Though we cannot fully reproduce the velocity centroids at large $R'$ with any
of our models, we can conclude that the observations are more consistent with
a ring model than with a disk model. The absence of low- and high-velocity
emission line wings suggests $R_{min} \ge 2$ kpc, while the location of the
velocity centroids in the PV plane suggests a larger $R_{min} = 6 - 7$
kpc. However, as the velocity centroids are not fully reproducable by this
model, we do not claim to robustly constrain $R_{min}$. Note that a true ring
model in which there is little eDIG within $R < R_{min}$ is more consistent
with observations than the quasi-ring produced by an isobaric gas layer in the
galactic gravitational potential (i.e., a model in which $h_{z}$ increases
with $R$). The latter model produces a slope that is steeper than observed in
the $I_{H\alpha}$, $z$ plane (see \S5.1.2).

Interestingly, \citet{Heald2006b} find evidence for a gas density profile that
peaks at $R \sim 6$ kpc and decreases to larger and smaller radii by modeling
the PV diagram of the eDIG layer in NGC 891 (see their Figure
3). \citet{Kamphuis2013} also argue for non-cylindrically symmetric HI layers
in edge-on galaxies from the HALOGAS survey. Thus, although cylindrical
geometry is often assumed for extraplanar gas layers, there is evidence for
non-cylindrical geometries in multiple galaxies.

In summary, the observed emission line intensities and profiles suggest that
the eDIG may fill only a small percent of the volume along a given
line of sight, and are qualitatively consistent with the eDIG being found in a
ring that is a few kpc thick between $R_{min} \le R \le 8$ kpc. It is likely that
any given line of sight intersects one or more eDIG clouds or filaments that
may be found over star-forming spiral arms, but are also subject to a certain
amount of randomization in their radial distribution and rotational velocity.

\subsubsection{Vertical Density Distribution}\label{sec_dist}

We now consider the vertical density distribution of the gas. The
H$\alpha$ intensity is related to the electron density along the line of sight
by:
\begin{equation}
  I_{H\alpha} = \frac{\int \phi n_{e}^{2} \mathrm{d}l}{2.75 T_{4}^{0.9}},   
\end{equation}
where $\phi$ is the volume filling factor, $n_{e}$ is the electron
density, and $T_{4}$ is the gas temperature expressed in units of $10^{4}$ K
(note that $I_{H\alpha}$ is in Rayleighs). We assume an electron density
distribution of the form:
\begin{equation}
  n_{e}(z) = n_{e,0}e^{-|z|/h_{z}},
\end{equation}
where $n_{e,0}$ is the electron density at the midplane and $h_{z}$ is the
exponential electron scale height defined in Equation (10). We choose a
distribution of this form for several reasons. First, there is empirical
evidence that this form provides a good fit to the data from past studies of
the eDIG layer in this galaxy \citep[e.g.,][]{Dettmar1990, Rand1990, Rand1997,
  Hoopes1999}. Second, above $|z| = 1$ kpc, we cannot confirm that $\sigma$ is
dependent on $z$ (see \S5.2), and $\frac{d\Phi}{dz}$ is only weakly
dependently on $z$ (see \S2 and the gravitational potential derived
therein). We see from Equation (9) that if these quantities are independent of
$z$, then an exponential profile with a constant scale height is a good
representation of the density distribution. We also tested a distribution of
the form $n_{e}(z) = n_{e,0}sech^{2}(z/h_{z})$, but the quality of the fit was
not improved.

We re-express the $H\alpha$ intensity as follows, assuming that the
temperature, filling factor, and pathlength through the eDIG layer are
independent of $z$ ($L = \int \mathrm{d}l$):
\begin{equation}
  I_{H\alpha} = \frac{\phi n_{e,0}^{2} L}{2.75 T_{4}^{0.9}}e^{-2|z|/h_{z}}.
\end{equation}
From Equation (13), we model $I_{H\alpha}$ for a range of $\phi n_{e,0}^{2}$
and $h_{z}$ values, and compare to the observed $I_{H\alpha}$ using a
$\chi^{2}$ minimization approach (i.e., minimizing the value of the reduced
$\chi^{2}$ defined by $\chi_{red}^{2} = \frac{1}{n}\sum \frac{(I_{H\alpha,obs}
  - I_{H\alpha,mod})^{2}}{\sigma_{obs}^{2}}$, where $n$ is the number of
degrees of freedom). A comparison of modeled and observed $I_{H\alpha}$ values
was made at every $R'$ and $z$ value where the eDIG was detected. This
comparison was performed separately for the East and West sides of the
galaxy. The observed $I_{H\alpha}$ values are given in online Table 3.

We consider both a single-component (thick disk) model as well as a
two-component (thick disk and halo) model. The two components are
distinguished by different scale heights as well as different electron
densities in the midplane. \citet{Rand1997} identifies a thick disk and a halo
component in long-slit spectroscopy of this system at $R' = 5$ kpc;
while we detect the eDIG to a height of $|z| \sim 3$ kpc above the midplane,
he detects the eDIG to the remarkable height of $|z| \geq 5$ kpc, and is thus
better able to characterize the halo properties. Therefore, we use his
parameterization of the halo in our two-component model (see Table 4), and
perform the $\chi_{red}^{2}$ minimization to parameterize the thick disk in
both our single- and multi-component models.

\begin{figure*}[h]
\epsscale{1.1}\plottwo{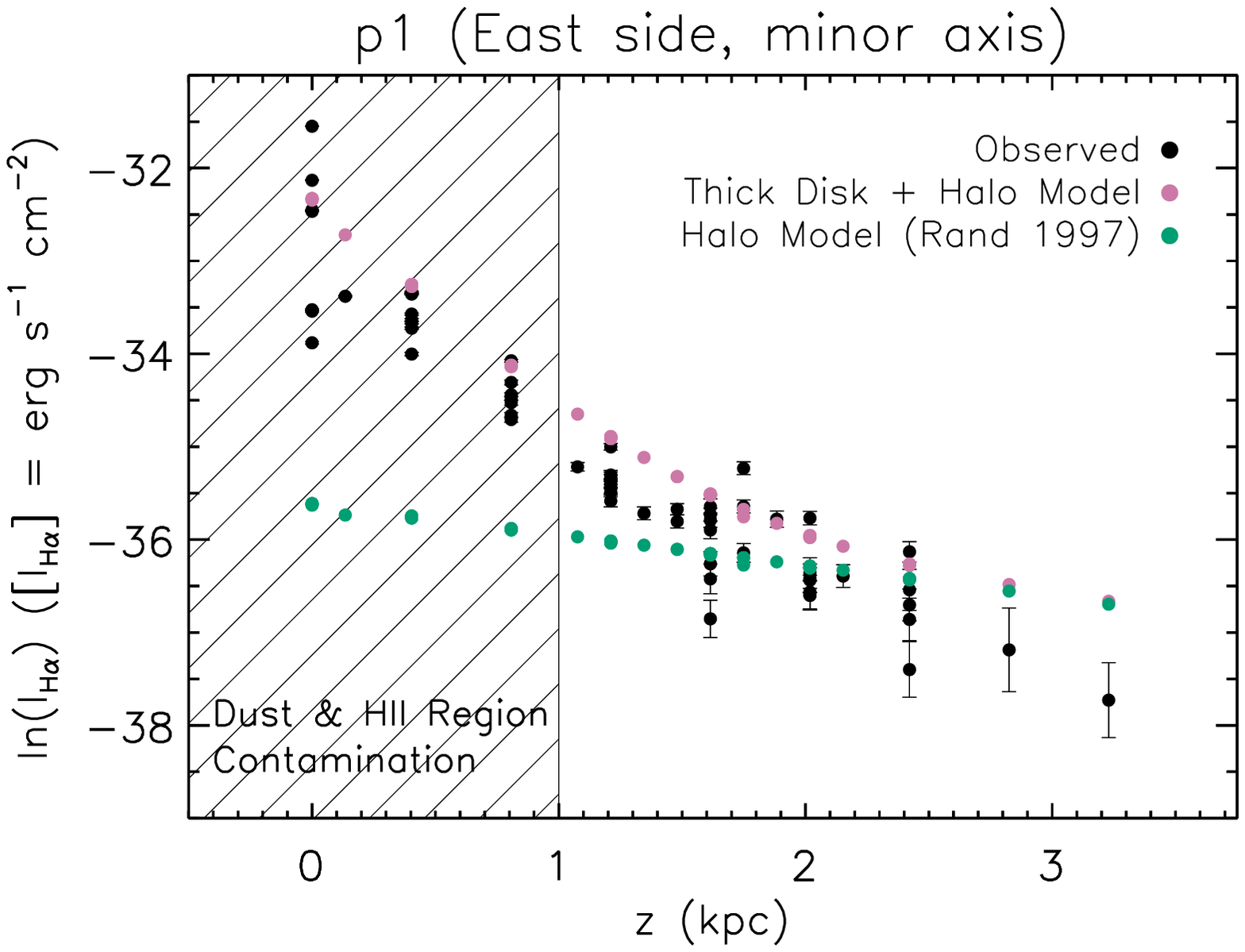}{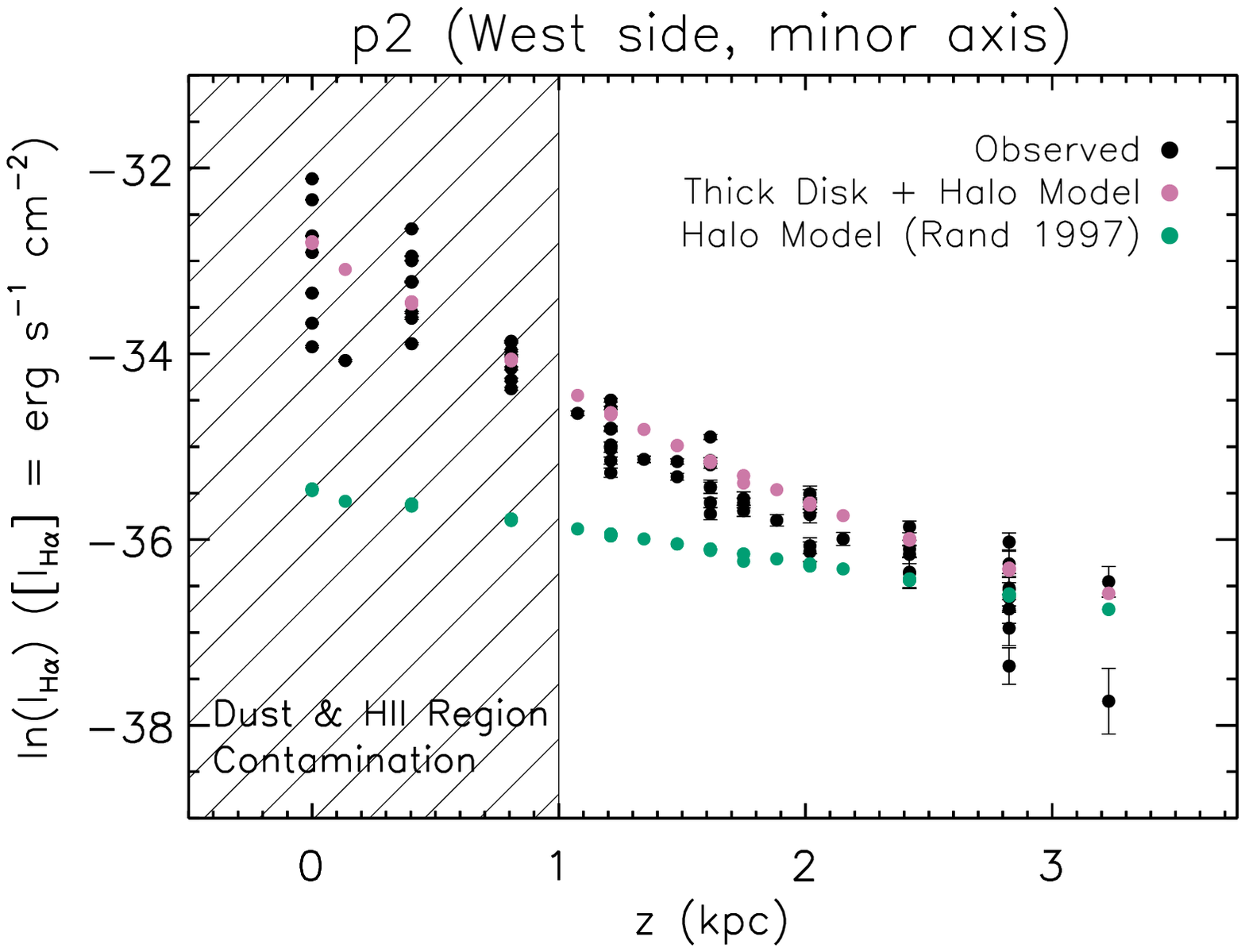}
\epsscale{1.1}\plottwo{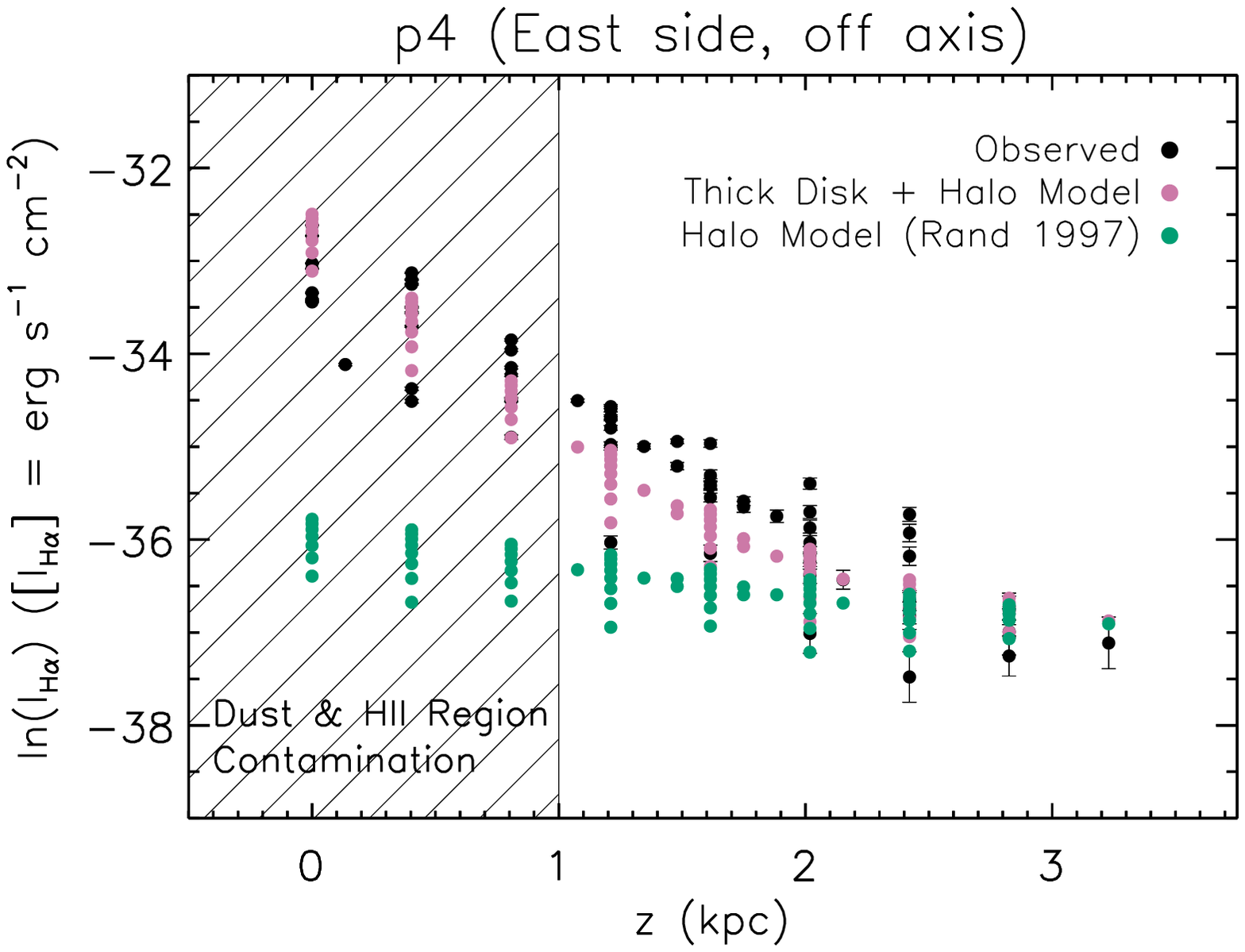}{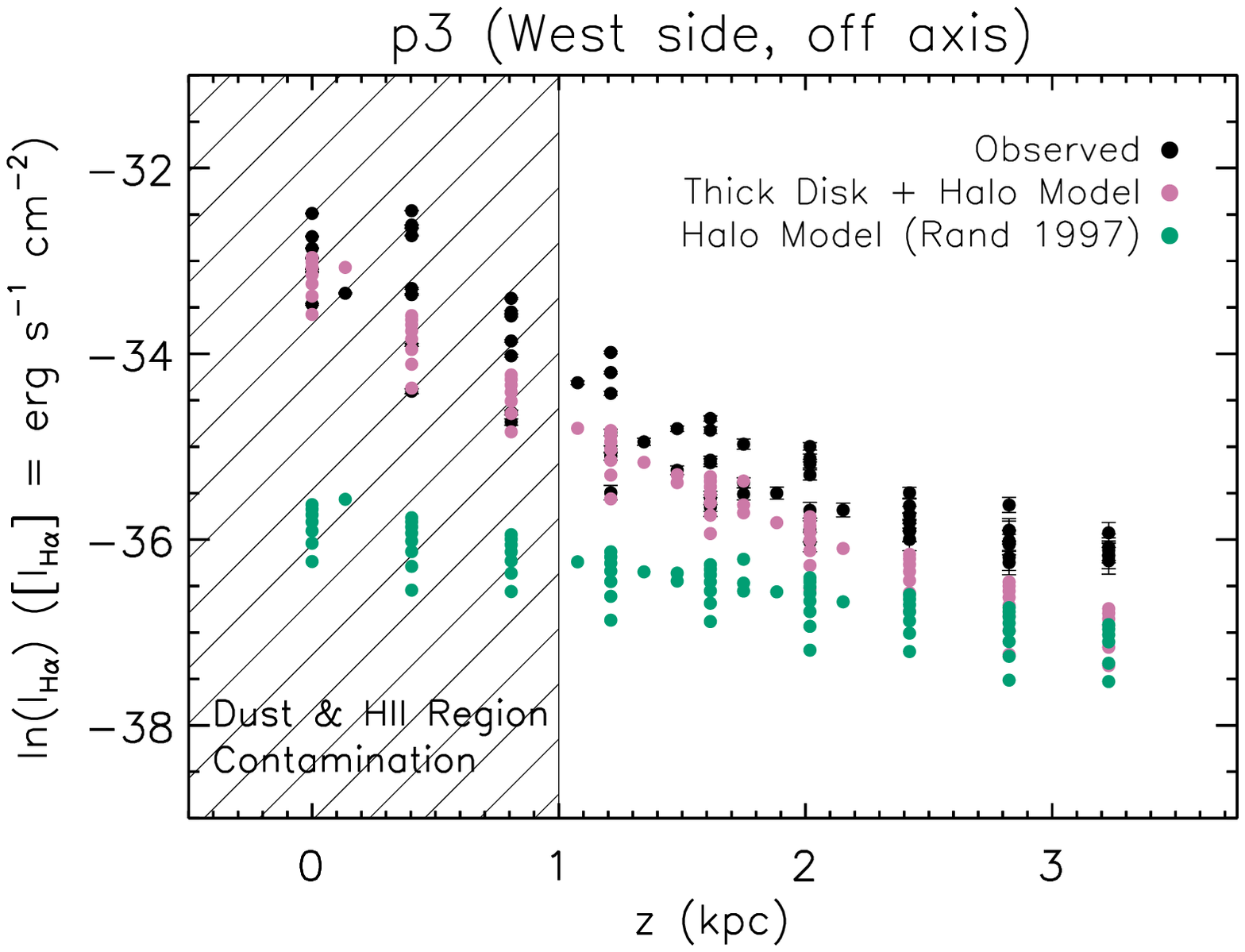}
\caption{The observed and modeled $I_{H\alpha}$ values are compared as a
  function of $z$ for each SparsePak pointing. A best fit model was determined
  for the East and West sides of the disk separately using a $\chi_{red}^{2}$
  minimization. A two-component (thick disk and halo) model is shown here,
  where the halo parameterization is taken from \citet{Rand1997}. The spread
  in the model $I_{H\alpha}$ at a given $z$ is due to different pathlengths
  through the gas sampled at different $R'$; additional spread in the observed
  $I_{H\alpha}$ is likely due to the clumpiness of the eDIG and the idealized
  geometry of our model.}
\end{figure*} 

We exclude all observations at $|z| < 1$ kpc to reduce the effects of dust
extinction and HII region contamination. Recently, the remarkable vertical
extent of extraplanar dust in NGC 891 has become clear; in addition to a dust
disk with a scale height of $h_{z,dust} = 0.2 - 0.25$ kpc
\citep{SchechtmanRook2013, Seon2014}, there is a dust halo with a scale height
determined by \citet{Seon2014} to be $h_{z,dust} = 1.2 - 2.0$ kpc, and by
\citet{Bocchio2016} to be $h_{z,dust} = 1.4$ kpc. As the eDIG layer and the
dust halo have comparable scale heights, dust extinction of the eDIG layer
must be considered.

\citet{Bocchio2016} model the $B$-band optical depth, $\tau_{B}$, as a function
of $z$ for NGC 891 (see their Figure 7). We consult the dust model of
\citet{Draine2011} to find $\tau_{H\alpha} \sim 0.6 \tau_{B}$ (see his Table
21.1). \citet{Bocchio2016} find $\tau_{B} = 1.0$ at $|z| = 1$ kpc, and thus we
find $\tau_{H\alpha} = 0.6$ at this height. Although this estimate is very
approximate, it suggests that we are detecting $\sim 50\%$ of the photons from
the far side of the disk above $|z| = 1$ kpc. This conservative cut also
reduces the contamination from HII region emission; Monte Carlo radiative
transfer simulations of diffuse and HII region photon propagation through the
dust disk of NGC 891 suggest that scattered HII region photons are only $10\%$
of diffuse photons by $|z| = 0.6$ kpc \citep{Ferrara1996}.

The values of $h_{z}$ and $\phi n_{e,0}^{2}$ that minimize $\chi_{red}^{2}$
for the one- and two-component models are given in Table 4. The minimum
$\chi_{red}^{2}$ exceeds unity by more than an order of magnitude in all
cases. This is likely due to the observed spread in $I_{H\alpha}$ at a given
$z$ that arises from the clumpiness of the eDIG as well as the uncertain
geometry of the layer (the modeled spread in $I_{H\alpha}$ is due to the
varying pathlength through the layer at different values of $R'$). 

For the East and West sides of the disk, respectively, the two-component fit
produces thick disk scale heights of $h_{z} = 0.8$ kpc and $h_{z} = 1.2$ kpc
that are consistent with past measurements \citep{Dettmar1990, Rand1990,
  Dettmar1991, Keppel1991, Rand1997, Hoopes1999}. Specifically, the thick disk
scale heights for both the one- and two-component fits are within $15\%$ of
those found by \citet{Hoopes1999} from H$\alpha$ imaging averaged over the
central 10 kpc of the disk, despite differences in the halo scale heights of a
factor of 2 - 3. This suggests that our thick disk scale heights are not
strongly sensitive to our choice of halo scale height, and our fit to a
sparsely sampled disk is consistent with that averaged over a fully imaged
disk. The two-component fit improves the $\chi_{red}^{2}$ value only
negligibly compared to the one-component fit.

The addition of a radial scale length to the electron density distribution
given in Equation (13) does not improve the quality of the fit in a
$\chi_{red}^{2}$ sense. However, we consider one variation on the radial
density distribution - a ring model in which the eDIG is excluded from the
inner part of the galaxy at $R < R_{min}$. We evaluate the $\chi_{red}^{2}$
values of our two-component model with $1 \leq R_{min} \leq 7$ kpc. An
improvement in $\chi_{red}^{2}$ is seen for both the East and West sides;
minimum $\chi_{red}^{2}$ values of 16.6 and 29.8 are found for $R_{min} = 4$
kpc for the East and West sides, respectively. The best-fit values of $\phi
n_{e,0}^{2}$ and $h_{z}$ remain within $25\%$ of those given in Table
4. Although this evidence is in favor of a ring model, we cannot robustly
constrain $R_{min}$ using this method or previous considerations of the line
profiles and PV diagram.

Thus, we have considered both one- and two-component fits to the H$\alpha$
intensity, as well as both disk and ring models. For the purposes of this
paper, we focus on the two-component disk model; the observed and model
$I_{H\alpha}$ values for this fit as a function of $z$ are shown in Figure
7. Since a disk model maximizes the pathlength through the gas, it minimizes
the value of $\phi n_{e,0}^{2}$, and thus a dynamical equilibrium model is
easier to satisfy for a disk than a ring model. Additionally, the smaller
scale height of the thick disk in the two-component model makes the dynamical
equilibrium model easier to satisfy than in the one-component model. For the
purposes of this paper, we focus on the dynamical state of the thick disk, for
which scale heights of $h_{z} = 0.8$ kpc and $h_{z} = 1.2$ kpc were found for
the East and West sides of the disk, respectively. We take the average of
these two values, $h_{z} = 1.0$ kpc, as well as the average of the electron
density in the disk, $\phi n_{e,0}^{2} = 0.013$ cm$^{-6}$, as the eDIG density
distribution that we seek to satisfy with our dynamical equilibrium model for
the remainder of this paper.

\begin{deluxetable*}{ccccccccccc}[t]
\tabletypesize{\scriptsize}
\tablecolumns{11}
\tablewidth{0pt}
\tablecaption{eDIG Electron Density Distribution}
\tablehead{
\colhead{} &
\colhead{} & 
\colhead{} & 
\colhead{East Side\tablenotemark{a}} & 
\colhead{} & 
\colhead{} & 
\colhead{} & 
\colhead{} & 
\colhead{West Side} & 
\colhead{} & 
\colhead{} \\  
\colhead{Model} &
\colhead{$h_{z,1}$} &
\colhead{$\phi n_{e,0,1}^{2}$} &
\colhead{$h_{z,2}$\tablenotemark{b}} &
\colhead{$\phi n_{e,0,2}^{2}$} &
\colhead{$\chi_{red}^{2}$} &
\colhead{$h_{z,1}$} &
\colhead{$\phi n_{e,0,1}^{2}$} &
\colhead{$h_{z,2}$} &
\colhead{$\phi n_{e,0,2}^{2}$} &
\colhead{$\chi_{red}^{2}$}\\
\colhead{} &
\colhead{(kpc)} &
\colhead{(cm$^{-6}$)} &
\colhead{(kpc)} &
\colhead{(cm$^{-6}$)} &
\colhead{} &
\colhead{(kpc)} &
\colhead{(cm$^{-6}$)} &
\colhead{(kpc)} &
\colhead{(cm$^{-6}$)} &
\colhead{}
}
\startdata
Thick Disk & 1.5 & $6.4 \times 10^{-3}$ & - & - & 77.5 & 1.8 & $6.4 \times
10^{-3}$ & - & - & 80.5 \\
Thick Disk + Halo & 0.8 & $1.6 \times 10^{-2}$ & 6.0 & $6.0 \times 10^{-4}$ &
77.4 & 1.2 & $9.0 \times 10^{-3}$ & 5.0 & $7.0 \times 10^{-4}$ & 79.2
\enddata 
\tablenotetext{a}{The East side of the disk (p1, p4 in Table 2) was fit
  separately from the West side (p2, p3 in Table 2).}  \tablenotetext{b}{The
  parameters of the second (halo) component are taken from \citet{Rand1997}.}
\end{deluxetable*}

\subsection{Thermal and Turbulent Support}\label{sec_thermturb}

\begin{figure}[h]
\epsscale{1.2}\plotone{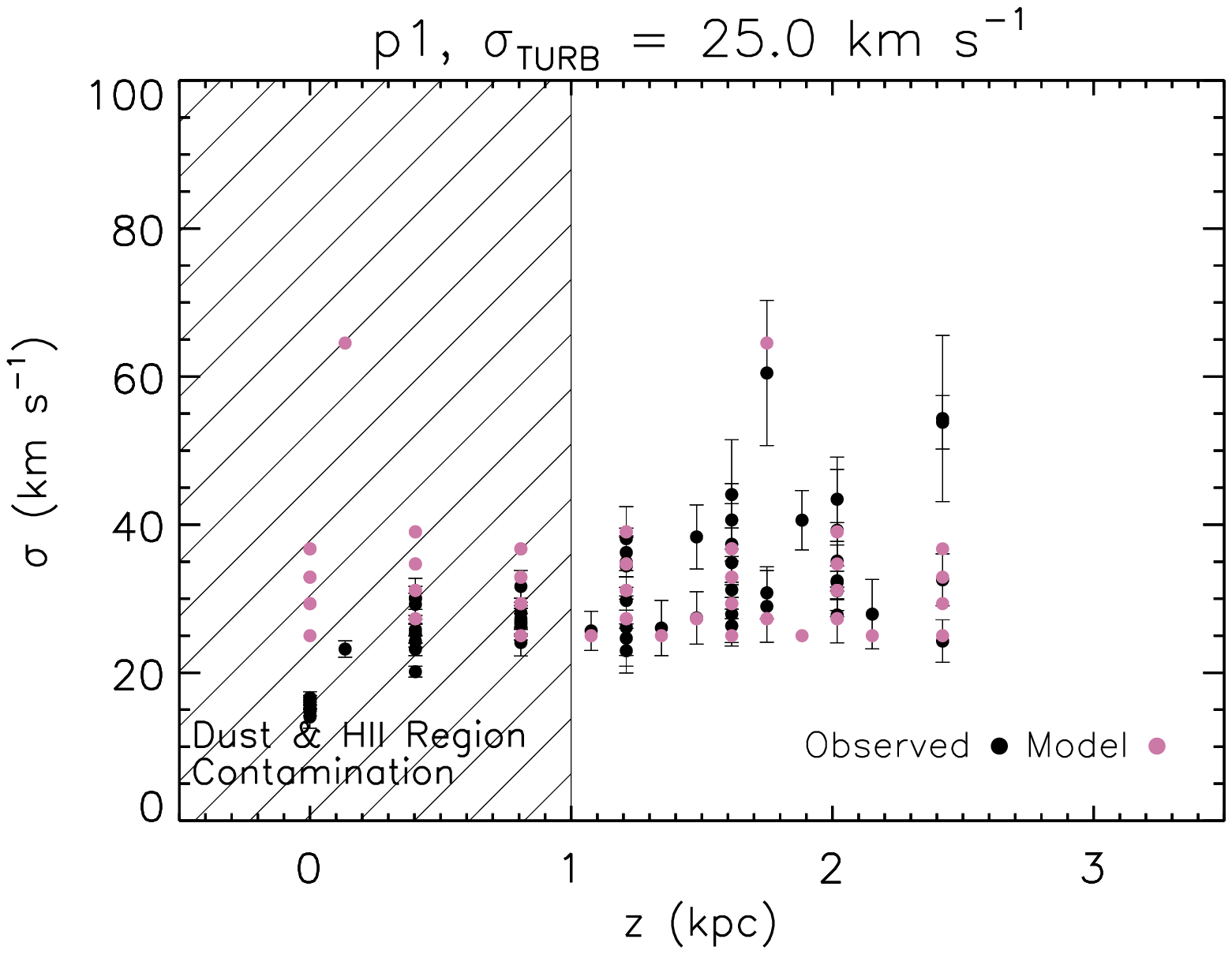}
\epsscale{1.2}\plotone{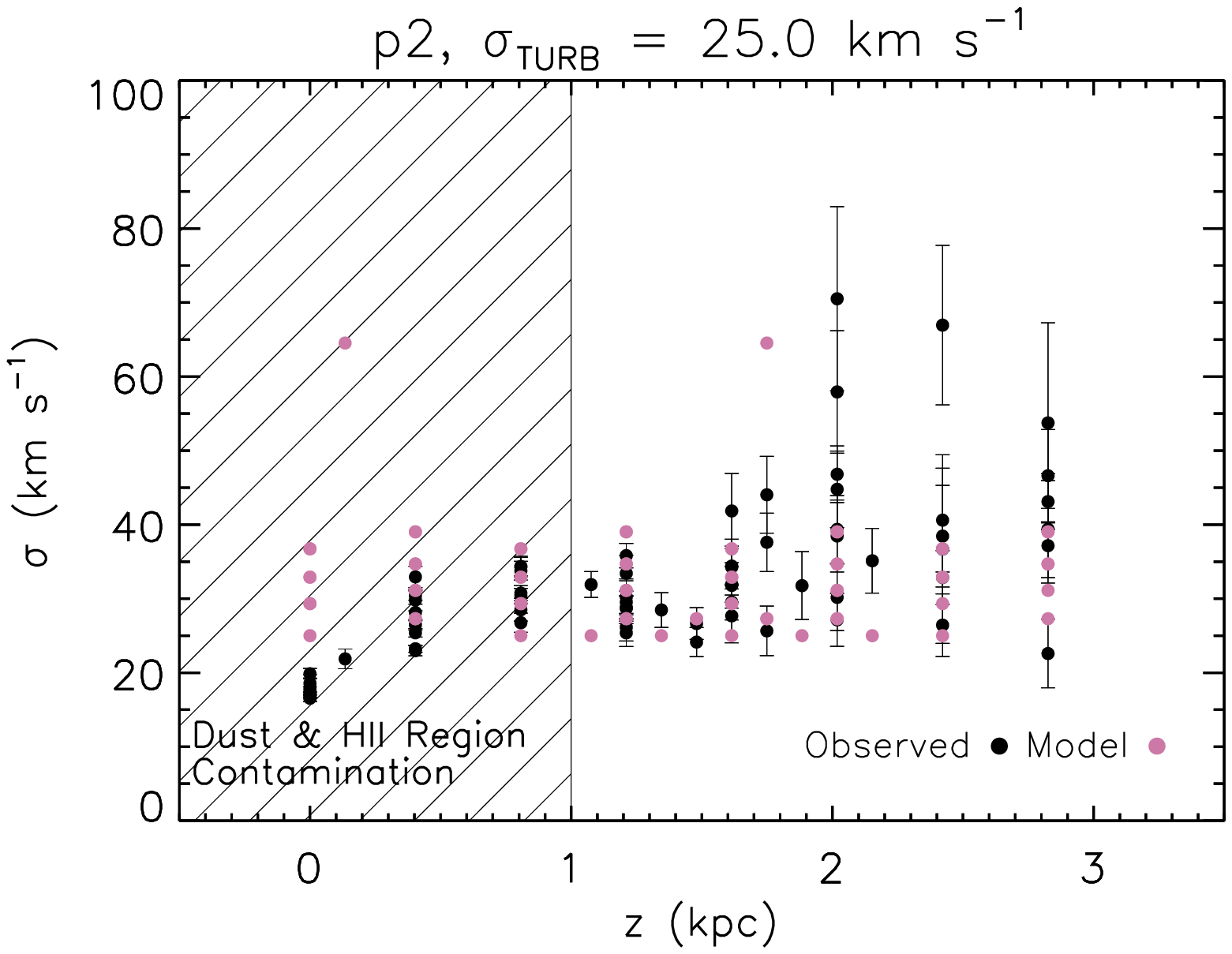}
\caption{The observed, non-thermal emission line widths along the minor axis
  of NGC 891 from p1 and p2 in Table 2. The observed line widths shown are
  median widths calculated from the $H\alpha$, [NII] $\lambda$6583, and [SII]
  $\lambda$6716, 6731 lines after correction for instrumental and thermal
  broadening. We model the emission line profiles and non-thermal line widths
  using the disk model presented in \S5.1.1. We compare the observed and model
  line widths for a range of $\sigma_{turb}$ values using a $\chi_{red}^{2}$
  minimization, and find that $\sigma_{turb} = 25$ km s$^{-1}$ gives a minimum
  $\chi_{red}^{2} \sim 4$ for both pointings. This velocity dispersion is an
  upper limit to the true turbulent velocity dispersion due to the possibility
  of line broadening by non-circular motions.}
\end{figure}

Our second goal is to quantify the thermal and turbulent (random) velocity
dispersions of the eDIG layer, and compare these velocity dispersions to those
required to satisfy the dynamical equilibrium model. We quantify the emission
line widths along the minor axis of the galaxy (p1 and p2 in Table 2), where the
contribution of rotation to the line width is minimized. Our data have the
highest spectral resolution with which the eDIG has been observed along the
minor axis of this galaxy, and thus our data are uniquely suited to
determining the velocity dispersion of the eDIG layer.

The width of an emission line that arises along a line of sight through an
edge-on eDIG layer is a result of several factors: thermal motions
($\sigma_{th}$), turbulent motions ($\sigma_{turb}$), galactic
rotation ($\sigma_{rot}$), and non-circular motions such as streaming along
spiral arms ($\sigma_{nc}$):
\begin{equation}
  \sigma_{tot}^{2} = \sigma_{th}^{2} + \sigma_{turb}^{2} + \sigma_{rot}^{2} + \sigma_{nc}^{2}.
\end{equation}
We consider each term in Equation (15). The term on the left hand side is an
observed quantity for each atomic species (H, N, S) as a function of $R'$ and
$z$. The first term on the right hand side follows for each atomic species
given our assumption of a gas temperature of $T = 10^{4}$ K. The third term
follows from the rotation curve of the galaxy, as well as the rotational
velocity gradient in $z$, given an assumption about the geometry of the eDIG
layer. We cannot robustly quantify the fourth term, which is due to bulk
motions that deviate from our simple assumptions about the geometry and
rotation of the eDIG layer. Thus, we combine the second and fourth terms into
a single $\sigma_{turb}$ term that is understood to represent an upper limit
on the turbulent velocity dispersion of the gas. We assume that the turbulent
motions in the gas are isotropic such that $\sigma_{turb,y} =
\sigma_{turb,z}$. This is the quantity that we desire to measure. The
observed $\sigma_{tot}$ values are given in online Table 3.

We measure an emission line width for all fibers with eDIG detections in
pointings p1 and p2. Although ideally we would only measure line widths for
fibers that fall strictly along the minor axis, the difficulty of precisely
aligning the fiber array with the true rotation axis necessitates the use of
all fibers in the minor axis pointings. For each fiber, we select all emission
lines that are detected at the 4$\sigma$ level or greater ([NII]
$\lambda$6548 is excluded due to its weak intensity). The line widths and
uncertainties are determined by fitting a Gaussian to the line profile using
the IDL function \texttt{mpfitfun} as described in \S4.3. These line widths
are then corrected for instrumental and thermal broadening, and a single,
non-thermal line width is determined by taking a median of the individual line
widths.

These observed, non-thermal line widths are shown in Figure 8 as a function of
$z$ for both minor axis pointings. The smallest line widths ($\sim 15$ km
s$^{-1}$) are observed in the disk, where dust extinction limits the distance
into the galaxy, and thus the rotational broadening, that we can observe. As
$|z|$ increases, so does the line width, and we transition by $|z| = 1$ kpc
from widths characteristic of extincted HII region emission to those
indicative of optically thin eDIG emission. Between $|z| = 1$ and $|z| = 3$
kpc, the line widths range largely between 25 and 40 km s$^{-1}$, with a
spread of $\sim 15$ km s$^{-1}$ observed at all $|z|$. Though there is no
strong evidence of a $z$-dependence in the line widths above $|z| = 1$ kpc,
there are several outliers at a range of $|z|$ values with line widths $\geq
50$ \kms. It is possible that some of these fibers sample gas with
intrinsically greater random motions; however, the low S/N at large $|z|$
means that we cannot distinguish between an intrinsically increased velocity
dispersion and statistical scatter.

To determine the turbulent velocity dispersion, we model the non-thermal
(turbulent and rotational) line widths for a range of turbulent velocity
dispersions, and compare them to the observed non-thermal line widths using a
$\chi_{red}^{2}$ minimization. To model the contribution of the rotational
velocity to the line width, we must make an assumption about the geometry of
the eDIG layer. Although we favor the ring model over the disk model presented
in \S5.1.1, we consider both models here due to our inability to definitively
determine the geometry of the layer. In this way, we consider the turbulent
velocity dispersion given the minimum (ring) and maximum (disk) amount of
rotational broadening.

The disk model results in the maximum contribution of rotation to the line
width. For this model, we create line profiles given the same assumptions as
in \S5.1.1, and compare the modeled line widths to the observed line widths
using a $\chi_{red}^{2}$ minimization. As shown in Figures 8 and 9, for both
minor axis pointings, a miniumum value of $\chi_{red}^{2} = 4$ is found for
$\sigma_{turb} = 25$ km s$^{-1}$. The $\chi_{red}^{2}$ value may exceed unity
by a factor of a few due to deviations from our idealized geometry and
rotation curve, as well as our inability to account for the effects of
non-circular motion. The rotational broadening over the fiber array is
consistent with the $15$ km s$^{-1}$ spread in line width observed between
$|z| = 1 - 3$ kpc, where the minimum observed line widths are consistent with a
purely turbulent origin, and the larger line widths are consistent with both a
turbulent and rotational origin.

The ring model results in a minimum (negligible) contribution of rotation to
the line width for $6 \le R \le 8$ kpc. The observed spread in emission line
width may then be attributed to either a spread in intrinsic velocity
dispersion, non-circular motions, or a combination of the two. Conservatively,
we take the minimum line width observed above $|z| = 1$ kpc to be the
turbulent velocity dispersion, which is again $\sigma_{turb} = 25$ \kms. It
may well be the case that larger observed line widths are due to larger
intrinsic velocity dispersions; however, due to the plausibility of
alternative explanations, we adopt $\sigma_{turb} = 25$ \kms for the purpose
of our dynamical equilibrium model. For a $T = 10^{4}$ K gas, this suggests
that the eDIG layer has at most a sonic Mach number of $M \sim 2 - 3$.

\begin{figure}[h]
\epsscale{1.2}\plotone{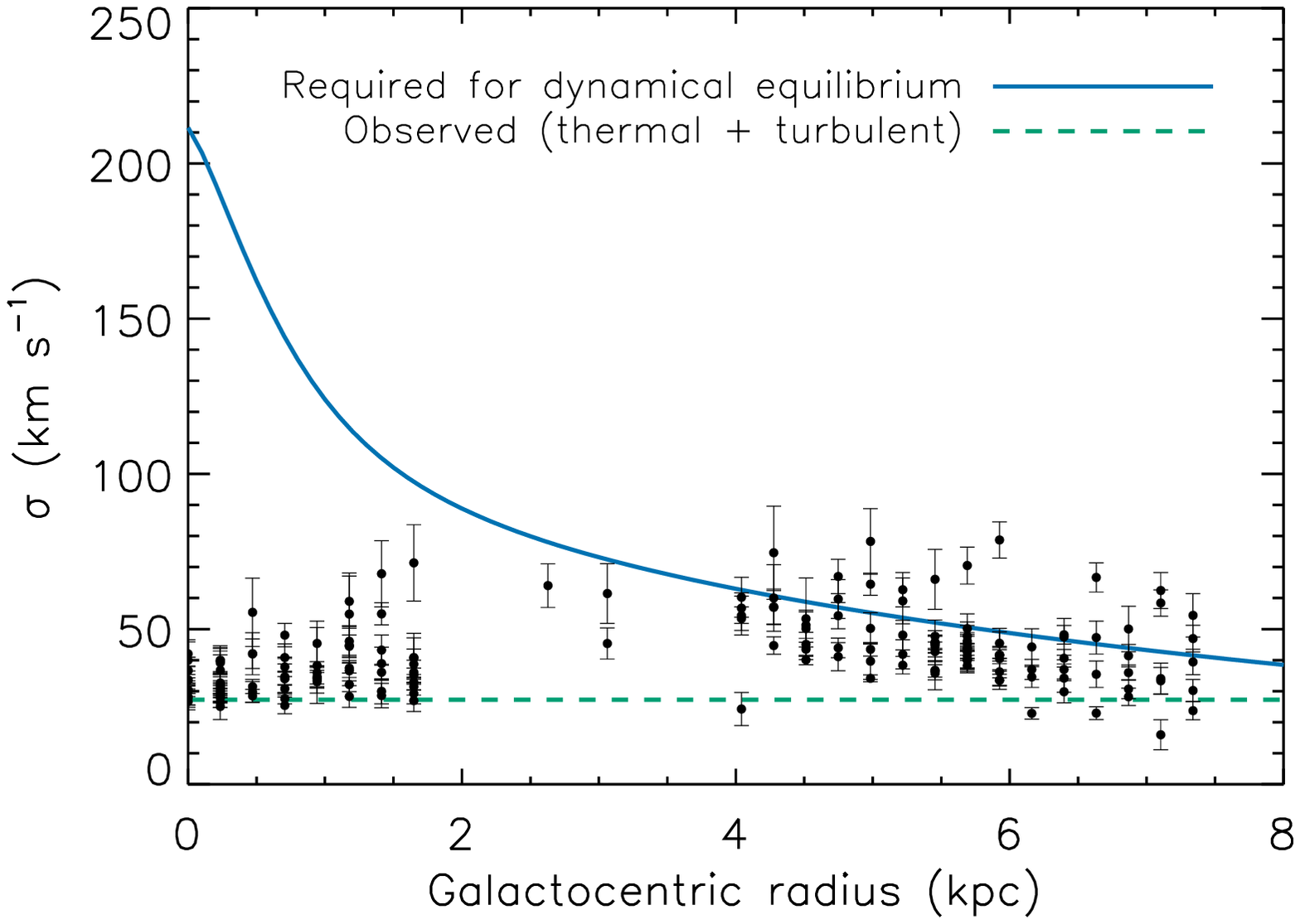}
\caption{We compare the observed velocity dipersion with that required to
  support the eDIG layer by thermal and turbulent pressure alone. The blue
  solid curve shows the required velocity dispersion as a function of
  galactocentric radius determined by solving Equation (5). The green dashed
  line indicates the observed total (thermal and turbulent) velocity
  dispersion, $\sigma = 27$ km s$^{-1}$, also as a function of galactocentric
  radius. Overplotted are the observed total (thermal and non-thermal) line
  widths for every eDIG detection determined as in Figure
  8. The radial dependence of the line widths at $R \ge 4$ kpc is
    likely due to rotational broadening. These points are plotted as a
  function of \emph{projected} radius, illustrating that a dynamical
  equilibrium model is viable only if the gas is found at much larger
  galactocentric radii than projected radii.}
\end{figure}

\subsection{Magnetic Field and Cosmic Ray Support}\label{sec_magcr}

Radio continuum observations reveal an extended synchrotron halo from
extraplanar magnetic fields and cosmic rays in NGC 891, as well as a spatial
correlation between radio continuum and H$\alpha$ emission
\citep{Dahlem1994}. Here, we evaluate the magnetic field and cosmic ray
pressure gradients inferred from the synchrotron emissivity as a function of
height above the disk. We refer to an analysis of 1.5 and 6 GHz observations
from the CHANG-ES survey performed by \citet[][in
preparation]{Schmidt2016}. This survey has obtained radio continuum
observations of 35 nearby edge-on disk galaxies using the Karl G. Jansky Very
Large Array \citep{Irwin2012, Wiegert2015}. We assess the vertical pressure
gradients given two different assumptions about energy equipartition between
the magnetic field and cosmic ray energy densities.

First, we consider the case where energy equipartition holds in the halo of
NGC 891. There is evidence from polarization studies of NGC 891 and other
edge-on galaxies for a magnetic field morphology that is largely parallel to
the midplane in the disk and increasingly perpendicular (``X-shaped'') in the
halo \citep[e.g.,][]{Krause2009}. We assume a magnetic field with a
simple, plane-parallel geometry of the form:
\begin{equation}
  B(z) = B_{0}e^{-|z|/h_{z,B}},
\end{equation}
where $B_{0}$ is the magnetic field strength in the disk and $h_{z,B}$ is the
magnetic field scale height. From the assumption of energy equipartition
($U_{cr} = U_{B}$), as well as that $P_{cr} = 0.45 U_{cr} = 0.45 U_{B}$
\citep{Ferriere2001}, we find:
\begin{equation}
  P_{B} + P_{cr} = U_{B} + 0.45 U_{B} = 1.45 \frac{B(z)^{2}}{8\pi},
\end{equation}
\begin{equation}
  P_{B} + P_{cr} = \frac{1.45}{8\pi} B_{0}^{2}e^{-2|z|/h_{z,B}}.
\end{equation}
\citet{Schmidt2016} finds a magnetic field strength in the disk of $B_{0} \sim
10$ $\mu$G, and an average non-thermal halo synchrotron scale height of
$h_{z,syn} = 1.3$ kpc at both 1.5 and 6 GHz averaged over the range of
galactocentric radii considered for the eDIG layer. This is broadly consistent
with past measurements; \citet{Dumke1998} find $h_{z,syn} = 1.8$ kpc excluding
the inner part of the galaxy at 4.85 GHz. \citet{Schmidt2016} shows that the
synchrotron scale height flares near the edges of the eDIG layer, but we
exclude these larger scale heights from our average because they are found in
radial bins centered at $R' > R_{max}$. The assumption of energy equipartition
allows the magnetic field scale height to be estimated from the synchrotron
scale height by $h_{z,B} = h_{z,syn}(3 - \alpha)$. Therefore, for a spectral
index $\alpha = -1$, we find $h_{z,B} = 4h_{z,syn} = 5.1$ kpc.

We now assess whether the available thermal, turbulent, magnetic field, and
cosmic ray pressure can support the eDIG layer at its observed scale
height. For each of these pressure terms, we calculate the vertical pressure
gradient $dP/dz$, and compare the sum of these pressure gradients to that
required to satisfy the hydrostatic equilibrium equation (i.e., to the product
of the density distribution found in \S5.1 and the gravitational acceleration
found in \S2). For $\phi = 1$, the pressure gradient is only sufficient to
satisfy the dynamical equilibrium model at a minimum galactocentric radius of
$R_{eq} = 8$ kpc. Due to the monotonically decreasing strength of the
gravitational field as a function of $R$, it is also sufficient to satisfy the
model at all larger galactocentric radii ($R \ge 8$ kpc). Where the pressure
gradient does satisfy the model, it does so largely due to the magnetic field
and cosmic ray pressure, with only small contributions from the thermal and
turbulent pressures. The relative contributions of these pressure terms are
shown in the top panel of Figure 10.

Now, we consider the case where energy equipartition no longer holds in the
halo of this galaxy. \citet{Schmidt2016} argues that energy equipartition is
not a good assumption in the halo due to discrepancies between the magnetic
field scale height found from equipartition arguments ($h_{z,eq} = 5.1$ kpc)
and from solving the cosmic ray transport equation ($h_{z,neq} = 3.2$
kpc). The latter value follows from modeling diffusive and advective cosmic
ray transport from the disk to the halo, and comparing observed and modeled
synchrotron emissivities and spectral indices as a function of height above
the disk. We then solve for the cosmic ray pressure as a function of height by
requiring that the synchrotron emissivity remain consistent as a function of
$z$ between the equipartition and non-equipartition cases (i.e., that the
product of the magnetic field and cosmic ray energy densities is consistent
between the models). Performing the same analysis as for the equipartition
case, we again find that the dynamical equilibrium model is satisfied at $R
\geq 8$ kpc, though this model relies even more heavily on the magnetic
pressure gradient to achieve dynamical equilibrium (see the bottom panel of
Figure 10).

\begin{figure}[h]
\epsscale{1.2}\plotone{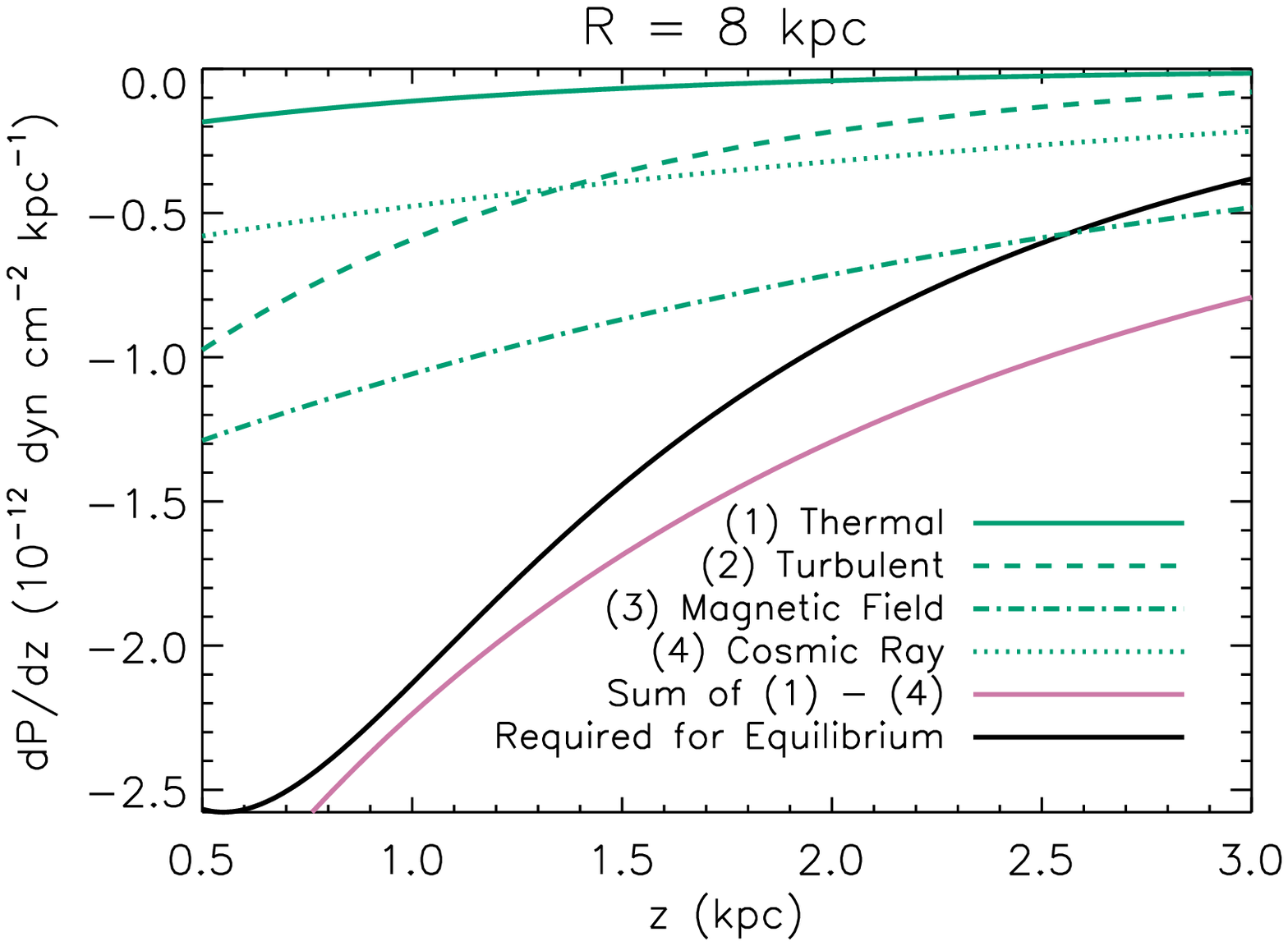}
\epsscale{1.2}\plotone{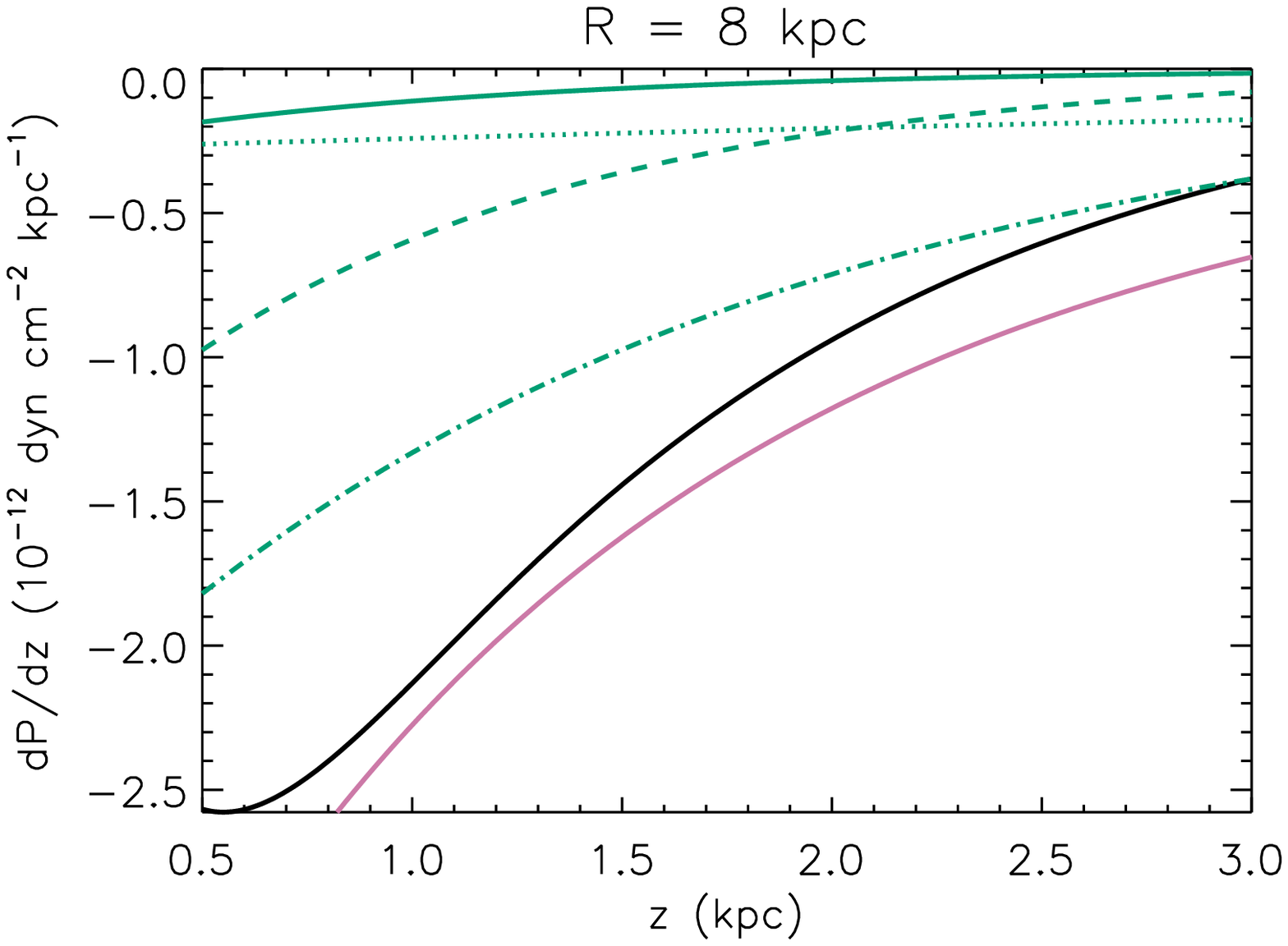}
\caption{The vertical pressure gradients $dP/dz$ are shown at $R = 8$ kpc for
  each source of vertical support considered in this paper: thermal ($T =
  10^{4}$ K) and turbulent motions ($\sigma_{turb} = 25$ km s$^{-1}$), as well
  as magnetic field and cosmic ray pressure inferred from radio continuum
  observations from the CHANG-ES survey \citep[][in
  preparation]{Schmidt2016}. The sum of these pressure gradients is compared
  to the required gradient for dynamical equilibrium determined by solving
  Equation (5) (i.e., to the product of the gas density and the gravitational
  acceleration). The dynamical equilibrium model is satisfied at $R \geq 8$
  kpc for $\phi = 1$ (i.e., $|dP/dz| \geq |dP/dz|_{eq}$). The top panel
  considers the case of energy equipartition between the magnetic field and
  the cosmic rays, and the bottom uses a magnetic field scale height
  determined by \citet{Schmidt2016} by solving the cosmic ray transport
  equation.}
\end{figure}

For the choice of eDIG filling factor, $\phi$, and magnetic field properties,
$B_{0}$ and $h_{z,B}$, considered so far, the dynamical equilibrium model is
not satisfied over the range of galactocentric radii where most of the eDIG is
found ($R \le 8$ kpc). However, the values of $\phi$, $B_{0}$, and $h_{z,B}$
are uncertain. It is likely that the gas is clumpy, reducing $\phi$ by as much
as an order of magnitude. If the magnetic field has an X-shaped rather than a
plane-parallel geometry, then the magnetic scale height may increase as a
function of galactocentric radius, as is observed by
\citet{Schmidt2016}. Additionally, if the synchroton-emitting region has a
ring geometry instead of the disk geometry assumed by \citet{Schmidt2016}, as
is likely the case for multiple phases of the ISM, then a stronger magnetic
field strength is required to reproduce the same synchrotron emissivity over a
shorter pathlength.

We consider how the success of the dynamical equilibrium model is affected by
variations in $\phi$, $B_{0}$, and $h_{z,B}$ in Figure 11. We vary $\phi$
between $0 \le \phi \le 1$, $h_{z,B}$ between $1 \le h_{z,B} \le 10$ kpc, and
$B_{0}$ by 50$\%$, and plot the minimum galactocentric radius at which the
dynamical equilibrium model is satisfied, $R_{eq}$, for both equipartition and
non-equipartition assumptions. As before, the model is satisfied at all $R \ge
R_{eq}$. In the $\phi$-$B_{0}$ plane (left panels), a smaller value of $\phi$
(e.g., denser eDIG clouds and filaments) requires a stronger magnetic field
(e.g., a steeper non-thermal pressure gradient) to satisfy the model at a
given value of $R$. In the $\phi$-$h_{z,B}$ plane (right panels), a smaller
value of $\phi$ requires a smaller value of $h_{z,B}$ to satisfy the model at
a given value of $R$ until the eDIG and non-thermal pressure scale heights
become comparable, at which point the non-thermal pressure can no longer
support the eDIG at large $z$.

Over much of parameter space, the dynamical equilibrium model again fails at
galactocentric radii where the eDIG is found ($R \le 8$ kpc). High magnetic
field strengths ($B_{0} \ge 12 \mu$G) and large eDIG filling factors ($\phi
\ge 0.5$) are required to satisfy the model at $R = 6 - 7$ kpc, and no choice
of parameters satisfies the model at $R < 6$ kpc. Thus, if the eDIG in NGC 891
is in dynamical equilibrium, the eDIG should be found in a ring over a limited
range of galactocentric radii ($6 \le R \le 8$ kpc).

\begin{figure*}[h]
\epsscale{1.0}\plottwo{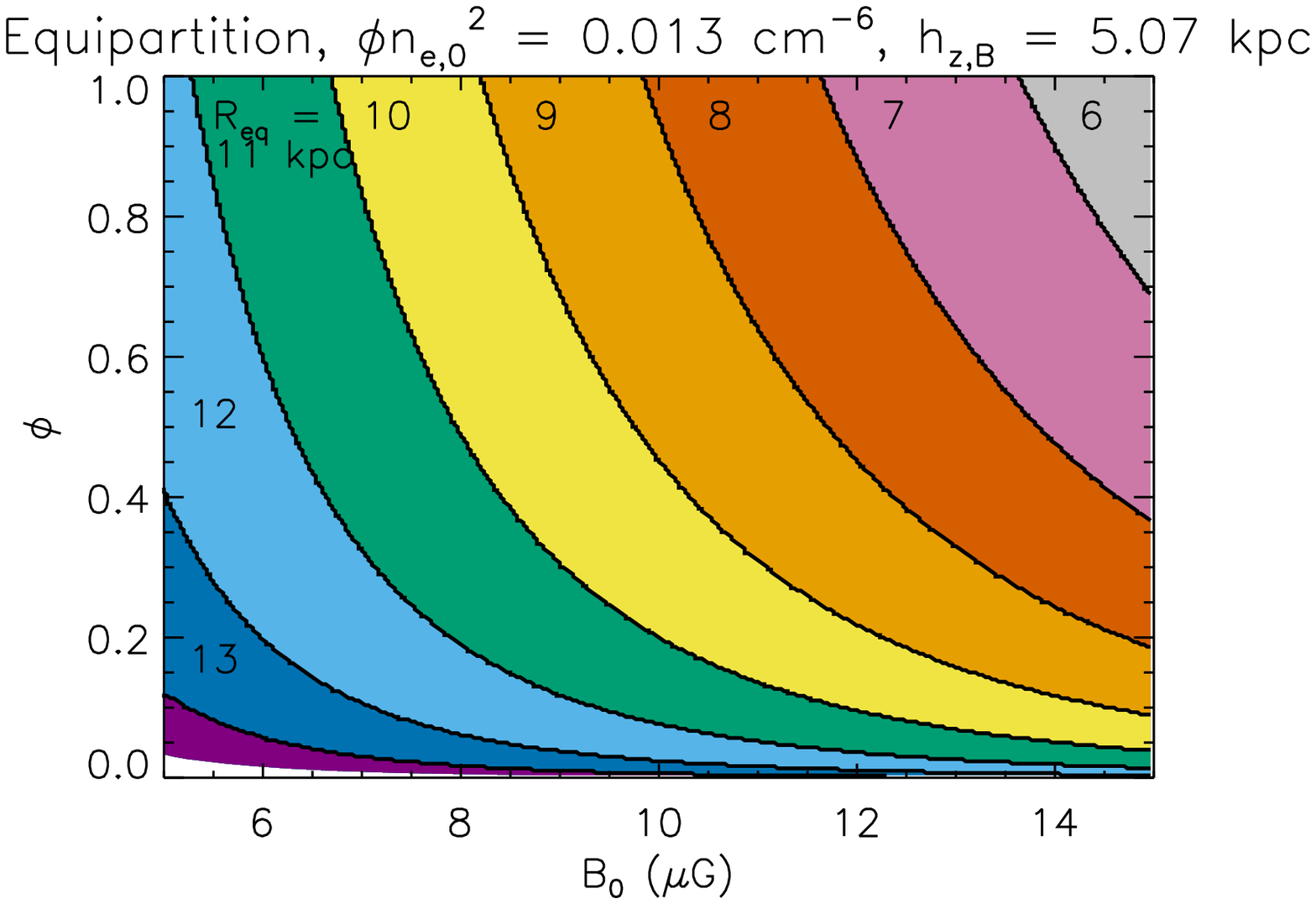}{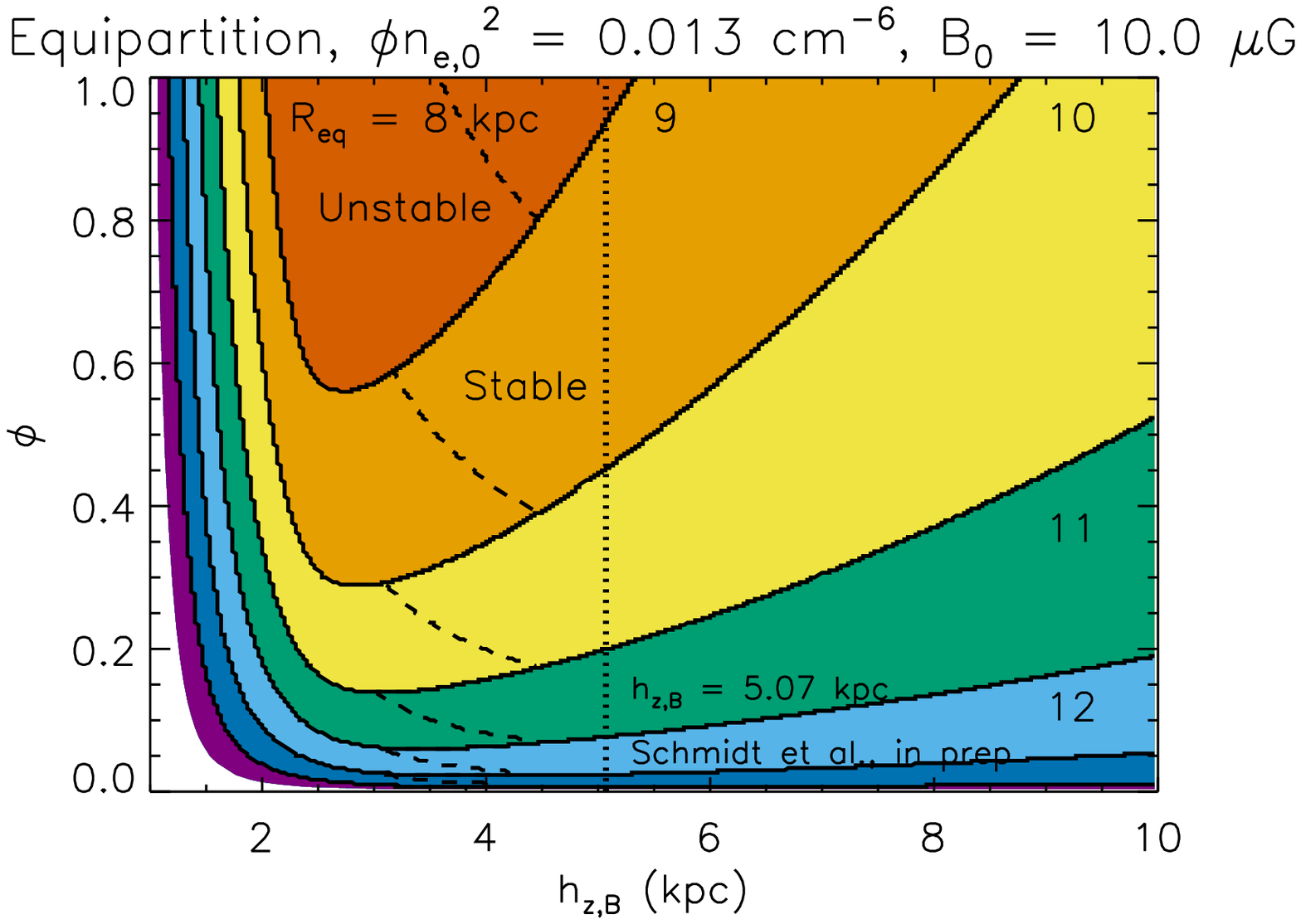}
\epsscale{1.0}\plottwo{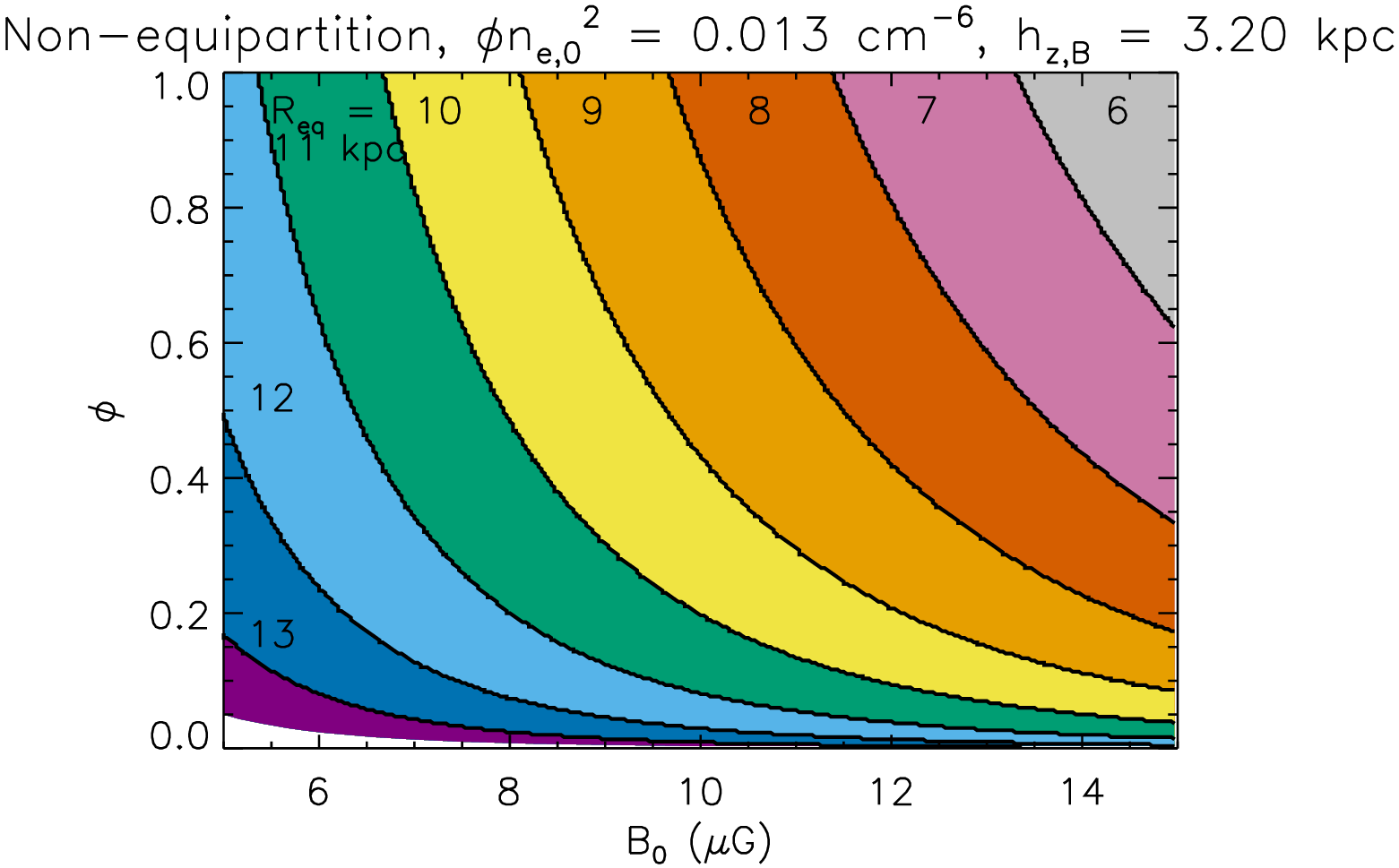}{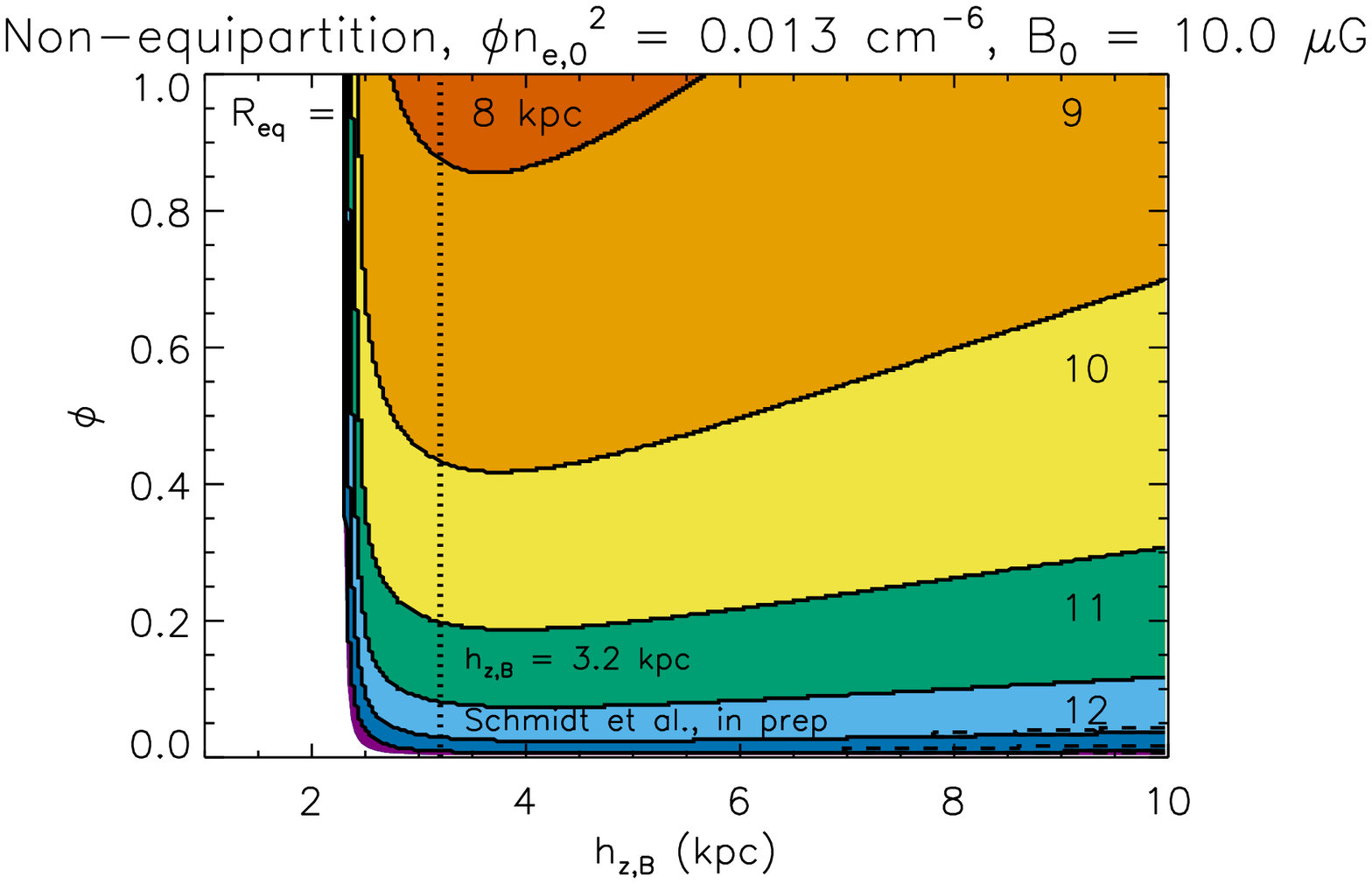}
\caption{Allowed radial range for dynamical equilibrium as a function of eDIG
  filling factor, $\phi$, magnetic field strength, $B_{0}$, and magnetic field
  scale height, $h_{z,B}$. We plot the minimum galactocentric radius at which
  the dynamical equilibrium model is satisfied, $R_{eq}$. The top and bottom
  panels consider the cases of magnetic field and cosmic ray energy
  equipartition and non-equipartition, respectively; the average values of
  $h_{z,B}$ from \citet[][in preparation]{Schmidt2016} are shown with dotted
  lines (see \S5.3). Even allowing for variations in these scale heights by a
  factor of 2 - 3, as well as variations in $B_{0} = 10 \mu$G by $\sim 50$\%,
  the dynamical equilibrium model is only satisfied at $R \geq 8$ kpc over
  most of parameter space. All models shown are Parker stable for $\gamma_{cr}
  = 1.45$ with the exception of those to the left of the dashed lines in the
  right panels; these models can be made to be stable by moving the gas to $R
  > R_{eq}$.}
\end{figure*}

\subsection{Stability of a Magnetized eDIG Layer}\label{sec_stab}

A magnetized plasma layer is subject to the well-known Parker instability
\citep{Parker1966}, and we assess the stability of our dynamical equilibrium
model here. The stability criterion for a horizontally magnetized plasma layer
modified to include cosmic rays is given by:
\begin{equation}
  -\frac{d \rho}{dz} > \frac{\rho^{2}g_{z}}{\gamma_{g} P_{g} + \gamma_{cr} P_{cr}},
\end{equation}
where $\gamma_{g}$ and $\gamma_{cr}$ refer to the adiabatic index of the
thermal gas and the cosmic rays, respectively \citep{Newcomb1961, Parker1966,
  Zweibel1975}. We evaluate whether the $\gamma_{g}$ value required for
stability is reasonable for two choices of $\gamma_{cr}$. Although
\citet{Zweibel1975} argue that values as large as $\gamma_{g} = 5/3 - 2$ are
appropriate for the turbulent, star-forming ISM, we choose the more
conservative value of $\gamma_{g} \leq 1$ for the eDIG \citep{Parker1966}. Our
choice of a plane-parallel magnetic field geometry is an over-simplification,
but has the advantage that the stability analysis is exactly
solvable. \citet{Asseo1978, Asseo1980} show similar results for more
complicated, curved configurations.

If the cosmic rays are coupled to the system via scattering ($\gamma_{cr} =
1.45$ for our choice of $P_{cr} = 0.45 U_{cr}$; \citealt{Zweibel2013}), they
contribute to stabilizing the layer for the motions parallel to the ambient
magnetic field characteristic of the Parker instability. In Figure 12, the
minimum $\gamma_{g}$ required for stability is shown as the dashed curves for
a range of galactocentric radii, the magnetic field parameters of
\citet{Schmidt2016} (see \S5.3), and no gas clumping ($\phi = 1$). If instead
the cosmic rays are not coupled to the system and simply diffuse along field
lines ($\gamma_{cr} = 0$), then they have the effect of destabilizing the
layer. The minimum $\gamma_{g}$ required for stability is again shown as the
solid curves in Figure 12. For both the equipartition and non-equipartition
cases, the stability criterion is only satisfied if $\gamma_{cr} = 1.45$, and
even then only at $R \geq 8$ kpc.

In Figure 11, we indicate the regions of parameter space where the dynamical
equilibrium model is unstable at $R_{eq}$ for $\gamma_{cr} = 1.45$. In these
cases, stability can be achieved by moving the gas to $R > R_{eq}$. Thus, our
dynamical equilibrium model is stable only if the cosmic rays are well-coupled
to the system, and, in some parts of parameter space, the minimum
galactocentric radius required for stability is greater than that required for
dynamical equilibrium.

\begin{figure}[h]
\epsscale{1.2}\plotone{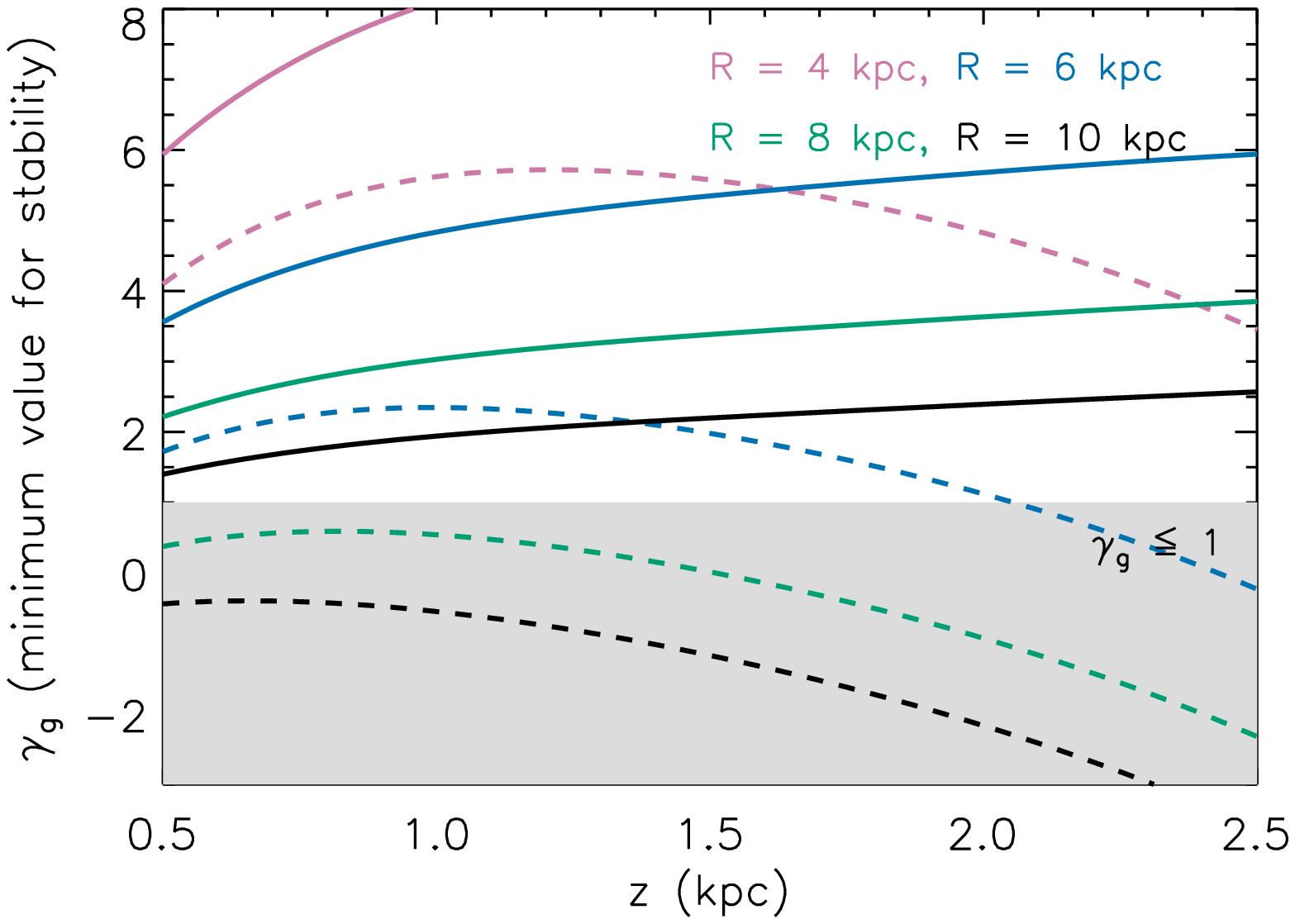}
\epsscale{1.2}\plotone{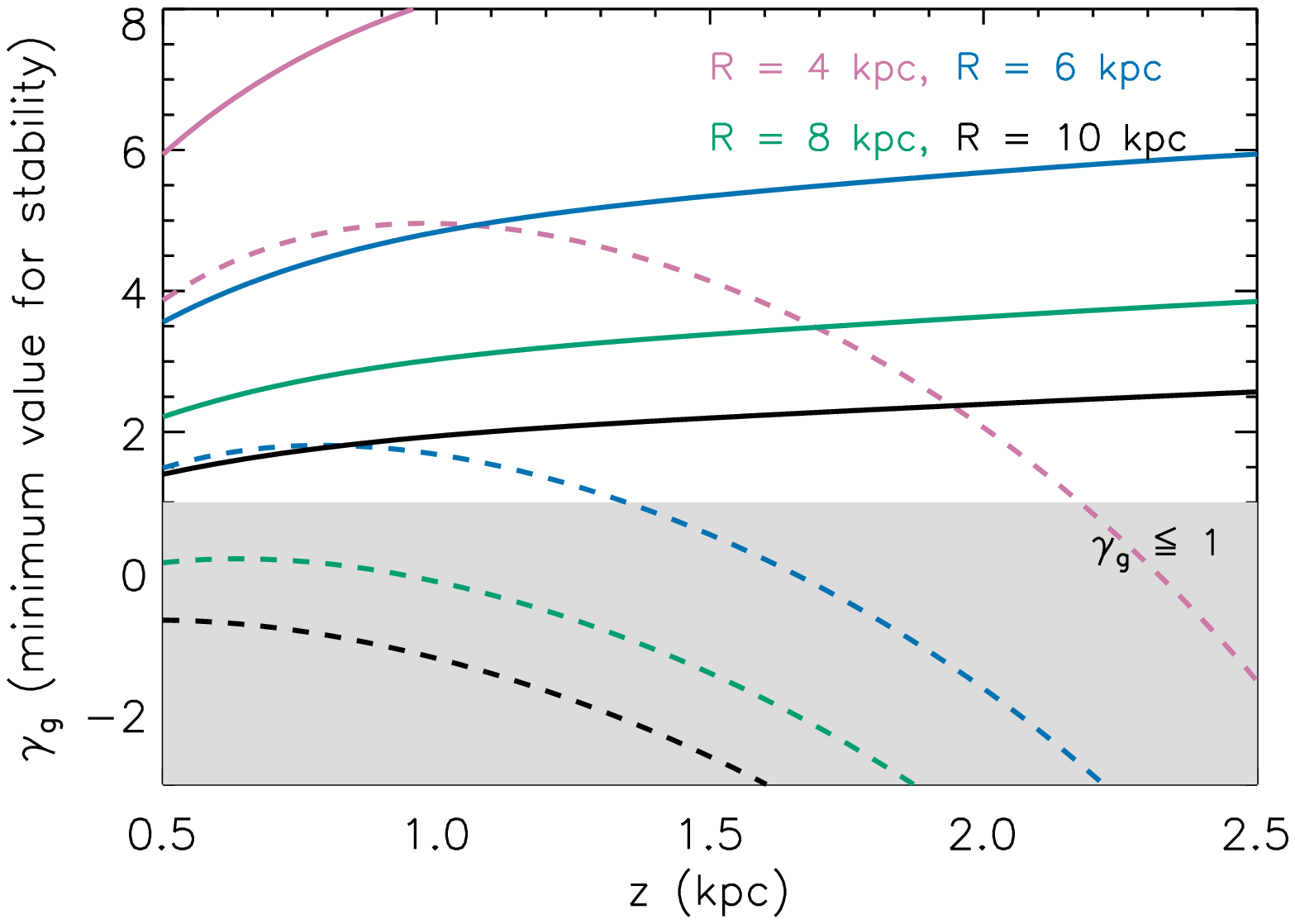}
\caption{We assess whether a magnetized eDIG layer in dynamical equilibrium is
  Parker stable by evaluating the minimum adiabatic index of the gas,
  $\gamma_{g}$, needed to satisfy the stability criterion of Equation (19). We
  do so for two choices of the adiabatic index of the cosmic rays; first, for
  the case in which the cosmic rays are coupled to the gas ($\gamma_{cr} =
  1.45$; dashed lines), and second, for the case in which the cosmic rays are
  not coupled to the system ($\gamma_{cr} = 0$; solid lines). The top panel
  considers the case where the magnetic field and cosmic ray energy densities
  are in equipartition, while the bottom uses a model where the magnetic field
  scale height is determined by solving the cosmic ray transport equation
  \citet[][in preparation]{Schmidt2016}. The stability criterion is only
  satisfied with reasonable values of $\gamma_{g} \leq 1$ for $\gamma_{cr} =
  1.45$ and $R \geq 8$ kpc. An eDIG filling factor of $\phi = 1$ and the
  magnetic field parameters of \citet{Schmidt2016} are assumed here (see
  \S5.3).}
\end{figure}

\section{Discussion}\label{sec_disc}

We tested a dynamical equilibrium model for the bright, vertically extended
eDIG layer in NGC 891. Using optical emission line spectroscopy, we
constrained the three-dimensional density distribution of the eDIG layer, and
found exponential electron scale heights of $h_{z} = 0.8$ kpc and $h_{z} =
1.2$ kpc on the East and West sides of the disk, respectively. We argue that
the symmetry of the emission line profiles, the location of the velocity
centroids in position-velocity space, and the comparable H$\alpha$ intensities
on and off of the minor axis all suggest that the eDIG is preferentially found
in a ring at moderate galactocentric radius ($R_{min} \le R \le 8$ kpc,
$R_{min} \ge 2$ kpc). This is qualitatively consistent with a picture in which
the eDIG is found in discrete clouds and filaments above star-forming spiral
arms where galactic chimneys and superbubbles can break out of the disk. The
small volume filling factor suggested by this picture, as well as the
likelihood of non-circular motions along the line of sight, are in qualitative
agreement with the scatter observed in both the emission line widths and
intensities.

From a mass model of NGC 891, we found that a velocity dispersion of $\sigma =
210 - 40$ \kms is required to support the layer in dynamical equilibrium
between $R = 0 - 8$ kpc. In the mid-disk, this is at least a factor of a few
larger than the thermal velocity dispersion of a $T = 10^{4}$ K gas
($\sigma_{th} = 11$ \kms), and is also larger than the turbulent velocity
dispersion along the minor axis ($\sigma_{turb} = 25$ \kms). Thus, we find
that the eDIG layer in this galaxy is not supported by thermal and turbulent
pressure gradients at any galactocentric radius.

Using radio continuum observations from the CHANG-ES survey \citep[][in
preparation]{Schmidt2016}, we demonstrate that the eDIG layer can be supported
by magnetic field and cosmic ray pressure gradients only at galactocentric
radii of $R \geq 8$ kpc for the magnetic field parameters of
\citet{Schmidt2016} and an eDIG filling factor of $\phi = 1$. As this radius
is comparable to the maximum radius at which the eDIG is observed, this model
is only viable if the eDIG is found in a very thin ring at $R = 8$ kpc. We
also explore variations in the eDIG filling factor, magnetic field strength,
and magnetic field scale height that suggest that a large filling factor
($\phi \ge 0.5$) and a strong magnetic field ($B_{0} \sim 14 \mu$G) are
required for the model to be viable over a larger range of radii ($6 \le R \le
8$) kpc.

A model in which the eDIG is found in a ring is qualitatively consistent with
a picture in which the gas is found primarily over star-forming spiral
arms. CO observations give some clues as to the nature of the spiral structure
in this galaxy. \cite{Sofue1993} argue for a spiral structure similar to that
in the Milky Way Galaxy: a compact nucleus, a bar with radius $R = 3$ kpc, a
ring at $R = 3 - 4$ kpc, and additional intensity peaks suggestive of spiral
arms beyond the ring. It is also possible that the ring at $R = 3.5$ kpc may
be a spiral arm viewed in projection \citep{GarciaBurillo1992,
  Scoville1993}. Given emission line spectroscopy with higher S/N and higher
spectral resolution, a stronger constraint on $R_{min}$ could be found for the
eDIG layer, and thus the geometry of the eDIG layer and the spiral structure
could be more closely compared. There is almost certainly eDIG found within $R
< R_{min}$, but this gas is likely to remain close to the disk due to the
kinematics of the bar and the depth of the potential well, and thus would be
obscured by the dust lane and HII region emission in our observations.

The success of the dynamical equilibrium model depends on our assumptions
about the vertical and radial distribution of the eDIG layer. Our assumption
of a symmetric density distribution, though necessary, is certainly a
simplification. Additionally, a single-component fit to the H$\alpha$
intensity produces a larger thick disk scale height than a two-component fit,
making the model more difficult to satisfy for the thick disk at a given value
of $R$. If the gas is found beyond $R_{max} = 8$ kpc, as is argued by
\citet{Heald2006b}, then our best-fit electron scale height decreases; for
instance, if $R_{max}$ increases by $50\%$, then $h_{z}$ decreases by $\sim
30\%$. Similarly, if the gas is not found within a large $R_{min}$, and the
pathlength through the gas decreases by a factor of a few, then the electron
density, $\phi n_{e,0}^{2}$, increases by the same factor. If the pathlength
through the gas decreases by an order of magnitude, then the dynamical
equilibrium model is only satisfied at $R_{eq} \ge 10$ kpc for the parts of
parameter space considered in Figure 11. Thus, more robustly constraining
$R_{max}$ through deep spectroscopic observations at $R' \ge 8$ kpc, as well
as $R_{min}$ and $R_{max}$ by looking for characteristic radial distributions
of eDIG in face-on disk galaxies, is important for future studies of the
dynamical state of the eDIG layers in these systems.

There are other phenomena that could affect the dynamical equilibrium
of the eDIG layer. First, it is possible that radiation pressure is a
supplemental source of vertical support not considered here. A full treatment
of radiation pressure in the eDIG layer would require knowledge of the
radiation field and the gas-to-dust ratio in the layer
\citep{Franco1991}. Second, any deviations from the simple, plane-parallel
magnetic field geometry considered here require the magnetic tension force to
be taken into account. \cite{Hill2012} unsuccessfully attempted to reproduce
an eDIG scale height of $h_{z} = 1$ kpc using magnetohydrodynamic simulations
of a turbulent, star-forming gas layer. Their simulations failed to produce
sufficient magnetic support for the gas layer due to the canceling of the
magnetic pressure by the magnetic tension force. However, it is unclear
whether this is simply a result of their periodic boundary conditions for the
magnetic field.

It is important that we consider the eDIG layer in NGC 891 as one phase of a
multi-phase gaseous halo. Studies of larger samples of galaxies have shown
correlations in the morphologies and luminosities at radio continuum,
far-infrared, H$\alpha$, ultraviolet, and soft X-ray wavelengths
\citep{Collins2000, Rossa2003a, Tullmann2006b, Tullmann2006a}. As multiple
phases are produced by the same star formation feedback processes, it is
likely that these phases are not only co-produced but also co-evolve over
time. The hot phase is of particular interest for eDIG layers, as extended
soft X-ray emission is spatially correlated with H$\alpha$ emission,
suggesting that the warm phase may exist as cool clouds, clumps, and filaments
embedded in a hotter phase \citep{Strickland2004, Tullmann2006b,
  Tullmann2006a}. \citet{HodgesKluck2013} detect an extended soft X-ray halo
in NGC 891 that shares broad morphological similarities with the H$\alpha$
emission. While the large scale height of the hot halo ($h_{z,hot} \sim 5 - 6$
kpc) suggests that the pressure gradient in this medium is too small to affect
the dynamical equilibrium of the eDIG, the presence of the hot halo may
influence the cloud-cloud velocity dispersion, equilibrium pressure, and
lifetime of eDIG clumps and filaments. In particular, line widths consistent
with a sonic Mach number of $M \sim 2 - 3$ in the warm phase are consistent
with subsonic turbulence in the hot phase, suggesting that the quantity we are
measuring is a cloud-cloud velocity dispersion of warm clumps embedded in a
hot halo.

\citet[][in preparation]{Schmidt2016} suggests that there is evidence for
cosmic ray advection by a galactic wind with advection speeds of a few hundred
\kms in the spectral index profile as a function of height above the
disk. Particularly, by solving the cosmic ray transport equation, they argue
that advective cosmic ray transport produces a spectral index profile more
consistent with observations than diffusive transport. If there is indeed a
galactic outflow of the hot ionized medium, and the warm and hot phases are
co-spatial, then there are implications for the dynamical equilibrium of the
eDIG. Both ram pressure associated with an outflow and magnetic tension
associated with a field that is anchored in the warm phase but advected in the
hot may be additional sources of support. Though a quantitative consideration
of this model is beyond the scope of this paper, the relationship between eDIG
layers and galactic outflows is of interest for further study.

\citet{Oosterloo2007} show the pervasive presence of neutral gas in NGC 891;
nearly $30\%$ of the HI mass is found in a halo with a scale height of
$h_{z,HI} = 1.3$ kpc between the radii of interest ($6 \le R \le 8$ kpc). The
neutral phase may affect the dynamical equilibrium of the ionized phase in
several ways. First, it increases the power requirement of the eDIG by
increasing the energy dissipation via shocks from cloud-cloud
collisions. Second, if both the neutral and ionized gas are supported in
dynamical equilibrium by magnetic field and cosmic ray pressure, then the
magnetic field and cosmic ray pressure gradients would need to be steeper than
required here to simultaneously support both phases. Additionally, satisfying
the stability requirement considered in \S5.4 is more difficult in the neutral
phase; due to ion-neutral damping of the Alfv\'{e}n waves generated by the
streaming instability \citep{Kulsrud1969}, the cosmic rays are not
well-coupled to the gas in the cold phase, and their effective adiabatic index
is lower than in the warm phase ($\gamma_{cr} = 0$). The dynamical state of
the neutral gas in NGC 891 is further complicated by extraplanar gas close to
the systemic velocity or counter-rotating that is suggestive of interaction
and/or accretion \citep{Oosterloo2007}. Thus, the impact of the neutral gas on
the dynamical state of the ionized gas is a nuanced question that cannot be
fully addressed here. A model that simultaneously considers the dynamical
state of the cold, warm, and hot phases in this galaxy is of interest for
future work.

\subsection{Implications for the Milky Way Galaxy}\label{sec_comp}

NGC 891 and the Milky Way Galaxy are often considered ``twin'' galaxies due to
similarities in structure and bolometric luminosity
\citep{vanderKruit1984}. As our internal vantage point presents challenges for
studying the eDIG layer in our own Galaxy (the ``Reynolds layer'';
\citealt{Reynolds1993, Haffner2009}), it is interesting to consider what
insights are gained by comparing the eDIG layers in these systems. In the
Galaxy, the warm, ionized medium (WIM) is observed along almost all lines of
sight in the Wisconsin H-Alpha Mapper (WHAM) survey \citep{Tufte1997}, which
has mapped the northern sky \citep{Haffner2003} and is currently concluding
mapping the southern \citep{Haffner2010}. With a one-degree beam, WHAM has
observed a complex WIM morphology of clumps and filaments superimposed on a
diffuse background. The physical conditions in the WIM are similar to those in
the eDIG of NGC 891, with a temperature $T = 6000 - 10^{4}$ K, an average
electron density $\langle n_{e,0} \rangle = 0.01 - 0.1$ cm$^{-3}$, an
ionization fraction $H^{+}/H > 90\%$ for $T = 10^{4}$ K \citep{Reynolds1998},
and a filling fraction $f \sim 0.1$ in the disk that increases to $f \geq 0.3$
by $|z| = 1$ kpc \citep{Reynolds1991, Haffner1999, Gaensler2008, Hill2008,
  Haffner2009}.

The remarkable feature that the exponential electron scale height greatly
exceeds the thermal scale height is observed in both NGC 891 and in the
Galaxy. A study of the H$\alpha$ intensity as a function of Galactic latitude
above the Perseus spiral arm suggests an exponential electron scale height of
$h_{z} = 1.0 \pm 0.1$ kpc \citep{Haffner1999}. In velocity space, the WIM
broadly traces the kinematics of nearby spiral arms, suggesting that the WIM
may be spatially associated with regions of star-formation activity. Towards
large galactic latitudes, the WIM is preferentially blueshifted, although it
is unclear whether this is indicative of a local feature or of a global bulk
flow \citep{Haffner2003}.

Thus, the dynamical state of the Reynolds layer is in question, and here we
consider a dynamical equilibrium model for this layer by qualitatively
comparing what is known about the thermal, turbulent, magnetic field, and
cosmic ray pressure gradients in NGC 891 and in the Galaxy. An H$\alpha$ line
width of $\sim 20$ \kms FWHM is observed toward the north Galactic pole
(L. M. Haffner, private communication). As this is comparable to the thermal
line width, this single, pencil-beam measurement suggests that the scale
height problem may be even more severe in the Galaxy than in NGC 891. The
non-thermal velocity dispersion of the WIM has also been measured by using the
difference in mass between H and S to separate the thermal and non-thermal
components of the H$\alpha$ and [SII] $\lambda$6716 emission lines
\citep{Reynolds1985}. The S line widths mainly range from $20 - 50$ \kms FWHM,
suggesting $0 \leq \sigma_{turb} \leq 20$ \kms for the WIM. These velocities
are likely a combination of an intercloud and a cloud-cloud velocity
dispersion. These results are consistent with (magneto)hydrodynamic
simulations of turbulence in the WIM. The spatial gradient of linearly
polarized emission is consistent with subsonic or transonic turbulence
\citep{Burkhart2012}, the electron density power spectrum suggests transonic
turbulence \citep{Kim2005}, and the emission measure distribution function
yields slightly supersonic Mach numbers of $M \sim 1.4 - 2.4$
\citep{Hill2008}. Therefore, available evidence suggests that neither the eDIG
layer in NGC 891 nor in the Galaxy are supported by a thermal or turbulent
pressure gradient.

In the Galaxy, the synchrotron halo implies extraplanar magnetic field
properties generally consistent with those of external quiescent galaxies. In
the solar neighborhood, the magnetic field strength is $B_{solar} = 5 - 6
\mu$G \citep{Burlaga2013}, and the magnetic field scale height is $h_{B} =
4.5 - 6$ kpc or larger from equipartition arguments \citep{Beuermann1985,
  Ferriere2001, Beck2016}. Optical polarization of starlight suggests
that the magnetic field lines are largely parallel to the midplane and follow
the structure of the nearest spiral arm in the disk \citep{Fosalba2002}. The
field also shows evidence of the extraplanar X-shaped morphology observed in
other edge-on galaxies \citep{Jansson2012}. Thus, the broad similarities
between the magnetic field strength, scale height, and morphology in NGC 891
and in the Galaxy suggest that magnetic field and cosmic ray pressure
gradients may play an important role in the vertical structure and support of
the Reynolds layer.

A full quantitative comparison of the eDIG layers in these galaxies is beyond
the scope of this paper, and requires considerations that range from the star
formation rate to the accretion and interaction history of each galaxy. There
are also clear differences between the warm, ionized phases in these systems;
compared to the Galaxy, the eDIG layer in NGC 891 has a greater vertical
extent, twice the surface density, and a higher ratio of ionized to neutral
gas, possibly due to the higher star formation rate in the latter system
\citep{Rand1990}. However, the success of the dynamical equilibrium model for
NGC 891 warrants the exploration of such a model for the Galaxy given the
broad similarities in the extraplanar gas and magnetic field properties of
these galaxies.

\section{Conclusions}\label{sec_conc}

We sought to determine the dynamical state of the eDIG layer in NGC 891. This
layer is remarkable due to an observed scale height that exceeds its thermal
scale height by a factor of a few. Specifically, we tested a dynamical
equilibrium model by quantifying the thermal, turbulent, magnetic field, and
cosmic ray pressure gradients in this galaxy. We summarize our results as
follows:

\begin{enumerate}
\item We obtained optical emission line spectroscopy of the eDIG layer using
  the SparsePak IFU on the WIYN telescope. We probed a wide range in height
  above and below the disk ($0 \leq |z| \leq 3.2$ kpc) and in projected radius
  ($-1.65 \leq R' \leq 7.35$ kpc) with moderate spectral resolution ($\sigma =
  17$ \kms at H$\alpha$). We found a thick disk exponential electron scale
  height of $h_{z} = 0.8$ kpc and $h_{z} = 1.2$ kpc on the East and West sides
  of the galaxy, respectively. This is consistent with past measurements in
  the literature.
\item Several pieces of evidence point to the eDIG being found preferentially
  at moderate galactocentric radius ($R_{min} \leq R \leq 8$ kpc, where
  $R_{min} \ge 2$ kpc). These include the comparable $I_{H\alpha}(z)$ on and
  off of the minor axis, the lack of low- and high-velocity emission line
  wings, and the location of the observed velocity centroids in
  position-velocity space.
\item We measured the H$\alpha$, [NII] $\lambda$6583, and [SII] $\lambda$6716,
  6731 emission line widths along the minor axis, and show that they are
  consistent with a turbulent medium with a sonic Mach number of $M = 2 -
  3$. The thermal ($\sigma_{th} = 11$ \kms) and turbulent ($\sigma_{turb} =
  25$ \kms) velocity dispersions are far below that required to support the
  eDIG layer by thermal and turbulent pressure gradients between $R = 0$ and
  the observed cutoff at $R = 8$ kpc ($\sigma = 210 - 40$ \kms). The observed
  turbulent velocity dispersion is supersonic for the warm phase, but is
  subsonic for the hot halo; this is consistent with the eDIG being a
  collection of cool clouds embedded in a hot, potentially outflowing phase.
\item We referred to an analysis by \citet[][in preparation]{Schmidt2016} of
  the synchrotron halo in NGC 891 using CHANG-ES radio continuum observations
  to determine the magnetic field and cosmic ray pressure gradients in this
  system. The combined thermal, turbulent, magnetic field, and cosmic ray
  pressure gradients are sufficient to support the eDIG layer at a scale
  height of $h_{z} = 1$ kpc at galactocentric radii of $R \geq 8$ kpc. The
  uncertainty on the eDIG filling factor, magnetic field strength, and
  magnetic scale height yield an uncertainty of a few kpc on this
  galactocentric radius, and thus it is possible that the eDIG is supported in
  dynamical equilibrium in a thin ring between $6 \le R \le 8$ kpc.
\item Our dynamical equilibrium model of a magnetized eDIG layer is Parker
  stable if the cosmic rays are sufficiently coupled to the system
  ($\gamma_{cr} = 1.45$).
\item Similarities between the thermal and turbulent properties of the warm
  ionized gas as well as the synchrotron halos in NGC 891 and the Milky Way
  Galaxy suggest that extraplanar magnetic fields and cosmic rays may play an
  important role in the dynamical state of the Reynolds layer.
\item In future work, a simultaneous treatment of the extraplanar cold, warm,
  and hot gas is desired to understand the dynamical state of the multi-phase
  gaseous halo.
\end{enumerate}  

Studying the eDIG layers of nearby edge-on disk galaxies has both advantages
and disadvantages. We are able to directly determine the vertical scale height
of the layer and quantify the gas, magnetic field, and cosmic ray pressure as
functions of height above the disk. However, we cannot directly measure the
vertical velocity dispersion or look for evidence of vertical inflow, outflow,
or other non-equilibrium phenomena. We also cannot definitively determine the
location of the gas along the line of sight, or search for spatial
correlations between the properties of the halo gas and the underlying
disk. Thus, fully characterizing the dynamical state of eDIG layers requires
observations of disk galaxies with a range of inclination angles. Future work
will focus on face-on disk galaxies at high spectral resolution to directly
measure the vertial velocity dispersion and any vertical bulk flows, determine
the radial distribution of the gas, explore connections between the gaseous
halo, the underlying stellar disk, and the intergalactic environment, and
assess the generality of the model presented here (Boettcher, Gallagher,
Zweibel, \& Benjamin, in preparation).

The future is bright for understanding the dynamics of extraplanar gas through
careful studies of small samples of nearby galaxies, as well as through
ongoing surveys studying the gaseous, magnetic field, and cosmic ray halos in
these systems. Large IFU surveys including the Mapping Nearby Galaxies at
Apache Point Observatory (MaNGA) project are enabling the statistical study of
extraplanar gas properties in hundreds of low-redshift galaxies as part of the
fourth-generation Sloan Digital Sky Survey (SDSS-IV)
\citep{Bundy2015}. Additionally, the CHANG-ES survey will allow the
extraplanar magnetic field and cosmic ray properties of tens of nearby
galaxies to be characterized, and thus the relationship between gaseous and
synchrotron halos to the studied in a statistical sense.

\acknowledgments Based on observations at Kitt Peak National Observatory,
National Optical Astronomy Observatory (NOAO Prop. ID: 2014B-0455; PI:
Boettcher), which is operated by the Association of Universities for Research
in Astronomy (AURA) under a cooperative agreement with the National Science
Foundation. This work has made use of NASA's Astrophysics Data System and of
the NASA/IPAC Extragalactic Database (NED) which is operated by the Jet
Propulsion Laboratory, California Institute of Technology, under contract with
the National Aeronautics and Space Administration. 

We thank the anonymous referee for helpful comments that improved the quality
and clarity of our manuscript. We are grateful to Philip Schmidt for sharing
his analysis of the CHANG-ES observations of NCC 891 in advance of
publication, and to Marita Krause, Ralf-J\"{u}rgen Dettmar, and the rest of
the CHANG-ES collaboration for useful discussions. We thank L. Matthew Haffner
for his insights on the WIM in the Milky Way Galaxy, and J. Christopher Howk,
Matthew Bershady, and Arthur Eigenbrot for helpful conversations about NGC
891.  We acknowledge the NOAO/WIYN support staff for their help during data
collection and reduction, Christy Tremonti for providing a Gaussian emission
line fitting code, and Masataka Okabe and Kei Ito for supplying the
colorblind-friendly color palette used in this paper (see
\url{fly.iam.u-tokyo.ac.jp/color/index.html}).

This material is based upon work supported by the National Science Foundation
Graduate Research Fellowship Program under Grant No. DGE-1256259. Any
opinions, findings, and conclusions or recommendations expressed in this
material are those of the author(s) and do not necessarily reflect the views
of the National Science Foundation. Support was also provided by the Graduate
School and the Office of the Vice Chancellor for Research and Graduate
Education at the University of Wisconsin-Madison with funding from the
Wisconsin Alumni Research Foundation.

\clearpage
\appendix

\begin{figure}[h]
\epsscale{1.0}\plottwo{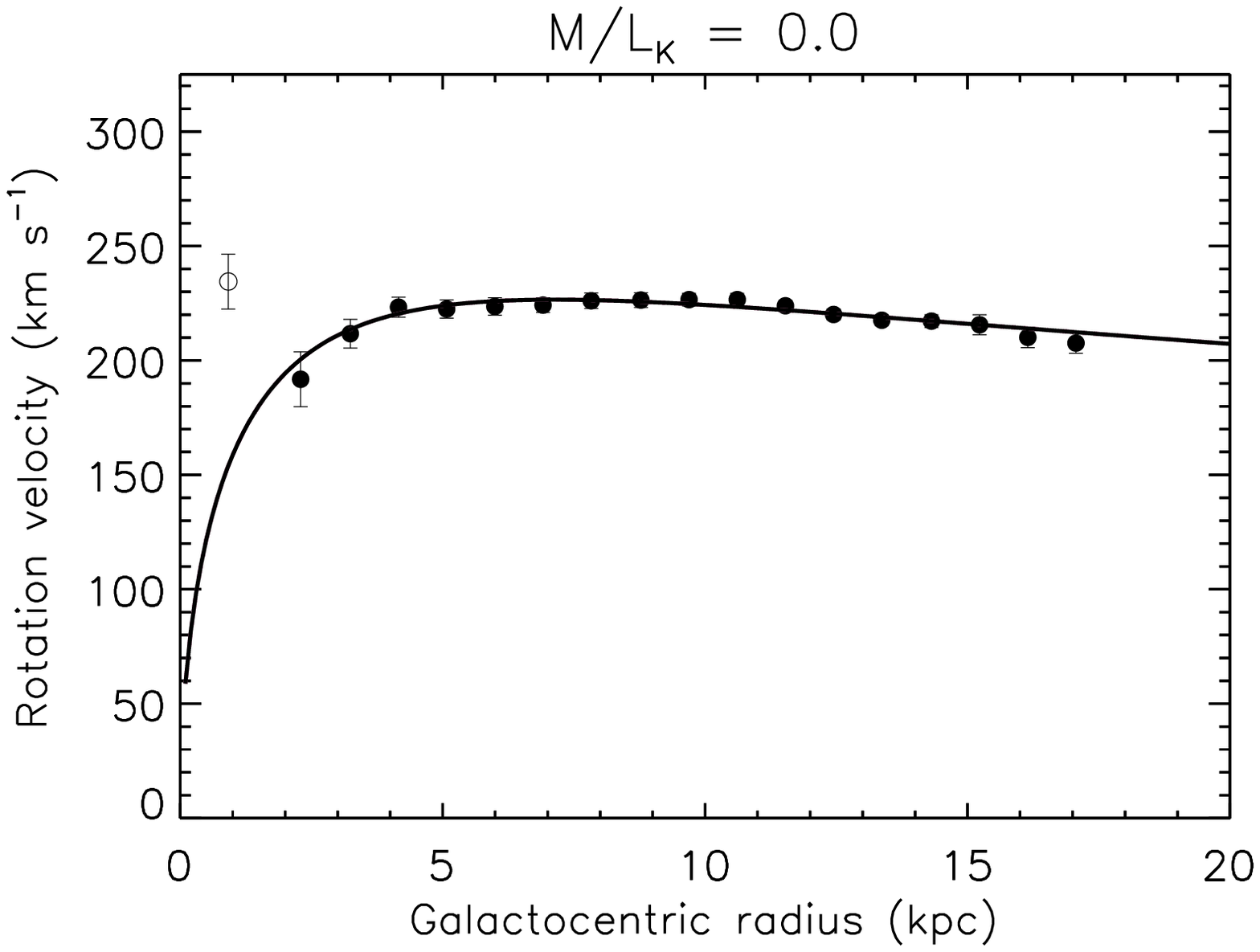}{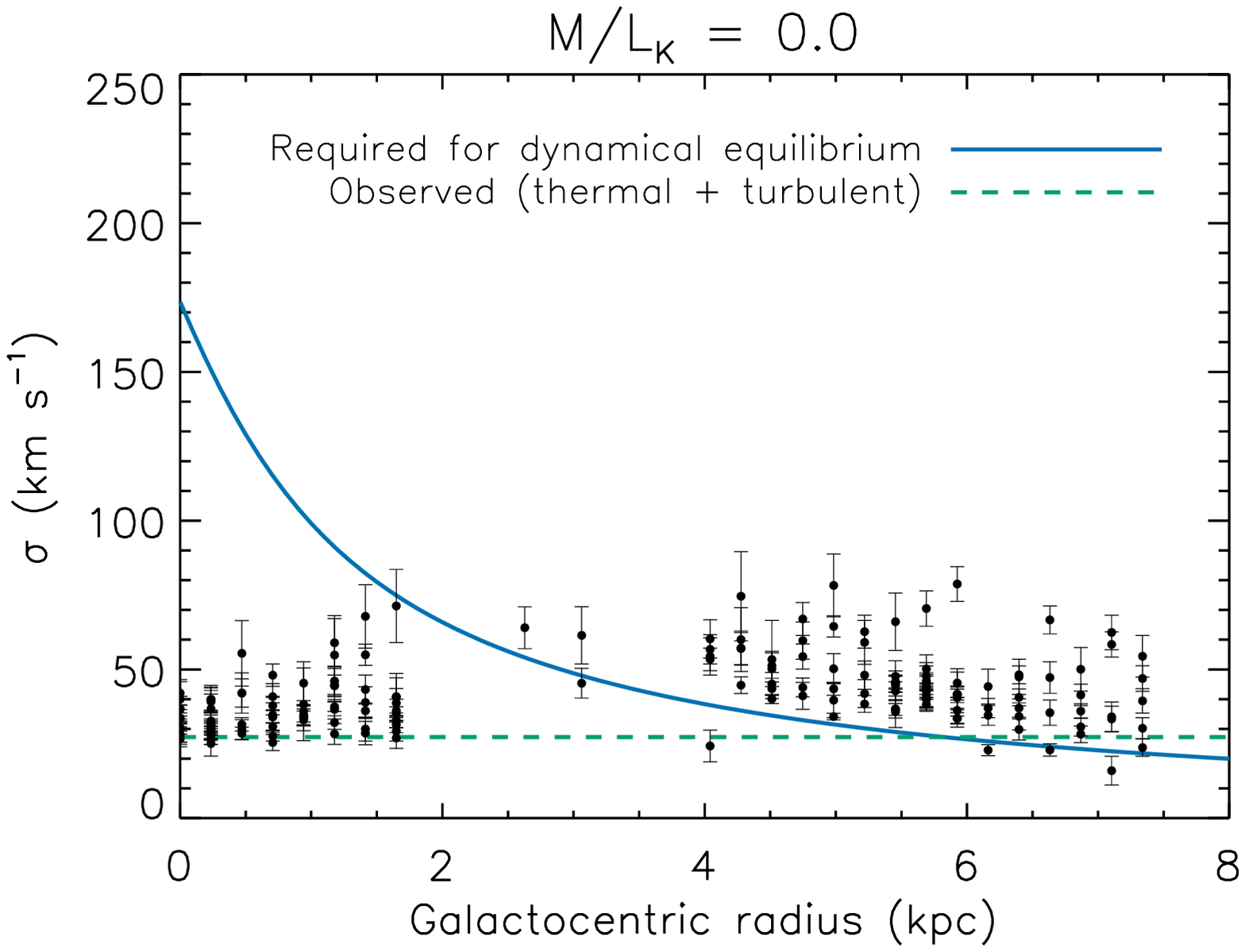}
\epsscale{1.0}\plottwo{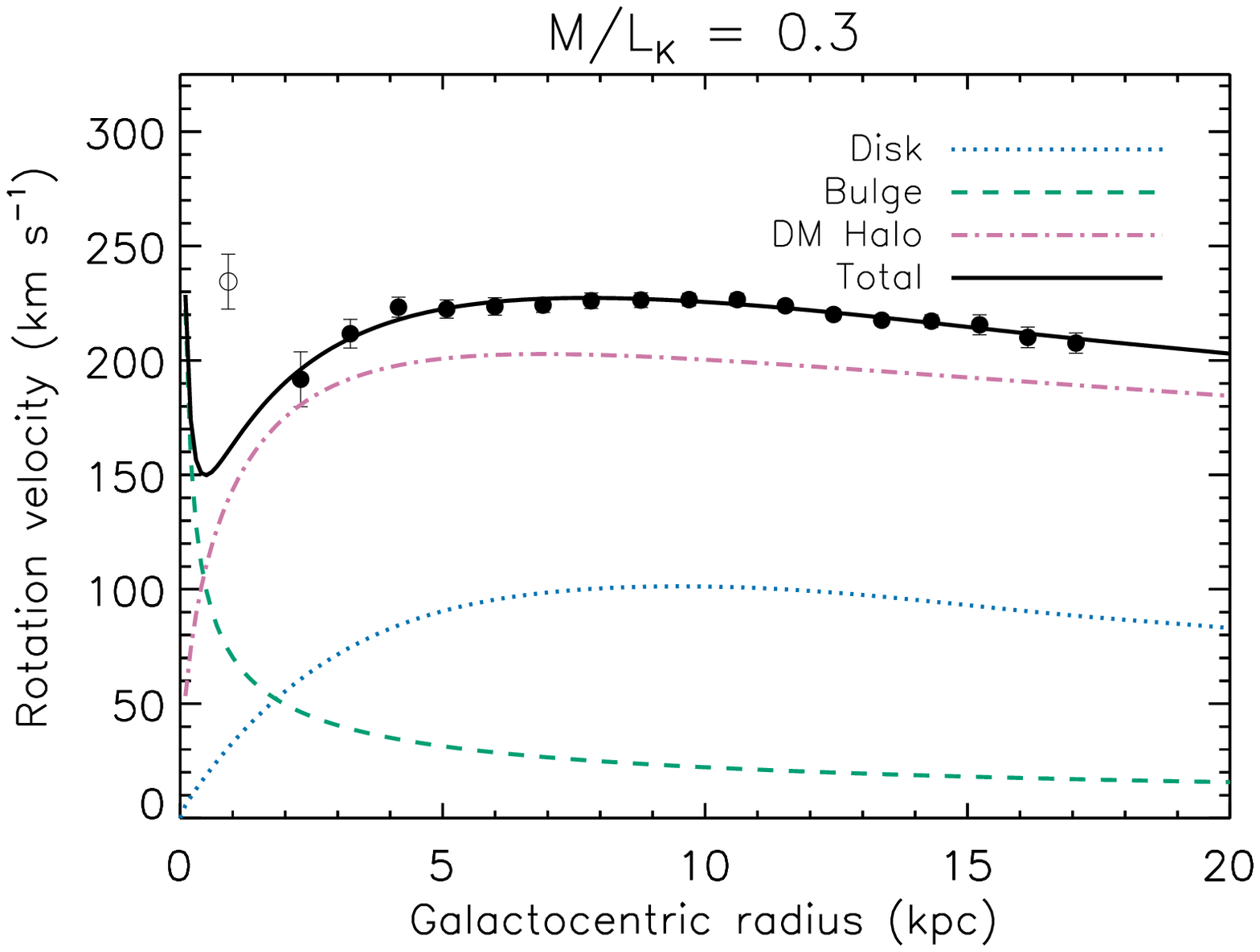}{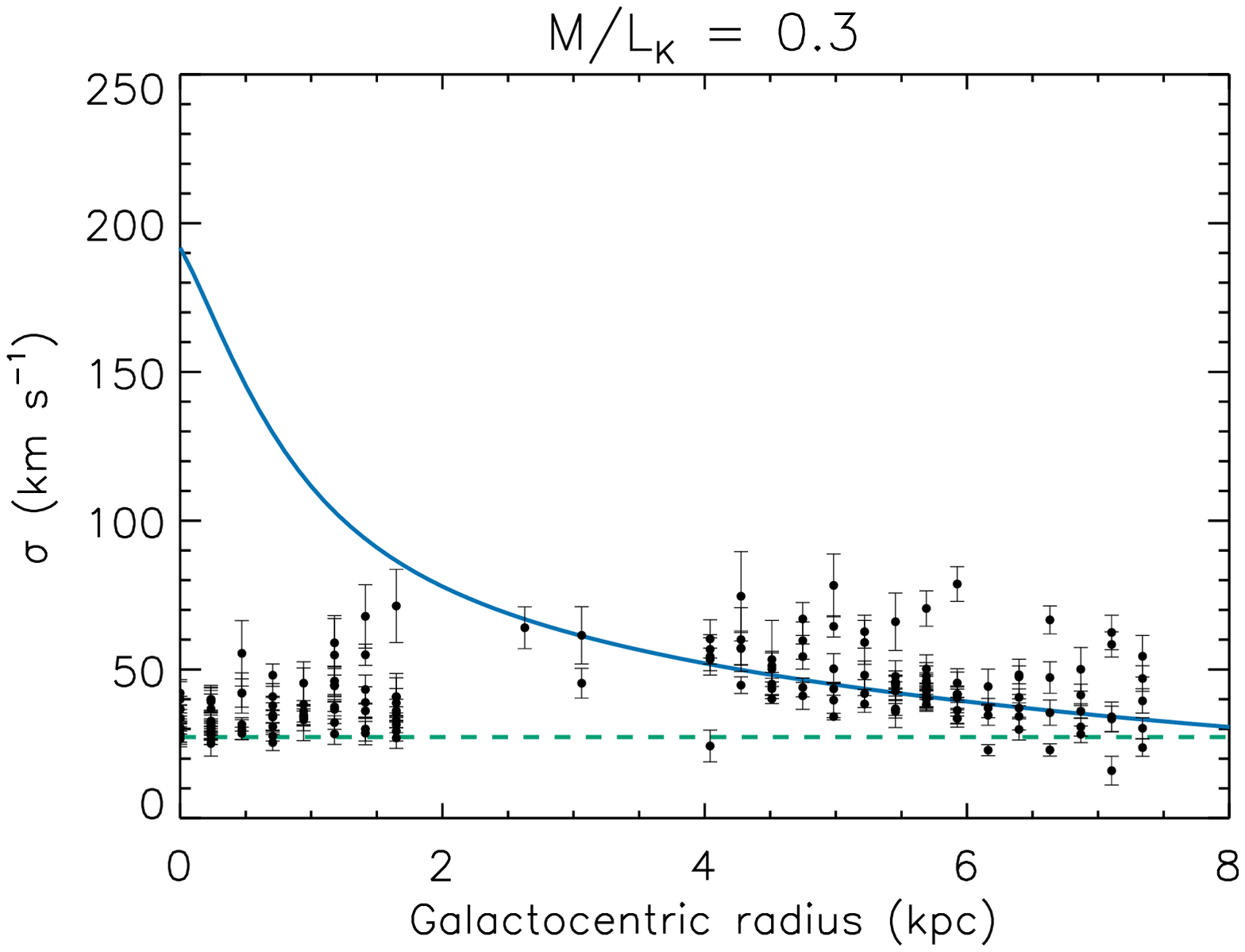}
\epsscale{1.0}\plottwo{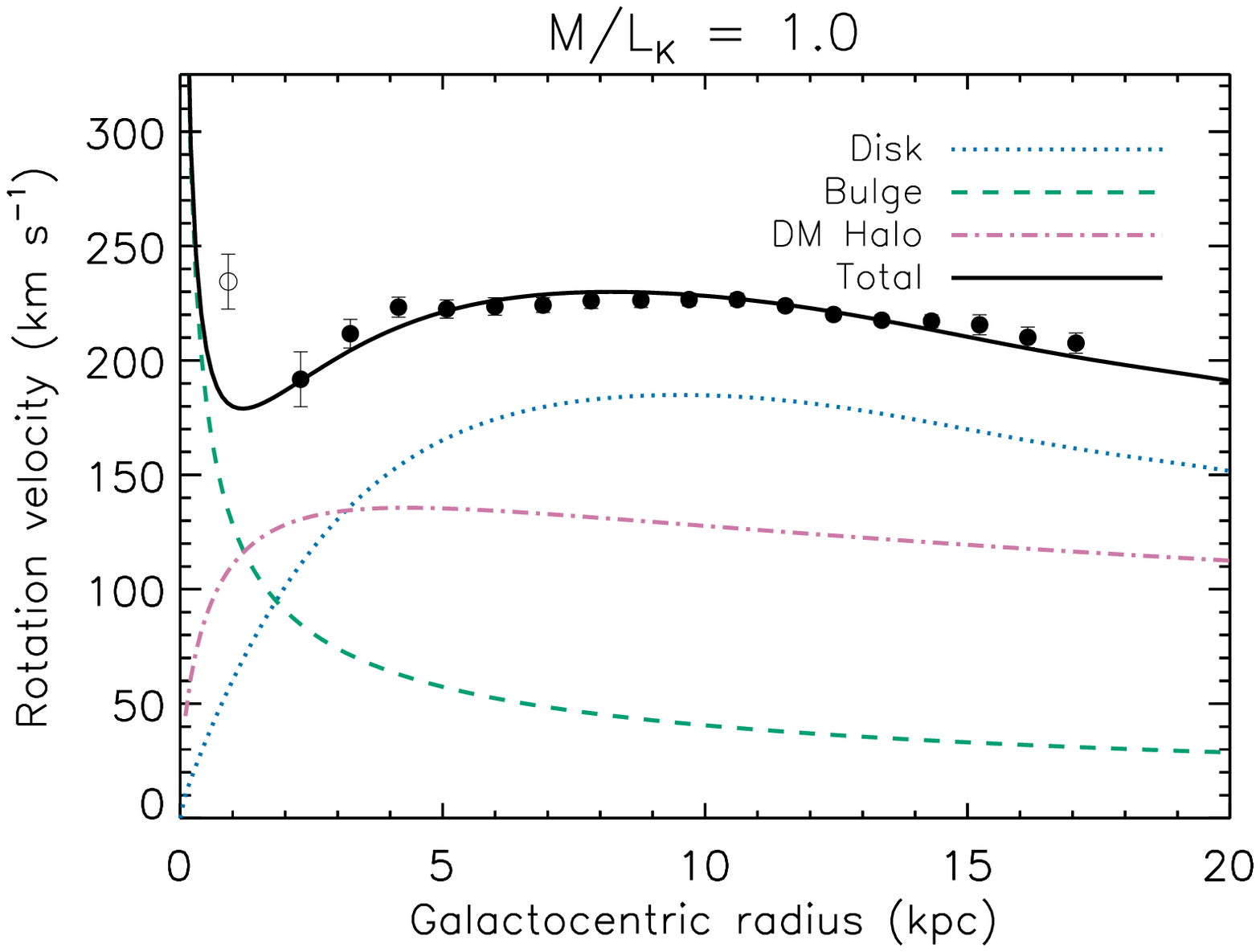}{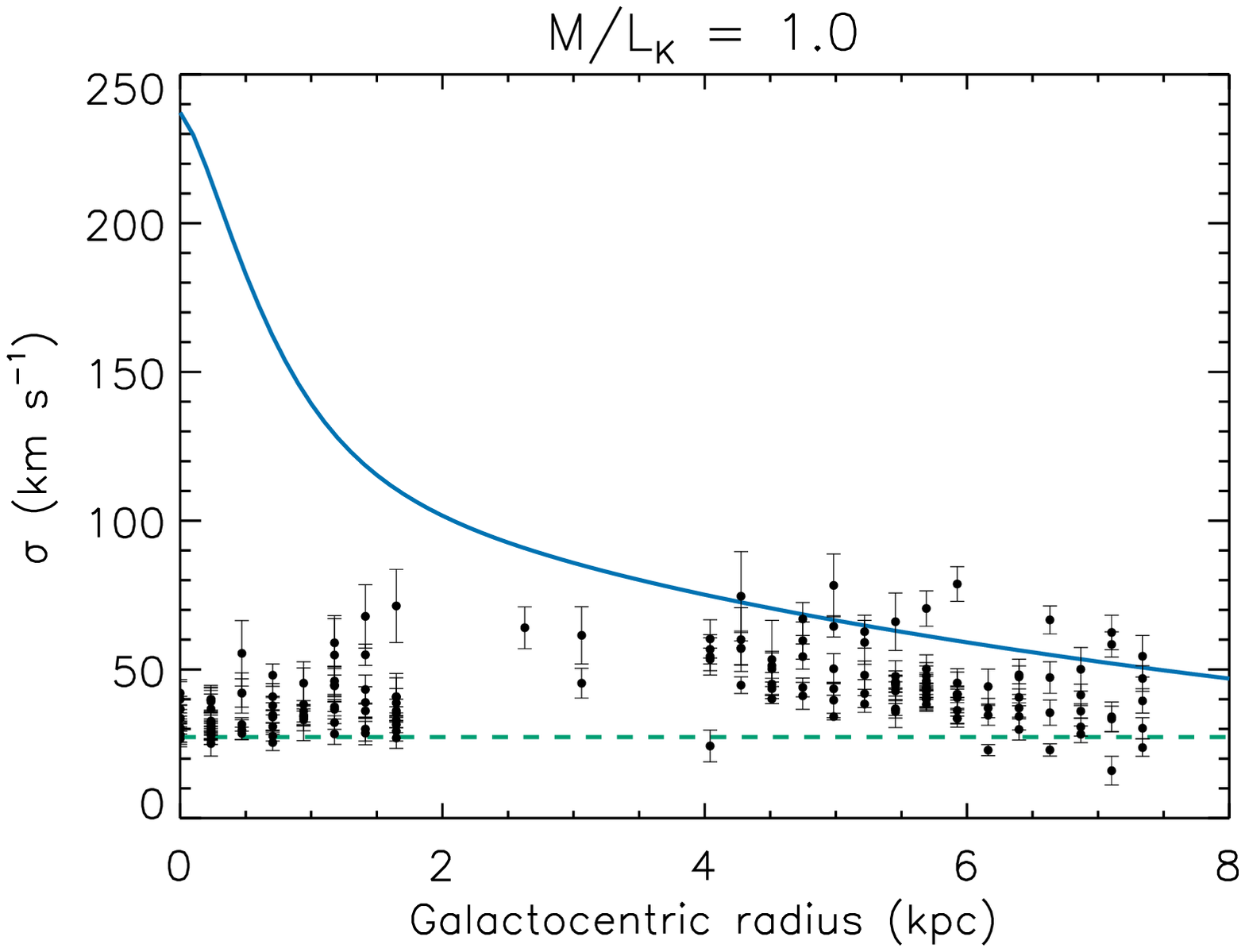}
\caption{We demonstrate that our results are not sensitive to our choice of
  $K$-band mass-to-light ratio by varying $M/L_{K}$ between $0 \le M/L_{K} \le
  1$. The rotation curves of these alternative mass models are shown on the
  left (as in Figure 1), and the observed and required velocity dispersions
  are shown on the right (as in Figure 9). For the unrealistic assumption of
  no baryonic mass ($M/L_{K} = 0$), the minimum galactocentric radius at which
  the dynamical equilibrium model is stably satisfied is $R_{eq} = 5$ kpc and
  $R_{eq} = 6$ kpc for the equipartition and non-equipartition cases,
  respectively. Additionally, this assumption is required to satisfy the model
  via thermal and turbulent pressure gradients alone within $R \le 8$ kpc. For
  $M/L_{K} = 0.3$ and $M/L_{K} = 1.0$, the dynamical equilibrium model is
  stably satisfied at $R_{eq} = 7$ kpc and $R_{eq} = 10$ kpc,
  respectively. Thus, the range of $R_{eq}$ that results from varying
  $M/L_{K}$ by $50\%$ is a few kpc, comparable to the range of $R_{eq}$ due to
  the uncertainty on the eDIG filling factor, magnetic field strength, and
  magnetic scale height (see Figure 11).}
\end{figure}

\begin{deluxetable}{lcccc}
\tabletypesize{\scriptsize}
\tablecaption{NGC 891 Secondary Mass Models}
\tablewidth{0pt}
\tablehead{ 
\colhead{$M/L_{K}$} &
\colhead{$a_{DM}$ (kpc)} &
\colhead{$\rho_{0,DM}$ (\Msun kpc$^{-3}$)}
}
\startdata
0.0 &  3.3 & $0.04 \times 10^{10}$\\
0.3 &  3.2 & $0.03 \times 10^{10}$\\
1.0 &  2.0 & $0.04 \times 10^{10}$
\tablecomments{The scale radius, $a_{DM}$, and the central density,
  $\rho_{0,DM}$, of the NFW dark matter model required to reproduce the HI
  rotation curve of \citet{Fraternali2011} are given for a range of $K$-band
  mass-to-light ratios, $M/L_{K}$.}
\enddata
\end{deluxetable}

\bibliographystyle{apj}

\end{document}